\documentclass[lettersize,conference]{IEEEtran}

\usepackage{amsmath,amsfonts}
\usepackage{algorithm}
\usepackage{array}
\usepackage{textcomp}
\usepackage{stfloats}
\usepackage{url}
\usepackage{verbatim}
\usepackage{graphicx}
\usepackage{cite}
\usepackage{amssymb}
\usepackage{paralist}
\usepackage{subcaption}
\usepackage{booktabs}
\usepackage[dvipsnames,table]{xcolor}
\usepackage{listings}
\lstdefinelanguage{myPython}{language = Python,
    deletekeywords=[2]{bool, input, len, list, range, set, type},
}
\lstdefinelanguage{aspectDSL}{language = Python,
    morekeywords = {traversal, fromTraversal, importAspect, sourceLabels, atPath, aspect, aspectType, utility, pointcut, mergeAspects, currentPoint, getAspect, enterLoop, getExprSymbs, getDescrSymbs, getExprDescrSymbs},
    deletekeywords=[2]{bool, input, len, list, range, set, type},
}
\lstdefinelanguage{myPython}{language = Python,
    deletekeywords=[2]{bool, input, len, list, range, set, type},
}

\definecolor{codewhite}{rgb}{1.0,1.0,1.0}
\lstdefinestyle{neutral}{
    backgroundcolor=\color{codewhite},
    frame=none,    
    numbers=none,
}

\definecolor{codegreen}{rgb}{0,0.6,0}
\definecolor{codegray}{rgb}{0.5,0.5,0.5}
\definecolor{codepurple}{rgb}{0.58,0,0.82}
\definecolor{backcolour}{rgb}{0.95,0.95,0.92}
\lstdefinestyle{mystyle}{
backgroundcolor=\color{backcolour},   
    commentstyle=\color{codegreen},
numberstyle=\tiny\color{black},
    stringstyle=\color{codepurple},
    basicstyle=\ttfamily\lst@ifdisplaystyle\footnotesize\fi,
    breakatwhitespace=false,         
    breaklines=true,                 
    captionpos=b,                    
    keepspaces=true,                 
    numbers=left,                    
showspaces=false,                
    showstringspaces=false,
    showtabs=false,                  
    tabsize=2,
    frame=single,
    framesep=4pt,
    boxpos=c
}
\usepackage{circledsteps}
\usepackage{algorithmicx}
\usepackage[noend]{algpseudocode}
    \algrenewcommand\alglinenumber[1]{\scriptsize #1:}\usepackage{multicol}
\usepackage{ifthen}\usepackage{xspace}\usepackage[numbers]{natbib}
\usepackage{siunitx}
\usepackage{tikz}
\usepackage{multirow}

\usepackage{hyperref}
\usepackage{cleveref}\crefname{subsection}{\S}{\S\S}
 \crefname{subsubsection}{\S}{\S\S}
 \crefname{algorithm}{algorithm}{algorithms}
\crefrangeformat{line}{lines~#3#1#4--#5#2#6}
\Crefrangeformat{line}{Lines~#3#1#4--#5#2#6}
 \newcommand{\cosmos}{COSMOS\xspace}

\newcommand{\aspectLg}{SABLE\xspace}
\newcommand{\aspectLgLong}{Static Aspect Building LanguagE\xspace}
\newcommand{\aspectTool}{SAGA\xspace}
\newcommand{\aspectToolLong}{Static Aspect General Analyzer\xspace}
\newcommand{\scfgPackage}{SCFG\xspace}
\newcommand{\scfgPython}{SCFGPython\xspace}

\newcommand{\github}{GitHub\xspace}

\newcommand{\java}{Java\xspace}
\newcommand{\python}{Python\xspace}
\newcommand{\bandit}{Bandit\xspace}
\newcommand{\semgrep}{Semgrep\xspace}
\newcommand{\pysa}{Pysa\xspace}
\newcommand{\joern}{Joern\xspace}
\newcommand{\joernScan}{\joern Scan\xspace}

\newcommand{\sourceTainting}{\specPredicate{Source tainting}}
\newcommand{\checkEndProc}{\specPredicate{Check endProc}}
\newcommand{\checkCalls}{\specPredicate{Check calls}}
\newcommand{\involvedSymbols}{\specPredicate{Involved symbols}}
\newcommand{\contextualValue}{\specPredicate{Contextual value}}

 \newcommand{\quotes}[1]{\text{``{#1}''}}
\newcommand{\code}[1]{\texttt{#1}}
\newcommand{\pyCode}[1]{\text{\lstinline[language=myPython]|#1|}}
\newcommand{\semPrim}[1]{\textsf{#1}}

\newcommand{\todoColorDefault}{Black}\newcommand{\todoColor}{\todoColorDefault}\newcommand{\todoSetColor}[1]{\ifthenelse{\equal{#1}{DB}}{
        \renewcommand{\todoColor}{Red}}{}\ifthenelse{\equal{#1}{JD}}{
        \renewcommand{\todoColor}{Violet}}{}\ifthenelse{\equal{#1}{YM}}{
        \renewcommand{\todoColor}{BurntOrange}}{}\ifthenelse{\equal{#1}{LB}}{
        \renewcommand{\todoColor}{Maroon}}{}}

\newcommand{\getIntToStringA}{wrongInput}
\newcommand{\setIntToStringA}[1]{\ifthenelse{\equal{#1}{1}}{\renewcommand{\getIntToStringA}{One}}{}\ifthenelse{\equal{#1}{2}}{\renewcommand{\getIntToStringA}{Two}}{}\ifthenelse{\equal{#1}{3}}{\renewcommand{\getIntToStringA}{Three}}{}\ifthenelse{\equal{#1}{4}}{\renewcommand{\getIntToStringA}{Four}}{}\ifthenelse{\equal{#1}{5}}{\renewcommand{\getIntToStringA}{Five}}{}\ifthenelse{\equal{#1}{6}}{\renewcommand{\getIntToStringA}{Six}}{}\ifthenelse{\equal{#1}{7}}{\renewcommand{\getIntToStringA}{Seven}}{}\ifthenelse{\equal{#1}{8}}{\renewcommand{\getIntToStringA}{Eight}}{}\ifthenelse{\equal{#1}{9}}{\renewcommand{\getIntToStringA}{Nine}}{}\ifthenelse{\equal{#1}{10}}{\renewcommand{\getIntToStringA}{Ten}}{}}
\newcommand{\getIntToStringB}{wrongInput}
\newcommand{\setIntToStringB}[1]{\ifthenelse{\equal{#1}{1}}{\renewcommand{\getIntToStringB}{One}}{}\ifthenelse{\equal{#1}{2}}{\renewcommand{\getIntToStringB}{Two}}{}\ifthenelse{\equal{#1}{3}}{\renewcommand{\getIntToStringB}{Three}}{}\ifthenelse{\equal{#1}{4}}{\renewcommand{\getIntToStringB}{Four}}{}\ifthenelse{\equal{#1}{5}}{\renewcommand{\getIntToStringB}{Five}}{}\ifthenelse{\equal{#1}{6}}{\renewcommand{\getIntToStringB}{Six}}{}\ifthenelse{\equal{#1}{7}}{\renewcommand{\getIntToStringB}{Seven}}{}\ifthenelse{\equal{#1}{8}}{\renewcommand{\getIntToStringB}{Eight}}{}\ifthenelse{\equal{#1}{9}}{\renewcommand{\getIntToStringB}{Nine}}{}\ifthenelse{\equal{#1}{10}}{\renewcommand{\getIntToStringB}{Ten}}{}}
\newcommand{\nameA}[2]{\begingroup\setIntToStringA{#2}\expandafter\endgroup
    \csname #1\getIntToStringA\endcsname
}
\newcommand{\nameAB}[3]{\begingroup\setIntToStringA{#2}\setIntToStringB{#3}\expandafter\endgroup
    \csname #1\getIntToStringA\getIntToStringB\endcsname
}

\newcommand{\D}[2]{\nameAB{D}{#1}{#2}}

\newcommand{\citeD}[2]{deliverable D#1.#2 ``\D{#1}{#2}''}

\DeclareMathOperator{\eqdef}{\overset{\text{\tiny def}}{=}}
\newcommand{\eqRule}[1]{\overset{\text{\tiny \Circled{#1}}}{=}}

\newcommand{\knowing}{\,|\,}
\newcommand{\set}[2]{\left\{{#1}\ifthenelse{\equal{\unexpanded{#2}}{}}{}{\knowing {#2}}\right\}}

\newcommand{\card}[1]{\left|{#1}\right|}
\newcommand{\abs}[1]{\left|{#1}\right|}

\DeclareMathOperator{\DomSymb}{Dom}
\newcommand{\Dom}[1]{\DomSymb\left({#1}\right)}
\newcommand{\tuple}[1]{\langle{#1}\rangle}
\newcommand{\dataList}[1]{\left[{#1}\right]}

\newcommand{\character}[1]{\text{`\texttt{#1}'}}
\newcommand{\bnfFunc}[1]{\character{(}\ {#1}\ \character{)}}
\DeclareMathOperator{\bnfEq}{\text{::=}}
\DeclareMathOperator{\bnfOr}{\code{|}}

\DeclareMathOperator{\bnfSymbStar}{\texttt{*}}

\newcommand{\bnfStmtQuestion}[1]{\left[{#1}\right]}
\newcommand{\bnfStmtStar}[1]{({#1})\!\bnfSymbStar}

\DeclareMathOperator{\algoTrue}{True}
\DeclareMathOperator{\algoFalse}{False}
 \newcommand{\srcSeqL}{\texttt{\{}}
\newcommand{\srcSeqR}{\texttt{\}}}
\newcommand{\srcSeqN}{\texttt{;}}
\newcommand{\srcSeqC}{\texttt{:}}
\newcommand{\srcSeq}[2]{{#1} \srcSeqN\, {#2}}

\newcommand{\srcName}{\mathit{name}}
\newcommand{\srcVar}{\mathit{var}}

\newcommand{\srcExpr}{\mathit{expr}}
\newcommand{\srcStmt}{\mathit{srcStmt}}

\newcommand{\srcAssign}{=}
\newcommand{\srcIf}{\code{if}}
\newcommand{\srcThen}{\code{then}}
\newcommand{\srcElse}{\code{else}}
\newcommand{\srcMatch}{\code{match}}
\newcommand{\srcCase}{\code{case}}
\newcommand{\srcWhile}{\code{while}}
\newcommand{\srcFor}{\code{for}}
\newcommand{\srcIn}{\code{in}}
\newcommand{\srcWith}{\code{with}}
\newcommand{\srcAs}{\code{as}}
\newcommand{\srcReturn}{\code{return}}
\newcommand{\srcPass}{\code{pass}}
\newcommand{\srcAsync}{\code{async}}
\newcommand{\srcDef}{\code{def}}
\newcommand{\srcImport}{\code{import}}
\newcommand{\srcFrom}{\code{from}}
\newcommand{\srcRaise}{\code{raise}}

\newcommand{\srcTry}{\code{try}}
\newcommand{\srcExcept}{\code{except}}
\newcommand{\srcFinally}{\code{finally}}
\newcommand{\srcContinue}{\code{continue}}
\newcommand{\srcBreak}{\code{break}}

\newcommand{\srcStmtAssign}[2]{{#1} \srcAssign {#2}}

\newcommand{\srcStmtCase}[2]{\srcCase\ {#1}\srcSeqC\, \srcSeqL{#2}\srcSeqR}
\newcommand{\srcStmtWhile}[2]{\srcWhile\ {#1}\srcSeqC\, \srcSeqL{#2}\srcSeqR}
\newcommand{\srcStmtFor}[3]{\srcFor\ {#1}\ \srcIn\ {#2}\srcSeqC\, \srcSeqL{#3}\srcSeqR}
\newcommand{\srcStmtReturn}[1]{\srcReturn\ {#1}}

\newcommand{\srcStmtImport}[1]{\srcImport\ {#1}}
\newcommand{\srcStmtFrom}[2]{\srcFrom\ {#1}\ \srcImport\ {#2}}
\newcommand{\srcStmtRaise}[1]{\srcRaise\ {#1}}
\newcommand{\srcStmtTry}[1]{\srcTry\srcSeqC\, \srcSeqL{#1}\srcSeqR}
\newcommand{\srcStmtElse}[1]{\srcElse\srcSeqC\, \srcSeqL{#1}\srcSeqR}
\newcommand{\srcStmtExcept}[2]{\srcExcept\ {#1}\srcSeqC\, \srcSeqL{#2}\srcSeqR}
\newcommand{\srcStmtFinally}[1]{\srcFinally\srcSeqC\, \srcSeqL{#1}\srcSeqR}

\newcommand{\srcStmtAs}[1]{\srcAs\ {#1}}

\newcommand{\progLine}{\ell}
\newcommand{\progLineStart}[1]{\progLine_{\mathit{start}}}
\newcommand{\progLineEnd}[1]{\progLine_{\mathit{end}}}
\newcommand{\symbState}{\sigma}
\newcommand{\symbStateStart}{\symbState_{\mathit{start}}}
\newcommand{\symbStateEnd}{\symbState_{\mathit{end}}}

\newcommand{\symbStateScope}{\symbState_{\mathit{loop}}}
\newcommand{\symbStateStmt}{\symbState_{\mathit{stmt}}}
\newcommand{\symbStateEndStmt}{\symbState_{\mathit{endStmt}}}
\newcommand{\symbStateNext}{\symbState_{\mathit{next}}}

\newcommand{\symbStateExcept}{\symbState^{\mathit{except}}}
\newcommand{\symbStateFinally}{\symbState^{\mathit{finally}}}
\newcommand{\symbStateCase}{\symbState^{\mathit{case}}}
\newcommand{\symbStateEndExcept}{\symbState_{\mathit{endExcept}}}
\newcommand{\progStmt}{\code{stmt}}
\DeclareMathOperator{\progPtName}{\ptProgName}
\newcommand{\progPt}[1]{\progPtName\left({#1}\right)}
\DeclareMathOperator{\progStmtLabelName}{\ptStmtName}
\newcommand{\progStmtLabel}[1]{\progStmtLabelName\left({#1}\right)}
\DeclareMathOperator{\progExprsName}{\ptExprsName}
\newcommand{\progExprs}[1]{\progExprsName\left({#1}\right)}
\DeclareMathOperator{\progLocName}{\ptLocName}
\newcommand{\progLoc}[1]{\progLocName\left({#1}\right)}
\DeclareMathOperator{\progLocEndName}{\ptEndLocName}
\newcommand{\progLocEnd}[1]{\progLocEndName\left({#1}\right)}
\DeclareMathOperator{\progSymbStateName}{symbState}
\newcommand{\progSymbState}[1]{\progSymbStateName\left({#1}\right)}
\DeclareMathOperator{\progSymbStatesName}{SymbStates}
\newcommand{\progSymbStates}[1]{\progSymbStatesName\left({#1}\right)}
\DeclareMathOperator{\edgesRecStateName}{recState}
\newcommand{\edgesRecState}[1]{\edgesRecStateName\left({#1}\right)}

\newcommand{\procedure}{p}
\DeclareMathOperator{\scfgName}{SCFG}
\newcommand{\scfg}[1]{\scfgName\left({#1}\right)}
\DeclareMathOperator{\procStmtsSeqName}{seqStmts}
\newcommand{\procStmtsSeq}[1]{\procStmtsSeqName\left({#1}\right)}
\DeclareMathOperator{\stmtsName}{Stmts}
\newcommand{\stmts}[1]{\stmtsName\left({#1}\right)}
\DeclareMathOperator{\procInputsName}{Inputs}
\newcommand{\procInputs}[1]{\procInputsName\left({#1}\right)}
\DeclareMathOperator{\verticesName}{V}
\newcommand{\vertices}[1]{\verticesName\left({#1}\right)}
\DeclareMathOperator{\progEndStateName}{endState}
\newcommand{\progEndState}[1]{\progEndStateName\left({#1}\right)}
\DeclareMathOperator{\edgesName}{E}
\newcommand{\edges}[1]{\edgesName\left({#1}\right)}
\DeclareMathOperator{\edgesRecName}{Edges}
\newcommand{\edgesRec}[3]{\edgesRecName\left({#1}, {#2}, {#3}\right)}
\newcommand{\stmtLabel}[1]{\quotes{#1}\xspace}
\newcommand{\labelEnterProcedure}{\stmtLabel{EnterProcedure}}
\newcommand{\labelExitProcedure}{\stmtLabel{ExitProcedure}}
\newcommand{\labelEnterContainer}{\stmtLabel{EnterContainer}}
\newcommand{\labelExitContainer}{\stmtLabel{ExitContainer}}
\newcommand{\labelExp}{\stmtLabel{Exp}}
\newcommand{\labelAssign}{\stmtLabel{Assign}}
\newcommand{\labelIf}{\stmtLabel{If}}

\newcommand{\labelElse}{\stmtLabel{Else}}
\newcommand{\labelMatch}{\stmtLabel{Match}}
\newcommand{\labelEndIf}{\stmtLabel{EndIf}}

\newcommand{\labelWhile}{\stmtLabel{While}}
\newcommand{\labelEndWhile}{\stmtLabel{EndWhile}}
\newcommand{\labelFor}{\stmtLabel{For}}
\newcommand{\labelEndFor}{\stmtLabel{EndFor}}
\newcommand{\labelReturn}{\stmtLabel{Return}}
\newcommand{\labelRaise}{\stmtLabel{Raise}}
\newcommand{\labelWith}{\stmtLabel{With}}

\newcommand{\labelTry}{\stmtLabel{Try}}
\newcommand{\labelEnd}{\stmtLabel{End}}
\DeclareMathOperator{\BranchLabels}{BranchLabels}
 \newcommand{\specPredicate}[1]{\textit{\small\textsf{#1}}\xspace}

\newcommand{\aspectSensitive}{\specPredicate{Sensitive}}

\newcommand{\aspectScopeSensitiveBranches}{\specPredicate{ScopeSensitiveBranches}}
\newcommand{\aspectSensitBranching}{\specPredicate{SensitiveBranching}}
\newcommand{\aspectConfidentialityViolation}{\specPredicate{ConfidentialityViolation}}
\newcommand{\aspectSinks}{\specPredicate{Sinks}}

\newcommand{\pyLiteral}{\mathit{pyLit}}
\newcommand{\pyStmt}{\mathit{pyStmt}}
\newcommand{\pyProg}{\mathit{pyProg}}

\newcommand{\aspectVar}{\mathit{var}}
\newcommand{\aspectPath}{\mathit{path}}
\newcommand{\aspectType}{\mathit{pyType}}
\newcommand{\aspectStmtLabel}{\mathit{stmtLabel}}
\newcommand{\aspectTraversal}{\code{traversal}}
\newcommand{\aspectUtility}{\code{utility}}
\newcommand{\aspectSrcLabels}{\code{sourceAnnotation}}

\newcommand{\aspectFromTraversal}{\code{fromTraversal}}
\newcommand{\aspectImportAspect}{\code{importAspect}}
\newcommand{\aspectAspect}{\code{aspect}}
\newcommand{\aspectAspectType}{\code{aspectType}}
\newcommand{\aspectPointcut}{\code{pointcut}}
\newcommand{\aspectMergeAspects}{\code{mergeAspects}}
\newcommand{\aspectStmtTrav}[2]{\aspectTraversal\ {#1}: \left\{{#2}\right\}}
\newcommand{\aspectStmtUtility}[1]{\aspectUtility: \left\{{#1}\right\}}
\newcommand{\aspectStmtSrcLabels}[2]{\aspectSrcLabels\ {#1}}

\newcommand{\aspectStmtAspect}[2]{\aspectAspect\ {#1}\ \aspectAspectType\ {#2}}

\newcommand{\aspectStmt}{\mathit{sableStmt}}

\newcommand{\aspectTypeTrav}{\mathit{sableTrav}}

\newcommand{\aspectTrigger}{\code{triggerFrom}}
\newcommand{\aspectTriggerVal}{\code{atValue}}
\newcommand{\aspectTriggerAt}[2]{\aspectTrigger\ {#1}\ \aspectTriggerVal\ {#2}}

\newcommand{\aspectLabel}[1]{\text{`{#1}'}}
\newcommand{\aspectRead}{\aspectLabel{use}}
\newcommand{\aspectWritten}{\aspectLabel{def}}
\newcommand{\aspectCalled}{\aspectLabel{call}}
\newcommand{\aspectAll}{\aspectLabel{all}}

\newcommand{\aspectCurrentPoint}{\code{currentPoint}}
\newcommand{\aspectGetAspect}[2]{\code{getAspect}\left({#1}, {#2}\right)}
\newcommand{\aspectEnterLoop}[1]{\code{enterLoop}\left({#1}\right)}
\newcommand{\aspectGetExprSymbs}[2]{\code{getExprSymbs}\left({#1}, {#2}\right)}
\newcommand{\aspectGetDescrSymbs}[2]{\code{getDescrSymbs}\left({#1}, {#2}\right)}

\newcommand{\ptTravName}{\mathit{travName}}
\newcommand{\ptAspectName}{\mathit{aspectName}}
\newcommand{\ptSrcLabelsName}{\mathit{srcAnnot}}
\newcommand{\ptStmtLabel}{\mathit{stmtLabel}}
\newcommand{\ptSymbLabel}{\mathit{symbLabel}}

\DeclareMathOperator{\ptProgName}{\semPrim{point}}
\newcommand{\ptProg}[1]{\ptProgName\left({#1}\right)}
\DeclareMathOperator{\ptStmtName}{\semPrim{label}}
\newcommand{\ptStmt}[1]{\ptStmtName\left({#1}\right)}
\DeclareMathOperator{\ptExprsName}{\semPrim{exprs}}
\newcommand{\ptExprs}[1]{\ptExprsName\left({#1}\right)}
\DeclareMathOperator{\ptLocName}{\semPrim{location}}
\newcommand{\ptLoc}[1]{\ptLocName\left({#1}\right)}
\DeclareMathOperator{\ptEndLocName}{\semPrim{endLoc}}
\newcommand{\ptEndLoc}[1]{\ptEndLocName\left({#1}\right)}
\DeclareMathOperator{\ptAnnotationName}{\semPrim{annotation}}
\newcommand{\ptAnnotation}[1]{\ptAnnotationName\left({#1}\right)}
\DeclareMathOperator{\ptEnterLoopName}{\semPrim{enterLoop}}
\newcommand{\ptEnterLoop}[1]{\ptEnterLoopName\left({#1}\right)}
\DeclareMathOperator{\ptNextsName}{\semPrim{Children}}
\newcommand{\ptNexts}[1]{\ptNextsName\left({#1}\right)}
\newcommand{\ptExit}[1]{\progEndState{#1}}

\DeclareMathOperator{\getAdviceName}{\semPrim{advice}}
\newcommand{\getAdvice}[1]{\getAdviceName\left({#1}\right)}

\newcommand{\srcSymbs}{\mathit{srcSymbs}}

\newcommand{\semStep}{\mathit{step}}
\DeclareMathOperator{\getTriggersName}{\semPrim{Triggers}}
\newcommand{\getTriggers}[1]{\getTriggersName\left({#1}\right)}
\DeclareMathOperator{\getAlarmsName}{\mathit{alarms}}

\DeclareMathOperator{\elementName}{\semPrim{elem}}
\newcommand{\element}[1]{\elementName\left({#1}\right)}

\DeclareMathOperator{\ptAnnotateName}{\semPrim{annotate}}
\newcommand{\ptAnnotate}[2]{\ptAnnotateName\left({#1}, {#2}\right)}
\DeclareMathOperator{\ptUpdEnterLoopName}{\semPrim{updEnterLoop}}
\newcommand{\ptUpdEnterLoop}[2]{\ptUpdEnterLoopName\left({#1}, {#2}\right)}
\DeclareMathOperator{\exeAdviceName}{exeAdvice}
\newcommand{\exeAdvice}[4]{\exeAdviceName\left({#1}, {#2}, {#3}, {#4}\right)}
\newcommand{\argAdvice}{\mathit{advice}}
\newcommand{\argExprs}{\mathit{exprs}}

\newcommand{\pyNone}{\mathit{None}}

\newcommand{\semTravMap}{\mathit{travMap}}
\DeclareMathOperator{\semMergeName}{mergeAspects}
\newcommand{\semMerge}[2]{\semMergeName\left({#1}, {#2}\right)}

\newcommand{\semStartPt}{\symbStateStart}
\newcommand{\semEndPt}{\symbStateEnd}

\newcommand{\semExitPt}{\symbState_{\mathit{join}}}

\DeclareMathOperator{\semVisitName}{\semPrim{visit}}
\newcommand{\semVisit}[3]{\semVisitName\left({#1}, {#2}, {#3}\right)}
\DeclareMathOperator{\semWeaveName}{\semPrim{weave}}
\newcommand{\semWeave}[2]{\semWeaveName\left({#1}, {#2}\right)}
\newcommand{\semOldAnnot}{\mathit{oldAnnot}}
\DeclareMathOperator{\semTransitionName}{\semPrim{transition}}
\newcommand{\semTransition}[4]{\semTransitionName\left({#1}, {#2}, {#3}, {#4}\right)}
\DeclareMathOperator{\semBranchingName}{\semPrim{branch}}
\newcommand{\semBranching}[4]{\semBranchingName\left({#1}, {#2}, {#3}, {#4}\right)}
\DeclareMathOperator{\semLoopBranchingName}{\semPrim{loopBranch}}
\newcommand{\semLoopBranching}[4]{\semLoopBranchingName\left({#1}, {#2}, {#3}, {#4}\right)}
\DeclareMathOperator{\semIsFixpointName}{\semPrim{isFixpoint}}
\newcommand{\semIsFixpoint}[2]{\semIsFixpointName\left({#1}, {#2}\right)}
\DeclareMathOperator{\semCondBranchingName}{\semPrim{condBranch}}
\newcommand{\semCondBranching}[3]{\semCondBranchingName\left({#1}, {#2}, {#3}\right)}
\newcommand{\semBranchPt}{\symbState_\mathit{branch}}
\newcommand{\semBranchMap}{\mathit{branchMap}}
\newcommand{\semBranches}{\mathit{BranchResults}}
\DeclareMathOperator{\semMergeMapsName}{\semPrim{mergeMaps}}
\newcommand{\semMergeMaps}[1]{\semMergeMapsName\left({#1}\right)}
\newcommand{\semTravMaps}{\mathit{TravMaps}}

\newcommand{\semTrav}{\mathit{trav}}
\newcommand{\semTravImport}{\mathit{travImport}}
\newcommand{\semTraversals}{\mathit{Traversals}}
\newcommand{\semDependencies}[1]{\mathit{Dependencies}\left({#1}\right)}
\newcommand{\semRemainingTraversals}{\mathit{RemTravs}}
\newcommand{\semOrderedTraversals}{\mathit{OrdTravs}}
\newcommand{\semValidTraversals}{\mathit{ValidTravs}}
\DeclareMathOperator{\setToListName}{list}
\newcommand{\setToList}[1]{\setToListName\left({#1}\right)}

\newcommand{\travSensitive}{\pyCode{travSensitive}}
\newcommand{\travConfidentiality}{\pyCode{travConfidentiality}}

 \newcommand{\vuln}{v}
\newcommand{\Vulns}{\textit{Vulns}}
\newcommand{\truthBefore}{\textit{truthBefore}}
\newcommand{\detectBefore}{\textit{detectBefore}}
\newcommand{\truthAfter}{\textit{truthAfter}}
\newcommand{\detectAfter}{\textit{detectAfter}}
\newcommand{\TP}{\textit{TP}}
\newcommand{\FN}{\textit{FN}}
\newcommand{\TN}{\textit{TN}}
\newcommand{\FP}{\textit{FP}}
\newcommand{\locStrict}{\textit{\,strict}}
\newcommand{\locRelax}{\textit{\,relax}}
\DeclareMathOperator{\precisionName}{\text{precision}}
\newcommand{\precision}[2]{\precisionName({#1}, {#2})}
\DeclareMathOperator{\specificityName}{\text{specificity}}
\newcommand{\specificity}[2]{\specificityName({#1}, {#2})}
\DeclareMathOperator{\sensitivityName}{\text{sensitivity}}
\newcommand{\sensitivity}[2]{\sensitivityName({#1}, {#2})}

\newcommand{\Fone}{F_1}

\newcommand{\metricRandomVar}{M}

\newcommand{\pValue}{p}
\newcommand{\effectSize}{E}
\newcommand{\tPlus}{R^{+}}
\newcommand{\tMinus}{R^{-}}
\DeclareMathOperator{\rankName}{\text{rank}}
\newcommand{\rank}[1]{\rankName({#1})}

\newcommand{\symbProba}{\text{P}}
\newcommand{\distrib}[1]{\ifthenelse{\equal{\unexpanded{#1}}{}}{\symbProba}{\symbProba_{#1}}}

\begin{document}

\title{SAGA: Detecting Security Vulnerabilities Using Static Aspect Analysis}

\author{Yoann~Marquer,
    Domenico~Bianculli,~\IEEEmembership{Member,~IEEE,}
    and~Lionel~C.~Briand,~\IEEEmembership{Fellow,~IEEE}\IEEEcompsocitemizethanks{\IEEEcompsocthanksitem Y. Marquer and D. Bianculli are with the Interdisciplinary Centre for Security, Reliability, and Trust (SnT) of the University of Luxembourg, Luxembourg, and L. Briand is with the School of Electrical and Computer Engineering of University of Ottawa, Canada, and with Lero SFI Centre for Software Research and University of Limerick, Ireland. E-mails: yoann.marquer@uni.lu, domenico.bianculli@uni.lu, lbriand@uottawa.ca}\thanks{Manuscript received \textbf{Month DD}, 2026; revised \textbf{Month DD}, 2026.}
}

\markboth{IEEE Transactions on Software Engineering,~Vol.~\textbf{XX}, No.~\textbf{XX}, \textbf{Month}~2026}{Marquer \MakeLowercase{\textit{et al.}}, SAGA: Detecting Security Vulnerabilities Using Static Aspect Analysis}

\maketitle

\begin{abstract}
Python is one of the most popular programming languages; as such, projects written in Python involve an increasing number of diverse security vulnerabilities.
However, existing state-of-the-art analysis tools for Python only support a few vulnerability types.
Hence, there is a need to detect a large variety of vulnerabilities in Python projects.
 
In this paper, we propose the SAGA approach for detecting and locating vulnerabilities in Python source code in a versatile way.
SAGA includes a source code parser that extracts control- and data-flow information and represents it as a symbolic control-flow graph, as well as a domain-specific language that defines static aspects of the source code and their evolution during graph traversals.
We have leveraged this language to define a library of static aspects for integrity, confidentiality, and other security-related properties. 
 
We have evaluated SAGA on a dataset of 108 vulnerabilities affecting 82 PyPi packages and covering 37 vulnerability types, obtaining 100\% sensitivity and 99.15\% specificity, with only one false positive, while outperforming four common security analysis tools.
This analysis was performed in less than 31 seconds, i.e., between 2.5 and 512.1 times faster than the baseline tools.
\end{abstract}


\section{Introduction}
\label{sec:introduction}

\label{sec:context}

\IEEEPARstart{P}{ython}
is the most popular programming language~\cite{IEEEspectrum}, being used by half of developers
on StackOverflow~\cite{StackOverflow}.
As the language grows in popularity, the number of security vulnerabilities in Python packages increases.
According to an empirical study of 1396 vulnerability reports from 2006 to 2021 affecting 698 PyPi packages, the number of security vulnerabilities in Python packages has increased over time~\cite{ACS23}.
Only 51.79\% of such vulnerabilities belong to the ten most common types.

Unfortunately, most analysis tools  for \python support only a single vulnerability
type, reducing the scope of their analysis.
For instance, some tools~\cite{ZZ13,HDM14,LK22} focus only on \emph{integrity} violations~\cite{Bib77,CW87,CMA17}, and not on \emph{confidentiality}~\cite{BL76,CMA17}).
Even tools that detect both integrity and confidentiality
vulnerabilities~\cite{TPC+13,TSBB18,FSBS21} only capture insecure
information flow from sources to sinks.
Such a restriction prevents them from detecting other vulnerability
types, e.g.,  side-channel attacks~\cite{Koc96,KJJ99,BCO04} do
not need a sink to leak sensitive information.

In this context, security engineers are required to use multiple tools
to detect the most common vulnerability types; however, configuring
several tools may be inconvenient.  Furthermore, in the case of
zero-day attacks~\cite{Rou21}, dedicated tools may not be available at
all.

The above situation warrants an analysis tool which is \emph{versatile
  enough} to detect many vulnerability types.  While automatically
detecting all possible security vulnerabilities using a
one-size-fits-all static analyzer seems out of reach, we propose the
\aspectToolLong (\aspectTool) approach, where a single tool can detect
many vulnerability types, \emph{given that each considered
  vulnerability is properly defined}.  \aspectTool allows a user to
precisely define, using the expressive power of a general-purpose
programming language like \python, the information flows that are
involved in a vulnerability.  More specifically, \aspectTool supports
the definition of \emph{static aspects}, i.e., aspects\footnote{Our concept of ``aspect'' is loosely inspired by the homonymous concept
  in aspect-oriented programming~\cite{KHH+01}.} of the
source code that may be independent from the business logic of the
program and that can be determined during static analysis.
Static aspects can be used, for example, to track a sequence of function calls or variables tainted by untrusted inputs. \aspectTool includes  \aspectLg (\aspectLgLong
        \footnote{In the public \citeD{4}{7} of the H2020 \cosmos
          European project (\url{https://www.cosmos-devops.org/results}, \aspectLg stood for
          Static-Aspect-Based Language Extension, the name being
          changed (but not the acronym) for this paper to better reflect \aspectLg as a
          standalone language.}),
        a domain-specific language enabling developers to express static analyses of
        source code, based on control- and data-flow information; it is embedded in \python to provide high expressive power.
We demonstrate the expressiveness of \aspectLg by providing a  library of  static aspect definitions for integrity, confidentiality, and other security properties, which developers or security experts can use on
their \python projects.

We aim to improve security vulnerability detection, compared to the state of the art, by
1) maximizing the number of detected vulnerabilities,
2) minimizing the number of false alarms, and
3) minimizing analysis time.
We have evaluated the effectiveness of \aspectTool in terms of vulnerability detection on a large (108 vulnerabilities) and diverse (37 vulnerability types) dataset, in comparison with four baseline tools, namely \bandit~\cite{Bandit}, \semgrep~\cite{SemGrep}, \pysa~\cite{Pysa}, and \joern~\cite{Joern}.
\aspectTool obtained 100\% sensitivity and 99.15\% specificity, with only one false positive, outperforming the baseline tools, which achieved up to 59.66\% sensitivity and 95.87\% specificity.
This analysis was performed in less than 31 seconds, i.e., between 2.5 and 512.1 times faster than the baseline tools.

To sum up, this work makes the following contributions:
\begin{itemize}
    \item  \aspectTool, a \emph{versatile, security-oriented static analyzer}, allowing a security expert to modularly define vulnerability types to be detected across multiple projects or to tailor the analysis to fit a particular vulnerability.
    \item \aspectLg,
        a domain-specific language to help developers express static analyses of source code, based on control- and data-flow information.
    \item A library of five static aspect definitions written in \aspectLg, expressing common security properties and provided in Appendix~\ref{sec:sable:sadLibrary} as supplementary material.
\item One of the first~\cite{ZWL+26} large, manually-verified, dataset of security vulnerabilities in \python projects. It comprises 108 vulnerabilities and ground-truth locations across 82 PyPI packages~\cite{ACS23} and covers 37 vulnerability types~\cite{Cwe}.
    \item An empirical evaluation of the effectiveness of \aspectTool in terms of vulnerability detection, in comparison with four state-of-the-art baseline tools. \end{itemize}

\label{sec:organization}

The rest of the paper is structured as follows.
We introduce the background information necessary to understand our approach
in \Cref{sec:background} and a running example in \Cref{sec:runningExample}.
\Cref{sec:overview} provides an overview of \aspectTool. In \Cref{sec:sourceCode}, we describe how data- and control-flow
information is captured by \aspectTool.
We illustrate the \aspectLg language in \Cref{sec:sable}.
We report on the empirical evaluation of the \aspectTool prototype implementation in
\Cref{sec:evaluation}.
We discuss our results in \Cref{sec:discussion} and the related work in \Cref{sec:related}, then draw some conclusions in \Cref{sec:conclusion}.

 \section{Background}
\label{sec:background}

In this section, we introduce the main background concepts required to define and illustrate our approach.
First, we give an overview of
security information policies, focusing on injection and side-channel
attacks (\Cref{sec:background:security}), because injection attacks provide widespread examples of source-to-sink vulnerable information flows and side-channel attacks provide examples of information leakages without sinks.
Then, we provide a summary of the main concepts related to
aspect-oriented programming (\Cref{sec:background:aop}). 
Finally, we describe a program representation based on symbolic control-flow graphs (\Cref{sec:background:scfg}).

\subsection{Security}
\label{sec:background:security}

\subsubsection{Information Flows}
\label{sec:background:security:infoFlow}

The \emph{confidentiality} information policy consists in preventing sensitive data from being leaked to untrusted users~\cite{BL76,CMA17},
while the \emph{integrity} information policy consists in preventing untrusted users from tampering with sensitive data~\cite{Bib77,CW87,CMA17}.
To detect confidentiality or integrity violations, security experts and tools track control- and data-flow information.

\begin{figure}[t]
\centering
\hfill
\begin{subfigure}{0.26\linewidth}
\begin{lstlisting}[language=Python,escapechar=@,basicstyle={\scriptsize \ttfamily},numbers=none]
x = secret + 42
broadcast(x)
\end{lstlisting}
\caption{Direct information flow}
\label{fig:code:infoFlowDirect}
\end{subfigure}\hfill
\begin{subfigure}{0.29\linewidth}
\begin{lstlisting}[language=Python,escapechar=@,basicstyle={\scriptsize \ttfamily},numbers=none]
if secret < 6.28:
    x = True
else:
    x = False
broadcast(x)
\end{lstlisting}
\caption{Indirect information flow}
\label{fig:code:infoFlowIndirect}
\end{subfigure}\hfill
\begin{subfigure}{0.29\linewidth}
\begin{lstlisting}[language=Python,escapechar=@,basicstyle={\scriptsize \ttfamily},numbers=none]
y = setupRequest(untrustedInput)@\label{line:tampered:directFlow}@
accessDataBase(y)@\label{line:tampered:database}@
\end{lstlisting}
\caption{Data\\tampering}
\label{fig:code:tampered}
\end{subfigure}\hfill
\hfill
\caption{Examples of information flows.}
\label{fig:code:infoFlow}
\end{figure}

A simple example of an information flow is provided in \Cref{fig:code:infoFlow}.
The code on the left demonstrates an example of \emph{direct information flow}, where the new value of the variable \pyCode{x} directly depends on the secret information stored in the variable \pyCode{secret}.
The code in the center demonstrates an example of \emph{indirect information flow}, where the value of a variable \pyCode{x} is determined by an instance of conditional branching depending on secret information stored in variable \pyCode{secret}.
More generally, a variable depending (through direct or indirect
information flow) on secret information is considered \emph{sensitive}.
In both examples, because variable \pyCode{secret} contains secret information, variable \pyCode{x} is sensitive.
In the same way, a variable (directly or indirectly) depending on untrusted data
has been \emph{tampered} with.
\Cref{fig:code:tampered} also provides a simple example where variable \pyCode{untrustedInput} has tampered with \pyCode{y} through a direct information flow (\Cref{line:tampered:directFlow}).

\subsubsection{Injection Attacks}
\label{sec:background:security:injection}

\emph{Sources} (resp. \emph{sinks)} are security-sensitive operations from (resp. to) which information can flow.
In the context of confidentiality, a source is usually a method that injects a secret input into a program, while a sink exposes information to public observers.
In the context of integrity, a source is typically a mechanism that injects untrusted input into a program, while a sink performs security-sensitive computations~\cite{TPC+13}.

An \emph{injection attack} involves an attacker supplying such an untrusted input to a program. 
This input is processed as part of a command or query, and that processing can alter the program's execution.
Injection attacks include cross-site scripting (XSS), SQL injection (SQLi), XML injection (XMLi), XPath injection (XPathi), and LDAP injection (LDAPi) attacks~\cite{FPBL13,TPC+13,ZZ13,HDM14,TSBB18,TSBB20}.
According to the CWE team, the two most common and impactful software weaknesses in 2025 were injection vulnerabilities\footnote{\url{https://cwe.mitre.org/top25/archive/2025/2025_cwe_top25.html}}.

Such injection vulnerabilities are exploited when external data, which can be manipulated by malicious users and is obtained through sources, is used in sinks without proper sanitisation or application of a validation mechanism.
If a tampered value is sent through a high-privilege communication channel, there is an integrity violation.
For instance, at \Cref{line:tampered:database} of \Cref{fig:code:tampered}, variable \pyCode{y} is sent through the high-privilege sink \pyCode{accessDataBase}, resulting in an integrity violation of the considered database.

Conversely, there is a confidentiality violation if a sensitive value
is sent through a low-privilege channel (e.g., a public
channel).
For instance, in the first two examples of \Cref{fig:code:infoFlow}, the sensitive variable \pyCode{x} is sent through the low-privilege sink \pyCode{broadcast}, which transmits information to the environment, resulting in a confidentiality violation of the secret.

\subsubsection{Side-Channel Attacks}
\label{sec:background:security:sca}

A \emph{side-channel} is a way of transmitting information (purposely or not) out of the intended communication channels.
Side-channel attacks rely on the relationship between information leaked through a side-channel and the secret data to obtain confidential information.
Such attacks usually aim to break cryptography by exploiting
information that is revealed by the algorithm's physical execution
like its execution time~\cite{Koc96} or its power consumption~\cite{KJJ99,BCO04}.
Hence, unlike injection vulnerabilities, a side-channel does not require a sink to leak sensitive information.

An instance of a loop or conditional branching with a condition depending on at least one sensitive value is called \emph{sensitive branching}.
For instance, different branches may not consist of the same sequence of instructions.
Hence, they may have different execution times or they may lead to different power profiles.
Thus, instances of conditional branching are primary targets for side-channel attacks.

\subsection{Aspect-Oriented Programming}
\label{sec:background:aop}

Hierarchical mechanisms of object-oriented languages cannot modularize all concerns of interest in complex systems.
\emph{Aspect-oriented programming} (AOP) has been proposed to improve the separation of concerns, crosscutting the modularity of the rest of the implementation by dynamically executing aspects~\cite{KHH+01}.
The basic concepts of AOP are join points, pointcuts, advice, and aspects. \emph{Join points} are well-defined points in the execution of the program, e.g., a method call or execution. They can be considered vertices in a call graph determined at runtime, with edges representing control-flow relations.
\emph{Pointcuts} are collections of join points defined using designators matching specific join points at runtime.
An \emph{advice} is a method-like construct that can be attached to a pointcut, and that is used to declare that some piece of code should execute at each of the join points in a pointcut.
Finally, \emph{aspects} are modular units of crosscutting implementation,
composed of pointcut and advice declarations.

\subsection{Symbolic Control-Flow Graphs}
\label{sec:background:scfg}

In traditional control-flow graphs (CFGs), directed edges represent control-flow statements corresponding to conditional or loop branching, while vertices represent basic blocks, i.e., sequences of statements without branching.

\citet{DR19} introduced \emph{Symbolic Control-Flow Graphs} (SCFGs) as a way to represent finer-grained statically computable information on a procedure of interest.
While a vertex of a CFG is a block that may contain several statements (e.g., a sequence of assignments), a vertex of an SCFG is a \emph{symbolic state}, i.e., the representation of an individual statement.
A directed edge connects two vertices of an SCFG if the corresponding statements can be executed in sequence, e.g., two assignments in a row, a for loop, and the first statement of the body of the loop.

These symbolic states encompass static information from the statements under consideration.
First, to identify distinct statements or distinct occurrences of the same statement in the procedure of interest, each symbolic state is associated with a unique integer identifier, called a \emph{program point}.
Given a statement $\progStmt$, we denote by $\progPt{\progStmt}$ its program point.
Second, a symbolic state captures static information about the usage of variable and function symbols at the considered statement.
The symbolic state associated with a program statement $\progStmt$ is a pair $\symbState = \tuple{\progPt{\progStmt}, m}$, where $m$ denotes a map from statement symbols to labels representing the usage of the symbols in the statement, e.g., \pyCode{x = y + f(z)} would be associated with a map $m$ such that $m(\pyCode{x}) = \quotes{changed}$, $m(\pyCode{y}) = m(\pyCode{z}) = \quotes{unchanged}$, and $m(\pyCode{f}) = \quotes{called}$.

 \section{Running Example}
\label{sec:runningExample}

Our running example (shown in \Cref{fig:runningExample}) illustrates a
simple communication protocol; we use it to demonstrate
two vulnerabilities: one based on a source-to-sink information flow and one that does not require a sink.

\begin{figure}[t]
\centering
\begin{minipage}{0.88\linewidth}
\begin{lstlisting}[language=myPython,escapechar=@,basicstyle={\scriptsize \ttfamily}]
def runningExample(keySize):@\label{line:runningExample:start}@
    p, g = genPublic()@\label{line:runningExample:genPublic}@# initialize communication
    broadcast([p, g])@\label{line:runningExample:noViolation}@# no confidentiality violation
    k = genPrivate(keySize)@\label{line:runningExample:source}@# source
    # modular exponentiation x = g**k % p
    x = 1@\label{line:runningExample:initVar}@
    for i in range(keySize - 1, -1, -1):@\label{line:runningExample:notSensitive}@# safe branching, i from leftmost to rightmost bit of k
        x = x**2 % p@\label{line:runningExample:square}@
        if (k & (1 << i)) == 1:@\label{line:runningExample:sensitBranching}@# sensitive branching (to be reported) on the value of bit i of k
            x = g*x % p@\label{line:runningExample:indirect}@# indirect flow from k
        else:
            y = g*x % p@\label{line:runningExample:dummy}@# dummy instruction
    # x is now sensitive, so it should not be sent through a low-level output without care
    broadcast([x])@\label{line:runningExample:violation}@# violation (to be reported)
\end{lstlisting}
 \end{minipage}
\caption{Source code of the running example.}
\label{fig:runningExample}
\end{figure}

This source code corresponds to the start of the Diffie–Hellman protocol~\cite{DH76}, which is used to exchange private keys between two parties over an untrusted communication channel.
Parties can only communicate using the \pyCode{broadcast} functions, which are sinks that transmit information to the environment; this includes the other party as well as potential attackers.
This code is subject to the \emph{confidentiality} information policy, which aims to prevent sensitive data from being leaked to untrusted users.
Hence, we must avoid transmitting sensitive information using the \pyCode{broadcast} sink.

However, there may be other ways to leak sensitive information, for instance through side-channels (which do not require a sink).
In the running example, branches in an instance of sensitive branching are padded with dummy instructions~\cite{Aga00}, so that an external observer cannot determine which branch was taken, thereby trading performance for security.

We now detail the source code of the running example.
Public parameters for the communication (prime modulus \pyCode{p} and generator \pyCode{g}) are randomly selected (\Cref{line:runningExample:genPublic}), then sent through a sink (\Cref{line:runningExample:noViolation}), so that the other party can use the same parameters.
Because \pyCode{p} and \pyCode{g} are not sensitive, this
information leakage satisfies the confidentiality policy.
The source of secret information in the running example is the \pyCode{genPrivate} (\Cref{line:runningExample:source}) function, which generates a private key \pyCode{k} from a fixed key size.
There is a direct information flow from the secret source
\pyCode{genPrivate} to \pyCode{k}, which is thus a sensitive
variable.

Then, the public key $x = g^k \bmod p$
is computed using a square-and-multiply-always algorithm~\cite{Cor99}.
The loop branching (\Cref{line:runningExample:notSensitive}) only depends on the key size, which is not sensitive.
So, this branching is not vulnerable to side-channel attacks, as opposed to the conditional branching (\Cref{line:runningExample:sensitBranching}) which depends on the sensitive variable \pyCode{k}.
Indeed, depending on the value \pyCode{k & (1 << i)} of the bit $i$ of the private key $k$, either the \code{then} or the \code{else} branch is taken.
As indicated before, the dummy instruction \pyCode{y = a*x \% n}
(\Cref{line:runningExample:dummy}) was added to prevent the attacker
from inferring sensitive information.
Without this dummy instruction, an attacker monitoring execution time could infer how many times the \code{then} branch was taken, and thus the number of 1s in $k$~\cite{Koc96}.
Moreover, by monitoring power consumption during execution, an attacker could directly read the $k$ bits using an oscilloscope~\cite{KJJ99}.

Unfortunately, such a simple countermeasure is not sufficient against more advanced timing (e.g., \pyCode{x} and \pyCode{y} may not have the same access time) or power~\cite{BCO04} side-channel attacks.
Because of the various (and hard to guard against) ways information may leak due to sensitive branching, they should always be reported by a static security analyzer.

Moreover, because of this sensitive conditional branching, there is an indirect information flow from the private key \pyCode{k} to the public key \pyCode{x} (\Cref{line:runningExample:indirect}), such that \pyCode{x} is also sensitive.
Finally, the public key \pyCode{x} is sent through a sink (\Cref{line:runningExample:violation}) so that the other party can compute the shared secret key.
Because \pyCode{x} is sensitive, depending on the selected parameters \pyCode{p} and \pyCode{g} (for instance, \pyCode{p} being too small~\cite{BGJ+2014}), there may be a confidentiality violation leaking the private key \pyCode{k} to the environment.
This demonstrates an insecure information flow from the source (\Cref{line:runningExample:source}) to the sink (\Cref{line:runningExample:violation}), which should also be reported.

 \section{Approach Overview}
\label{sec:overview}

We aim to track source code control and data-flow information to detect various types of security vulnerabilities.
For instance, to detect insecure information flows or side-channel
vulnerabilities (\Cref{sec:runningExample}), one needs to identify, for each program point, which functions are sources or sinks and which variables or branching instances are sensitive.
Other vulnerabilities may involve a sequence of function calls or the verification of the properties of certain inputs. For each vulnerability type, such concepts can be expressed by defining primitives in a dedicated language, but the large variety of security vulnerabilities warrants a more versatile approach.

Control- and data-flow properties that characterize security vulnerabilities cannot be simply extracted from a variable's value in the source code.
They are aspects of the source code that crosscut the modularity of the implementation, as they do not neatly match methods or attributes defined by the developer and can affect the whole code.
This is compounded by the fact that developers often write code without considering security properties, as they prioritize other concerns.
Hence, these security properties are analogous to aspects found in
AOP, except that in AOP \emph{dynamic} aspects are determined during source code execution, while the security properties we investigate are determined during a static analysis of the source code; hence, we call them \emph{static aspects}.

We identified three main requirements for our approach:
\begin{enumerate}[R1:]
  \item Since we focus on security vulnerabilities like sensitive
    information leakage or data tampering, our approach shall be able
    to capture data- and control-flow information from the source code.
  \item  Our approach shall allow security engineers to express security aspects in a way that is sufficiently versatile to capture a wide variety of security vulnerabilities.
  \item Our approach shall enable the detection and localization of security vulnerabilities in all statements of the source code.
\end{enumerate}

\begin{figure}[t]
    \centering
    \includegraphics[width=1.0\linewidth]{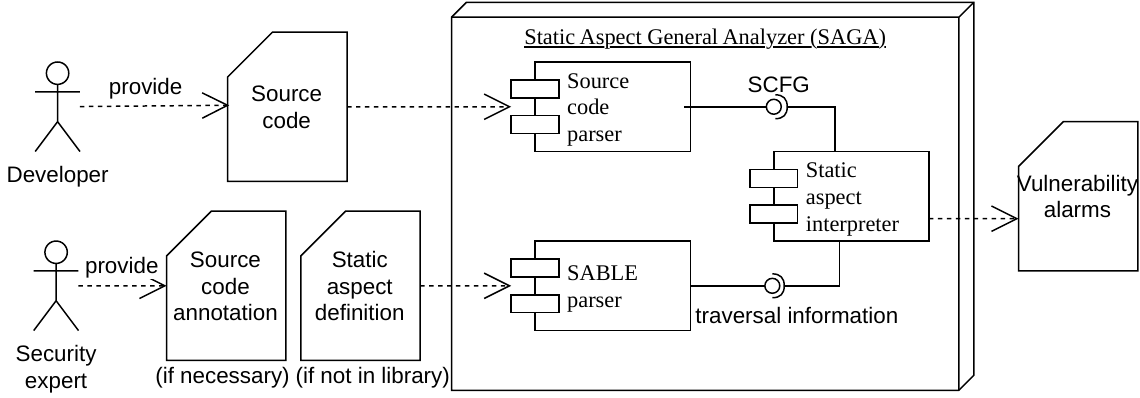}
    \caption{Toolflow of \aspectTool.}
    \label{fig:staticAspects}
\end{figure}

Our \aspectToolLong (\aspectTool) approach for security is shown in
\Cref{fig:staticAspects}.
It takes as input the source code provided by a developer and,
optionally, the source code annotation (if needed) and a static aspect definition (from our library or provided by a security expert); it returns vulnerability alarms.
It includes a parser for the source code of interest to build its representation, a parser for the static aspect definition, and an interpreter that annotates the source code representation with static aspect information.

To address R1, \aspectTool parses the source code and builds its SCFG to represent relevant control- and data-flow information.
To address R2, we propose \aspectLg (\aspectLgLong), a domain-specific language for defining static aspects in a versatile way.
It enables a security expert to define how static aspect values should be updated for each statement during the SCFG traversal.
To address R3, the static aspect interpreter first performs the SCFG traversal, then leverages the resulting static aspect information to report detected security vulnerabilities with their locations in the source code.

In the following, we provide an overview of the main steps of the \aspectTool approach, illustrated using our running example.

First, \aspectTool parses the source code to produce the corresponding SCFG.
We recall that the vertices of an SCFG are \emph{symbolic states}.
Each symbolic state is associated with a unique statement, and several symbolic states may share the same \emph{statement label} like \labelAssign, \labelFor, \labelIf.
Hence, in the context of an SCFG, symbolic states are similar to AOP join points, and statement labels are similar to pointcuts.
For instance, our example source code (\Cref{fig:runningExample}) involves two statements (\Cref{line:runningExample:noViolation,line:runningExample:violation}) where the \pyCode{broadcast} function is called, both being associated with the same statement label \labelExp, standing for expression.

Moreover, variables and functions may have security roles (e.g., sources or sinks) that cannot be determined solely from the source code.
In this case, some domain knowledge from the developer or a security expert is required, which can be provided as \emph{source code annotation}.
In our running example, the \pyCode{genPrivate} function may be annotated as a source and the \pyCode{broadcast} function as a sink.

Static aspects are written in a \emph{static aspect
  definition} using the \aspectLg domain-specific language; we detail its syntax in \Cref{sec:sable}. \aspectTool comes with a library of static aspect definitions for various vulnerability types, but a security expert can also use \aspectLg to define custom static aspects.
Static aspect definitions contain advice code
written in \python to precisely describe the information flow involved
in a vulnerability, making \aspectTool a versatile, security-oriented
static analyzer.

In our running example, the \aspectSensitive static aspect is a set of
symbols (i.e., variables or functions) that are sources or are
(directly or indirectly) tainted by a source or another sensitive
symbol.
The advice code associated with this aspect (and further detailed in
\Cref{sec:sable}, specifically \Cref{fig:sad:travSensitive}) works as follows. \aspectSensitive is initialized with the source symbols from the source code annotation, then updated at each \labelAssign and \labelFor statements, depending on whether a sensitive symbol is used (direct information flow) or if the statement belongs to an instance of sensitive conditional branching (indirect information flow).
If this is the case, the defined symbols are added to
\aspectSensitive; otherwise, their new value is safe, and they are
removed from \aspectSensitive.

Continuing with the example,  \aspectConfidentialityViolation is a Boolean value
indicating an information flow between a source and a sink; its advice
code (Figure~\ref{fig:sad:confidentiality} in Appendix~\ref{sec:secondSadExample}, provided as supplementary material) works as follows. 
\aspectConfidentialityViolation is initialized to $\algoFalse$ and becomes $\algoTrue$ each time a sink is involved in a statement involving sensitive symbols. 

The considered source code annotation and static aspect definition are parsed to extract the information required during the SCFG traversal to determine static aspect values.
More precisely, traversal information determines which piece of advice code is executed for which statement label; this information is used by the interpreter during SCFG traversal to annotate it with static-aspect information.
For instance, at \Cref{line:runningExample:source}, the advice code for \labelAssign is executed; since there is a direct information flow from source \pyCode{genPrivate} to variable \pyCode{k}, the value of the \aspectSensitive static aspect is updated from $\set{\pyCode{genPrivate}}{}$ to $\set{\pyCode{genPrivate}, \pyCode{k}}{}$.
Similarly, at \Cref{line:runningExample:violation}, the advice code for \labelExp is executed; since \pyCode{broadcast} is a sink, \pyCode{x} has been tainted during the execution, and both are involved in the statement, the value of the \aspectConfidentialityViolation static aspect is updated to $\algoTrue$.

During an SCFG traversal, each time a static aspect value in an SCFG annotation matches specific values from the static aspect definition, e.g., the static aspect \aspectConfidentialityViolation being $\algoTrue$, a vulnerability is detected at the corresponding symbolic state.
In that case, the location of the program statement corresponding to this symbolic state, along with other relevant information, is gathered as a \emph{vulnerability alarm}.
Finally, when the SCFG traversal is complete, \aspectTool reports all the gathered vulnerability alarms.

 \section{Source Code Representation}
\label{sec:sourceCode}

\aspectTool is language-agnostic; however, it relies on a specific source-code representation.
More precisely, \aspectTool is made of several components and only one of them, \scfgPython, depends on the source language; \aspectTool could support other imperative languages than \python, if a way to build SCFGs (\Cref{sec:scfg}) for each considered language is provided.
We choose to focus on \python because it was ranked as the top-most
programming language by IEEE Spectrum in 2025~\cite{IEEEspectrum} and used by 57.9\% of developers (according to the 2025
Stack Overflow survey~\cite{StackOverflow}).

In this section, we first present the source language supported by our implementation (\Cref{sec:srcSyntax}). Then, we describe our extension of symbolic states
(\Cref{sec:symbStates}) and SCFGs (\Cref{sec:scfg}) to
represent relevant statically-computable information of the source code.
Finally, we introduce SCFG annotations (\Cref{sec:staticAspects:annotations}) to represent static aspects.

\subsection{Source Language}
\label{sec:srcSyntax}

We support Python versions from 3.8 to 3.12, i.e., all officially
supported Python versions at the time this work was developed.
For the sake of conciseness, in the context of \python, the compact notation \character{\srcSeqL} means adding a new line and incrementing the indentation, 
\character{\srcSeqR} means adding a new line and decrementing the indentation, and
\character{\srcSeqN} means adding a new line (if it was not preceded by \character{\srcSeqR}) and (in every case) preserving the current indentation.

\subsection{Symbolic State Extension}
\label{sec:symbStates}

We use our running example to illustrate various concepts related to the static representation of the source code we use in this work.
For the sake of simplicity and in the context of this example, we
denote by $\progStmt_{\progLine}$ the statement at line $\progLine$ of the source code example (\Cref{fig:runningExample}).

Symbolic states were initially introduced by \citet{DR19} (\Cref{sec:background:scfg}).
We use and extend them with new attributes to extract sufficient statically computable information to perform our static aspect analysis and detect vulnerabilities.

First, \emph{program points}, i.e., unique identifiers for each
program statement, are computed as in the original work~\cite{DR19}.
We assume they start at $1$ and are incremented for each new statement.
For our running example, $\progPt{\progStmt_{\ref*{line:runningExample:genPublic}}} = 1$, $\progPt{\progStmt_{\ref*{line:runningExample:noViolation}}} = 2$, $\progPt{\progStmt_{\ref*{line:runningExample:source}}} = 3$, etc.

Since we aim to locate vulnerabilities in source code, we extract \emph{program locations} from the AST.
For each statement $\progStmt$, its program location, denoted $\progLoc{\progStmt}$, is an integer corresponding to the line number where $\progStmt$ starts.
For control-flow statements corresponding to conditional or loop branching, we also gather the line number $\progLocEnd{\progStmt}$ where the scope of $\progStmt$ ends.
We use program points instead of line numbers as identifiers because, in general, several statements can be written on the same line.
Nevertheless, for our running example, there is no ambiguity, and we have $\progLoc{\progStmt_{\progLine}} = \progLine$ for each statement.

Then, from the AST we also extract \emph{statement labels}, which are strings corresponding to \python statement types.
Each statement $\progStmt$ is always associated with a unique statement label, denoted $\progStmtLabel{\progStmt}$.
For our running example, $\progStmtLabel{\progStmt_{\ref*{line:runningExample:genPublic}}} = \labelAssign$, $\progStmtLabel{\progStmt_{\ref*{line:runningExample:noViolation}}} = \labelExp$, $\progStmtLabel{\progStmt_{\ref*{line:runningExample:notSensitive}}} = \labelFor$, etc.

Finally, we gather information about the symbols that occur in each statement.
As opposed to the original work~\cite{DR19} where symbols are considered together for each statement, we split each statement into relevant \emph{statement expressions}.
For instance, we consider in the assignment statement \pyCode{k = genPrivate(keySize)} its left-hand side \pyCode{k} and its right-hand side \pyCode{genPrivate(keySize)} as distinct expressions.
Note that statements like $\srcPass$ and $\srcBreak$ contain no statement expression, statements like $\srcStmtReturn{\srcExpr}$ and $\srcWhile\ \srcExpr$ contain one statement expression $\srcExpr$, and statements like $\srcStmtAssign{\srcVar}{\srcExpr}$ and $\srcFor\ \srcVar\ \srcIn\ \srcExpr$ contain two statement expressions $\srcVar$ and $\srcExpr$.
Given a statement $\progStmt$, we denote by $\progExprs{\progStmt}$ the list of statement expressions occurring in that statement.

We represent the use of variable and function symbols at the expression level. Each statement expression is represented by a triple $\tuple{\text{Def}, \text{Use}, \text{Call}}$, where $\text{Def}$ denotes the set of the variables defined in the statement expression, $\text{Use}$ the set of used variables or literals, and $\text{Call}$ the set of called functions.
For instance, in our running example, the assignment statement \pyCode{k = genPrivate(keySize)} is represented by statement expressions $\progExprs{\progStmt_{\ref*{line:runningExample:source}}} = \dataList{\tuple{\set{\pyCode{k}}{}, \set{}{}, \set{}{}}, \tuple{\set{}{}, \set{\pyCode{keySize}}{}, \set{\pyCode{genPrivate}}{}}}$.
The use of symbols is determined as in the original work:
\begin{itemize}
\item \emph{Called} functions are obtained by the \python parser and
  are available in the AST of the source code.
\item Variables are distinguished using their position in a statement.
A variable is \emph{defined} if it is on the leftmost position of the left-hand side of an assignment statement or between the $\srcFor$ and the $\srcIn$ keywords of a for statement.
Variables which are not defined are \emph{used}.
\end{itemize}
Thus, each statement $\progStmt$ is associated with a \emph{symbolic state}, which is a quadruple consisting of the program point, the program location, the statement label, and the statement expressions of $\progStmt$:
$$
\begin{array}{r@{}l}
    \progSymbState{\progStmt} \eqdef
        {}& \langle \progPt{\progStmt}, \progLoc{\progStmt},\\
        {}&\progStmtLabel{\progStmt}, \progExprs{\progStmt} \rangle\\
\end{array}
$$

In the context of our running example, we denote by $\symbState_\progLine
\eqdef \progSymbState{\progStmt_\progLine}$ the symbolic state
associated with the statement at line $\progLine$ of the source code
example. For instance, for $\progLine=4$ we have:
\begin{itemize}
    \item $\progPt{\progStmt_{\ref*{line:runningExample:genPublic}}} = 1$,
    \item $\progLoc{\progStmt_{\ref*{line:runningExample:genPublic}}} = \ref*{line:runningExample:genPublic}$,
    \item $\progStmtLabel{\progStmt_{\ref*{line:runningExample:genPublic}}} = \labelAssign$,
    \item and $\progExprs{\progStmt_{\ref*{line:runningExample:genPublic}}} =[\tuple{\set{\pyCode{p}, \pyCode{g}}{}, \set{}{}, \set{}{}}$, $\tuple{\set{}{}, \set{}{}, \set{\pyCode{genPublic}}{}}]$,
\end{itemize}
resulting in $
\symbState_{\ref*{line:runningExample:genPublic}} = \progSymbState{\progStmt_{\ref*{line:runningExample:genPublic}}} = \tuple{1, \ref*{line:runningExample:genPublic} , \labelAssign, \dataList{\tuple{\set{\pyCode{p}, \pyCode{g}}{}, \set{}{}, \set{}{}}, \tuple{\set{}{}, \set{}{}, \set{\pyCode{genPublic}}{}}}}$.

For a given symbolic state $\symbState$, we denote by $\ptProg{\symbState}$ its program point, $\ptLoc{\symbState}$ its program location, $\ptStmt{\symbState}$ its statement label, and $\ptExprs{\symbState}$ its statement expressions.

\subsection{SCFG Extension}
\label{sec:scfg}

Symbolic control-flow graphs (SCFGs) were initially introduced by \citet{DR19} (\Cref{sec:background:scfg}).
We compute them in a manner similar to the original work; however, because \aspectTool covers more constructs than the original source language, we detail the procedure for constructing SCFGs.

We consider a procedure $\procedure$ defined using the declaration
$\bnfStmtQuestion{\srcAsync} \srcDef\ \procedure(\srcVar_1, \dots, \srcVar_n) \srcSeqC$ $\srcSeqL \srcStmt \srcSeqR$.
We denote $\procInputs{\procedure}$ the set of variables $\set{\srcVar_1, \dots, \srcVar_n}{}$, $\procStmtsSeq{\procedure}$ the sequence of statements $\srcStmt$ in the declaration of $\procedure$, and $\progLineStart{\procedure}$ (resp. $\progLineEnd{\procedure}$) the line number where the scope of the declaration of $\procedure$ starts (resp. ends).
For our running example, $\procInputs{\pyCode{runningExample}} = \set{\pyCode{keySize}}{}$, $\procStmtsSeq{\pyCode{runningExample}}$ consists of the sequence of statements from \Cref{line:runningExample:genPublic} to \Cref{line:runningExample:violation}, $\progLineStart{\pyCode{runningExample}} = \ref{line:runningExample:start}$, and $\progLineEnd{\pyCode{runningExample}} = \ref{line:runningExample:violation}$.

The SCFG of $\procedure$, denoted $\scfg{\procedure}$, is a directed
graph, where vertices are symbolic states and edges represent the
control flow between program points of $\procedure$.
Formally, it is a pair $\scfg{\procedure} \eqdef \tuple{\vertices{\procedure}, \edges{\procedure}}$, where the set of vertices $\vertices{\procedure}$ and the set of edges $\edges{\procedure}$ are defined below.

First, we denote by $\stmts{\procedure}$ the set of statements occurring in the sequence $\procStmtsSeq{\procedure}$, including statements in the scope of other statements.
For each statement $\progStmt \in \stmts{\procedure}$, we consider the corresponding symbolic state $\progSymbState{\progStmt}$ as part of the vertices of $\scfg{\procedure}$:
$$\progSymbStates{\procedure} \eqdef \set{\progSymbState{\progStmt}}{\progStmt \in \stmts{\procedure}}$$
We also consider two special symbolic states, which respectively indicate the start and the end of the SCFG:
$$
\begin{array}{l@{}l}
    \symbStateStart
        &{} \eqdef \tuple{0, \progLineStart{\procedure}, \labelEnterProcedure, \dataList{\tuple{\procInputs{\procedure}, \set{}{}, \set{}{}}}}\\
    \symbStateEnd
        &{} \eqdef \tuple{\card{\stmts{\procedure}} + 1, \progLineEnd{\procedure}, \labelExitProcedure, \dataList{}}\\
\end{array}
$$
Finally, for each symbolic state representing
an instance of a branching statement (e.g., with label \labelIf, \labelWhile), we introduce the corresponding \emph{ending symbolic state} (e.g., with label \labelEndIf, \labelEndWhile):
$$\progEndState{\symbState} \eqdef \tuple{\ptProg{\symbState}, \ptEndLoc{\symbState}, \labelEnd + \ptStmt{\symbState}, \dataList{}}$$
where the operator $+$ denotes string concatenation and
$\ptEndLoc{\symbState} = \progLocEnd{\progStmt}$ such that $\progStmt$ is the unique statement in $\stmts{\procedure}$ with program point $\ptProg{\symbState}$.

Thus, the vertices of $\scfg{\procedure}$ are defined as follows:
$$
\begin{array}{l@{}l}
    \vertices{\procedure}
        &{} \eqdef \progSymbStates{\procedure} \cup \set{\symbStateStart, \symbStateEnd}{}\\
        &{} \quad \cup \{ \progEndState{\symbState} \knowing\symbState \in \progSymbStates{\procedure}\\
        &{} \quad \land \ptStmt{\symbState} \in \BranchLabels \}\\
\end{array}
$$
where $\BranchLabels \eqdef \{\labelIf,$ $\labelTry,$ $\labelElse,$ $\labelMatch,$ $\labelWith,$ $\labelWhile,$ $\labelFor\}$.

An oriented edge of $\scfg{\procedure}$ is represented by a pair $\tuple{\symbState_1, \symbState_2}$, where $\symbState_1$ and $\symbState_2$ are symbolic states such that $\symbState_2$ is a successor of $\symbState_1$ regarding the control-flow information extracted from $\procedure$.
The edges of $\scfg{\procedure}$ are defined as:
$$
\begin{array}{l@{}l}
    \edges{\procedure}
        &{} \eqdef \set{\tuple{\symbStateStart, \progSymbState{\progStmt_1}}}{} \\
        &{} \quad\cup \edgesRec{\procStmtsSeq{\procedure}}{\symbStateStart}{\symbStateEnd}
\end{array}
$$
where $\progStmt_1$ is the unique statement in $\stmts{\procedure}$ such that $\progPt{\progStmt} = 1$, and $\edgesRecName(\progStmt, \symbStateScope,$ $\symbStateNext)$ is a function returning a set of edges.
It is defined by induction on $\progStmt$ in Appendix~\ref{sec:scfgEdges}, provided as supplementary material.

A symbolic state $\symbState \in \vertices{\procedure}$ is \emph{reachable} if there exists symbolic states $\symbState_0, \symbState_1, \dots, \symbState_n \in \vertices{\procedure}$ such that $\symbState_0 = \symbStateStart$, $\symbState_n = \symbState$, and for each $0 \le i \le n - 1$ we have $\tuple{\symbState_i, \symbState_{i + 1}} \in \edges{\procedure}$.
While \python enables one to write statements corresponding to unreachable
states (e.g., any statement after a $\srcReturn$ statement), such statements cannot be executed and are hence useless to detect security vulnerabilities.
Thus, before performing the rest of the analysis, unreachable states
are removed from $\vertices{\procedure}$.

Finally, to ease the description of SCFG traversals, we define \emph{children} of each symbolic state $\symbState_1 \in \vertices{\procedure}$, i.e., symbolic states which are successors of $\symbState_1$ in $\scfg{\procedure}$:
$$\ptNexts{\symbState_1} \eqdef \set{\symbState_2 \in \vertices{\procedure}}{\tuple{\symbState_1, \symbState_2} \in \edges{\procedure}}$$

\subsubsection*{Application to the Running Example}

\begin{figure}[t]
\centering
\includegraphics[width=1.0\linewidth]{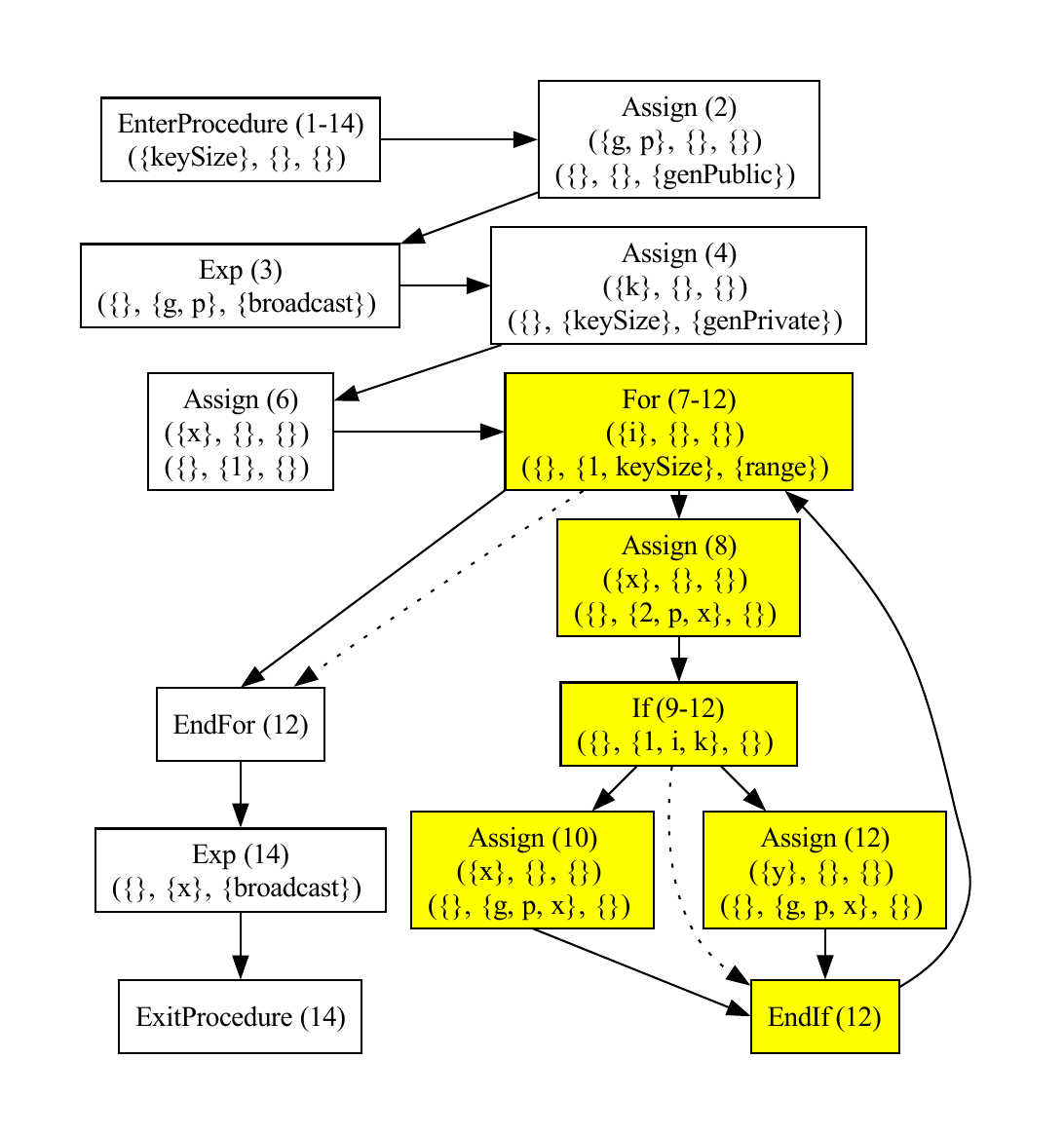}
\caption{SCFG of the source code example from \Cref{fig:runningExample} (the yellow vertices are detailed in the text).}
\label{fig:code:runningExampleSCFG}
\end{figure}

The SCFG for our running example is depicted in \Cref{fig:code:runningExampleSCFG}.
Each symbolic state is represented as a box containing the statement
label, followed by the program location (between parentheses), and the statement expressions.
In case of control-flow statements, the line number where the scope of
the statement ends is also indicated (as in ``For (7--12)'').
Each oriented edge $\tuple{\symbState_1, \symbState_2}$ is represented by a plain arrow $\symbState_1 \rightarrow \symbState_2$.
Moreover, since each ending symbolic state is only represented by its
statement label,
our SCFG figures include dotted arrows between each state $\symbState$ (for which $\ptStmt{\symbState} \in
\BranchLabels$) and its corresponding ending state $\progEndState{\symbState}$; 
Note that this is only a graphical convenience for simply mapping each ending state to its corresponding control-flow state in SCFG figures, but it does not reflect an existing edge in $\scfg{\procedure}$.

To illustrate how SCFGs are generated, we describe how the SCFG subgraph represented in yellow in \Cref{fig:code:runningExampleSCFG} can be obtained for the fragment
\begin{lstlisting}[style=neutral]
x = x**2 \end{lstlisting}
from \Cref{line:runningExample:square} to
\Cref{line:runningExample:dummy} of the source code example, denoted $\progStmt$ in the rest of the section.

As the considered fragment corresponds to the body of the loop, the starting and ending points of the subgraph are the same symbolic state $\symbState_{\ref*{line:runningExample:notSensitive}}$, which corresponds to the \code{for} statement at \Cref{line:runningExample:notSensitive}.
This fragment involves four other statements, one initial assignment,
one for the conditional branching, and one assignment per branch,
respectively associated with the symbolic states
$\symbState_{\ref*{line:runningExample:square}}$,
$\symbState_{\ref*{line:runningExample:sensitBranching}}$,
$\symbState_{\ref*{line:runningExample:indirect}}$, and
$\symbState_{\ref*{line:runningExample:dummy}}$, with each symbolic state index corresponding to the statement line number.
Moreover, the symbolic state $\symbState_{\ref*{line:runningExample:sensitBranching}}$ for the conditional branching is associated with an ending symbolic state $\progEndState{\symbState_{\ref*{line:runningExample:sensitBranching}}}$, which we denote $\symbState'_{\ref*{line:runningExample:sensitBranching}}$ for the sake of conciseness.
Hence, according to the definition of $\vertices{\procedure}$, the vertices for this subgraph are $\symbState_{\ref*{line:runningExample:notSensitive}}$, $\symbState_{\ref*{line:runningExample:square}}$, $\symbState_{\ref*{line:runningExample:sensitBranching}}$, $\symbState_{\ref*{line:runningExample:indirect}}$, $\symbState_{\ref*{line:runningExample:dummy}}$, and $\symbState'_{\ref*{line:runningExample:sensitBranching}}$.

According to the definition of $\edges{\procedure}$, the first edge is $\tuple{\symbState_{\ref*{line:runningExample:notSensitive}}, \symbState_{\ref*{line:runningExample:square}}}$, connecting the start of the subgraph with the state of the first statement of the fragment.
The other edges are determined by induction on the fragment $\progStmt$ using $\edgesRec{\progStmt}{\symbState_{\ref*{line:runningExample:notSensitive}}}{\symbState_{\ref*{line:runningExample:notSensitive}}}$.
We detail the computation in Appendix~A and obtain $\tuple{\symbState_{\ref*{line:runningExample:square}}, \symbState_{\ref*{line:runningExample:sensitBranching}}}$, $\tuple{\symbState_{\ref*{line:runningExample:sensitBranching}}, \symbState_{\ref*{line:runningExample:indirect}}}$,
$\tuple{\symbState_{\ref*{line:runningExample:sensitBranching}}, \symbState_{\ref*{line:runningExample:dummy}}}$, $\tuple{\symbState_{\ref*{line:runningExample:indirect}}, \symbState'_{\ref*{line:runningExample:sensitBranching}}}$, $\tuple{\symbState_{\ref*{line:runningExample:dummy}}, \symbState'_{\ref*{line:runningExample:sensitBranching}}}$, and $\tuple{\symbState'_{\ref*{line:runningExample:sensitBranching}}, \symbState_{\ref*{line:runningExample:notSensitive}}}$.

Finally, the assignment at \Cref{line:runningExample:square} has only one successor, i.e., the conditional branching at \Cref{line:runningExample:sensitBranching}, so $\ptNexts{\symbState_{\ref*{line:runningExample:square}}} = \set{\symbState_{\ref*{line:runningExample:sensitBranching}}}{}$.
This conditional branching has two branches, one at \Cref{line:runningExample:indirect} and one at \Cref{line:runningExample:dummy}, so $\ptNexts{\symbState_{\ref*{line:runningExample:sensitBranching}}} = \set{\symbState_{\ref*{line:runningExample:indirect}}, \symbState_{\ref*{line:runningExample:dummy}}}{}$.

\subsection{SCFG Annotation}
\label{sec:staticAspects:annotations}

We want to perform a static aspect analysis of a procedure $\procedure$, i.e., to determine various static aspects (defined in \Cref{sec:sable}) for each statement in $\procedure$; for instance, to determine which variables are sensitive or if there is a confidentiality violation.

Such an analysis is performed on each symbolic state of $\scfg{\procedure}$.
Since various aspects can be associated with any given symbolic state, we enrich each symbolic state $\symbState \in \vertices{\procedure}$ with a new attribute called an \emph{SCFG annotation}, denoted $\ptAnnotation{\symbState}$, which maps a string (the static aspect name) to any value (the static aspect value).
For instance, our static aspect analysis may annotate $\scfg{\procedure}$ at symbolic state $\symbState$ such that
$\ptAnnotation{\symbState}$ maps
$\aspectSensitive$ to $\set{\pyCode{k}, \pyCode{x}}{}$ and
$\aspectConfidentialityViolation$ to $\algoTrue$.

 \section{\aspectLg}
\label{sec:sable}

In this section, we present \aspectLg (\aspectLgLong), a
domain-specific language for defining security-relevant static aspects of the source
code in a versatile way.
After an initial overview (\Cref{sec:staticAspects:overview}, we present its syntax (\Cref{sec:staticAspects:syntax}) and its usage (\Cref{sec:staticAspects:usage}) --- including how security experts can provide domain knowledge to the static aspect analysis and an example of static aspect definition.
As supplementary material, we present
another example of static aspect definition in Appendix~\ref{sec:secondSadExample},
a description of our static aspect library in Appendix~\ref{sec:sable:sadLibrary}, and 
 \aspectLg semantics in Appendix~\ref{sec:staticAspects:semantics}.
The static aspect library consists of five static aspect definitions that capture, respectively, taint analysis, conditions involving critical data, sequences of function calls, the presence or absence of critical variables or functions, and the setup of values in relevant contexts (e.g., permissions before opening a file).

\subsection{\aspectLg Overview}
\label{sec:staticAspects:overview}

\aspectLg enables security experts to provide domain knowledge on vulnerabilities by providing \emph{source annotations} (e.g., a source or a sink).
Also, to leverage data- and control-flow information from source code for vulnerability detection, \aspectLg defines \emph{SCFG traversals}, in which each source code statement is visited and annotated.
For the sake of modularity, \aspectLg also enables users to define different aspects of the code in separate traversals and to retrieve, during a traversal, information obtained in previous traversals.

\aspectLg expresses \emph{static aspects} of a traversal in a versatile way, as \python variables are updated each time a statement is visited, based on source annotations, the SCFG symbolic state, and (possibly) static aspect values from previous traversals.
As security properties depend on statement types (e.g., conditional, loop), \aspectLg takes inspiration from AOP (\Cref{sec:background:aop}) by performing these updates using \emph{advice code} controlled by \emph{pointcuts} depending on the statement label of the visited symbolic state.
For the sake of modularity, \aspectLg also enables users to define utility functions as standard \python functions that can be called in advice code.
Since utility functions and advice code pieces are written in \python, \aspectLg benefits from the large expressive power of a general-purpose language to define static aspects.

Finally, to enable vulnerability detection during SCFG traversals, \aspectLg provides \emph{triggers}, i.e., values that, when taken by a static aspect, trigger alarms that are then reported in a security report.

\subsection{\aspectLg Syntax}
\label{sec:staticAspects:syntax}

The syntax of \aspectLg is defined in \Cref{table:aspectLg:syntax}.
In this grammar,
\character{\srcSeqL},
\character{\srcSeqR},
and \character{\srcSeqN} have the same meaning as in \Cref{sec:srcSyntax},
$\bnfStmtQuestion{.}$ and $\bnfStmtStar{.}$ have the same meaning as in \Cref{sec:scfg},
$\srcName$ indicates a \python object name,
$\pyLiteral$ a \python literal,
$\aspectType$ a \python type,
$\pyStmt$ a standard \python statement,
$\ptAspectName$ an aspect name,
$\ptTravName$ a name identifying a particular SCFG traversal,
and $\aspectStmtLabel$ a source statement label as introduced in \Cref{sec:symbStates}.

The grammar is organized as follows.
Non-terminal $\pyProg$ captures standard \python code.
Non-terminal $\aspectStmt$ captures statements used in a traversal
definition; the following statements are supported:
\begin{itemize}
  \item[$\aspectUtility$] introduces the \python code for the current traversal, e.g., to import libraries or to define procedures used in advice code.
\item[$\aspectImportAspect$] imports a list of static aspects from a previous traversal, whose name is provided after \aspectFromTraversal.
\item[$\aspectSrcLabels$] stores the source code annotation provided by the developer or a security expert (\Cref{sec:staticAspects:usage}) in the provided variable.
\item[$\aspectAspect$] declares a new static aspect
  for the current traversal, having as type the one provided after \aspectAspectType; the parser checks that aspect names in a traversal are distinct.
\item[$\aspectTrigger$] triggers an alarm each time the static aspect with the provided name has a value equal to the \python literal indicated after \aspectTriggerVal.
For instance,
\code{triggerFrom Vulnerability atValue True}
raises an alarm each time static aspect $\mathit{Vulnerability}$ becomes $\algoTrue$.
This construct can be used several times for different static aspects or for the same static aspect, but with different values.
Hereafter, we denote by $\getTriggers{\ptAspectName}$ the set of values
triggering an alarm for the static aspect $\ptAspectName$; for instance, in the previous example, $\getTriggers{\mathit{Vulnerability}} = \set{\algoTrue}{}$.
\item[$\aspectPointcut$] executes the provided advice code each time the traversal visits a symbolic state with the provided statement label. The provided variables store the expressions used at that statement. The parser checks that each statement label is used only once per traversal.
Hereafter, we denote by $\getAdvice{\ptStmtLabel}$ the piece of advice
code associated with statement label $\ptStmtLabel$ by a
$\aspectPointcut$ keyword.
For instance, one can initialize a static aspect $\mathit{Tainted}$ as an empty set using
\code{pointcut(EnterProcedure, inputs): \{Tainted = set()\}}.
In this example, $\getAdvice{\labelEnterProcedure}$ is \code{Tainted = set()}.
\item[$\aspectMergeAspects$] specifies how static aspect values are determined
when two branches merge. If this keyword is not used, then the default function is used, whose code is provided in
\Cref{fig:sad:merge} of Appendix~\ref{sec:staticAspects:semantics}, provided as supplementary material.
The parser checks that at most one merge function is defined per traversal.
\end{itemize}

Non-terminal $\aspectTypeTrav$ captures a sequence of traversal
definitions, each one defined by the keyword $\aspectTraversal$. \aspectLg allows users to define distinct instructions for several SCFG traversals, each with a unique name; the parser ensures that each traversal name is distinct.

\begin{figure}[t]
\centering
\footnotesize
$$\begin{array}{r@{}l}

{}& \textbf{\python Code}\\
    \pyProg \bnfEq
        {}&\srcStmtImport{\srcName}\\
    \bnfOr
        {}&\srcStmtFrom{\srcName}{\srcName \bnfStmtStar{, \srcName} }\\
    \bnfOr
{}&\srcDef\ \srcName\ \bnfFunc{\bnfStmtQuestion{\srcName \bnfStmtStar{, \srcName} }}\srcSeqC \\
        {}&\srcSeqL \pyStmt \bnfStmtStar{\srcSeqN \pyStmt} \srcSeqR\\
    \bnfOr
        {}&\srcSeq{\pyProg}{\pyProg}\\[0.2cm]

{}& \textbf{Statements in Traversals}\\
\aspectStmt \bnfEq
    {}&\aspectStmtUtility{\pyProg}\\
\bnfOr
{}&\aspectFromTraversal\ \ptTravName\ \\
    {}&\aspectImportAspect\ \ptAspectName \bnfStmtStar{, \ptAspectName}\\
\bnfOr
    {}&\aspectStmtSrcLabels{\aspectVar}{\aspectPath}\\
\bnfOr
    {}&\aspectStmtAspect{\ptAspectName}{\aspectType}\\
\bnfOr
    {}&\aspectTriggerAt{\ptAspectName}{\pyLiteral}\\
\bnfOr
{}&\aspectPointcut\bnfFunc{\aspectStmtLabel\bnfStmtStar{, \aspectVar}}: \\
    {}&\left\{\pyStmt \bnfStmtStar{\srcSeqN \pyStmt} \right\}\\
\bnfOr
{}&\aspectMergeAspects \bnfFunc{\aspectVar, \aspectVar}: \\
    {}& \left\{\pyStmt \bnfStmtStar{\srcSeqN \pyStmt} \right\}\\
\bnfOr
    {}&\srcSeq{\aspectStmt}{\aspectStmt}\\[0.2cm]

{}& \textbf{Traversals}\\
\aspectTypeTrav \bnfEq
    {}&\aspectStmtTrav{\ptTravName}{\aspectStmt}\\
\bnfOr
    {}&\srcSeq{\aspectTypeTrav}{\aspectTypeTrav}\\

\end{array}$$
   \caption{Syntax of the \aspectLg domain-specific language.}
\label{table:aspectLg:syntax}
\end{figure}

\python statements $\pyStmt$ used in $\aspectUtility$, $\aspectPointcut$ or $\aspectMergeAspects$ constructs may contain, on top of standard \python primitives, \aspectLg-specific primitives used for manipulating static aspects or source code symbols:
\begin{itemize}
\item[$\aspectCurrentPoint$] Used only in the scope of a $\aspectPointcut$, returns the matching symbolic state.
\item[$\aspectGetAspect{\symbState}{\ptAspectName}$] Returns the value of the static aspect $\ptAspectName$, defined during a previous traversal, at the symbolic state $\symbState$.
\item[$\aspectEnterLoop{\symbState}$] Returns a Boolean, which is true only for a symbolic state $\symbState$ 1) representing a loop statement and 2) visited during SCFG traversal when entering the loop.
Notice that the value is false each time the loop statement is visited because its body was visited just before.\item[$\aspectGetExprSymbs{\ptSymbLabel}{\srcExpr}$] returns the
  symbols in the expression $\srcExpr$ matching the symbol label
  $\ptSymbLabel$. If another symbol label apart from $\aspectRead$, $\aspectWritten$, $\aspectCalled$, or $\aspectAll$ is used, then an error is raised.
This primitive is used with expressions stored by the $\aspectPointcut$ keyword.
For instance, one may capture in a variable $\mathit{Symbs}$ the symbols defined in the LHS and the symbols used in the RHS of an assignment using advice code like
\code{pointcut(Assign, left, right): \{Symbs = getExprSymbs('def', left) | getExprSymbs('use', right)\}}.
\item[$\aspectGetDescrSymbs{\ptSymbLabel}{\ptSrcLabelsName}$] returns the symbols in the source annotation $\ptSrcLabelsName$ matching the symbol label $\ptSymbLabel$. If $\ptSymbLabel$ is not in $\ptSrcLabelsName$, then an error is raised.
\end{itemize}

\subsection{\aspectLg Usage}
\label{sec:staticAspects:usage}

\aspectLg enables developers to define how to process source code information during one or more SCFG traversal(s), using a static aspect definition.
To write a static aspect definition, one starts by writing a sequence of traversal definitions, depending on the steps developers have in mind to analyze the source code.
Each traversal definition must comprise, in any order,
\begin{inparaenum}[(a)]
  \item the declaration of the static aspects used during the traversal;
  \item the \aspectPointcut{} constructs used to update the static
    aspect values during the traversal;
  \item (optional) utility functions used in the \aspectPointcut{}
    constructs (and defined separately, for the sake of modularity);
  \item  (when relevant) triggers depending on static aspect values,
    which generate alarms to be reported at the end of the traversals.
\end{inparaenum}

\subsubsection{Domain Knowledge from the Security Expert}
\label{sec:staticAspects:expert}

Since not all relevant information can be inferred from the source code, \aspectLg enables security experts to provide domain knowledge to static aspect analysis through \emph{source (code) annotations}.
More precisely, for each procedure of interest, an expert can write a source annotation that maps variable or function names to security labels.
In our running example, \code{\{"source\_code.py:runningExample": \{"source": ["genPrivate"], "sink": ["broadcast"]\}\}}
indicates that, in the procedure \code{runningExample} from the \code{source\_code.py} file, the function \code{genPrivate} is annotated as a source and the function \code{broadcast} is annotated as a sink.

The source annotation is then made available to the static aspect definition using the \aspectSrcLabels{} construct (\Cref{sec:staticAspects:syntax}).
For instance, the \code{sourceAnnotation labeled\_symbols} statement stores the content of the source annotation in the variable \code{labeled\_symbols}.
This variable can then be used in other constructs, e.g., \code{Tainted = getDescrSymbs('source', labeled\_symbols)} retrieves the set of symbols annotated as  sources and assigns it to the variable \code{Tainted}.

\subsubsection{Example of Static Aspect Definition}
\label{sec:sad:example}

We demonstrate the use of \aspectLg with the traversal definition shown in \Cref{fig:sad:travSensitive}.

This traversal considers sensitive sources annotated by the expert, tracks symbols tainted by sources or other tainted symbols, and raises an alarm when a branching instance is sensitive and thus may be vulnerable to side-channel attacks (\Cref{sec:background:security:sca}).
More precisely, traversal \code{travSensitive} is declared at \Cref{line:travSensitive:traversal} and is structured in three main parts: 1) declarations, 2) utility functions, and 3) pointcuts.

\begin{figure}[t!]
\centering
\begin{minipage}{0.95\linewidth}
\begin{lstlisting}[language=aspectDSL,escapechar=@,basicstyle={\scriptsize \ttfamily}]
traversal travSensitive:@\label{line:travSensitive:traversal}@
	# 1) source annotation, aspect, and alarm declarations
	sourceAnnotation labeled_symbols# from security expert@\label{line:travSensitive:sourceAnnotation}@
	aspect Sensitive aspectType set@\label{line:travSensitive:Sensitive}@
	aspect ScopeSensitiveBranches aspectType list@\label{line:travSensitive:ScopeSensitiveBranches}@
	aspect SensitiveBranching aspectType bool@\label{line:travSensitive:SensitiveBranching}@
	triggerFrom SensitiveBranching atValue True# may trigger an alarm@\label{line:travSensitive:triggerFrom}@
	
	utility:# 2) functions used during the traversal@\label{line:travSensitive:utility}@
		def isSensitExpr(expr, Sensitive):@\label{line:travSensitive:isSensitExpr}@
			Symbs = getExprSymbs('use', expr) | getExprSymbs('call', expr)
			return len(Symbs & Sensitive) > 0
		def infoFlow(expr_written, expr_read, Sensitive, ScopeSensitiveBranches):@\label{line:travSensitive:infoFlow}@
			WrittenSymbs = getExprSymbs('def', expr_written)
			if isSensitExpr(expr_read, Sensitive) or True in ScopeSensitiveBranches:
				Sensitive = Sensitive | WrittenSymbs
			else:
				Sensitive = Sensitive - WrittenSymbs
			return Sensitive
		def append_ScopeSensitiveBranches(expr, Sensitive, ScopeSensitiveBranches):@\label{line:travSensitive:append_ScopeSensitiveBranches}@
			sensitBranch = isSensitExpr(expr, Sensitive)
			ScopeSensitiveBranches.append(sensitBranch)@\label{line:travSensitive:endUtility}@
	
	# 3) pointcuts and advice code
	pointcut(EnterProcedure, inputs)@\label{line:travSensitive:EnterProcedure}@:# initialize static aspect values
		Sensitive = getDescrSymbs('source', labeled_symbols)# source annotation is used@\label{line:travSensitive:EnterProcedure:start}@
		ScopeSensitiveBranches = []
		SensitiveBranching = False@\label{line:travSensitive:EnterProcedure:end}@
	pointcut(Assign, left, right):@\label{line:travSensitive:Assign}@
		Sensitive = infoFlow(left, right, Sensitive, ScopeSensitiveBranches)
		SensitiveBranching = False
	pointcut(If, cond):
		append_ScopeSensitiveBranches(cond, Sensitive, ScopeSensitiveBranches)
		SensitiveBranching = isSensitExpr(cond, Sensitive)
	pointcut(While, cond):
		if enterLoop(currentPoint):
			append_ScopeSensitiveBranches(cond, Sensitive, ScopeSensitiveBranches)
		SensitiveBranching = isSensitExpr(cond, Sensitive)
	pointcut(For, index, bound):@\label{line:travSensitive:For}@
		Sensitive = infoFlow(index, bound, Sensitive, ScopeSensitiveBranches)
		if enterLoop(currentPoint):
			append_ScopeSensitiveBranches(bound, Sensitive, ScopeSensitiveBranches)
		SensitiveBranching = isSensitExpr(bound, Sensitive)
	pointcut(EndIf):
		ScopeSensitiveBranches.pop(-1)
		SensitiveBranching = False
	pointcut(EndWhile):
		ScopeSensitiveBranches.pop(-1)
		SensitiveBranching = False
	pointcut(EndFor):
		ScopeSensitiveBranches.pop(-1)
		SensitiveBranching = False
	pointcut(Return, output):
		ScopeSensitiveBranches = []
		SensitiveBranching = False
	pointcut(ExitProcedure):
		SensitiveBranching = False
\end{lstlisting}
 \end{minipage}
\caption{Instructions to determine the values of the \aspectSensitive and \aspectSensitBranching static aspects during an SCFG traversal.}
\label{fig:sad:travSensitive}
\end{figure}

The first part contains source annotation, aspect, and alarm declarations.
The \aspectSrcLabels{} construct (\Cref{line:travSensitive:sourceAnnotation}) stores the content of a source annotation in the variable \code{labeled\_symbols}.
Static aspects are declared at
\Cref{line:travSensitive:Sensitive,line:travSensitive:ScopeSensitiveBranches,line:travSensitive:SensitiveBranching}:
\begin{itemize}
\item \aspectSensitive is a set representing the source symbols or
  symbols tainted by other sensitive symbols.
\item \aspectScopeSensitiveBranches is an intermediary static aspect, used
only to determine how \aspectSensitive should be updated: it is a
stack (represented by a list)
of Booleans;
they represent the nested branchings whose scope includes the visited symbolic state.
A Boolean from this stack is $\algoTrue$ if the branching is sensitive, and $\algoFalse$ otherwise.
\item \aspectSensitBranching is a Boolean indicating if the current symbolic state corresponds to a sensitive branching.
\end{itemize}
The $\aspectTrigger$ construct declares that an alarm is raised when  \aspectSensitBranching takes the value $\algoTrue$ (\Cref{line:travSensitive:triggerFrom}).

The second part (Lines~\ref{line:travSensitive:utility}--\ref{line:travSensitive:endUtility}) contains \python functions used during the traversal.
They take static aspects as inputs and treat them as variables, allowing them to use static aspect values and update them.
Function \code{isSensitExpr} takes an expression and the \code{Sensitive} static aspect as inputs;
it returns $\algoTrue$ if the expression uses or calls a symbol in \code{Sensitive}.
Function \code{infoFlow} takes written and read expressions as well as the \code{Sensitive} and  \code{ScopeSensitiveBranches} static aspects as inputs.
Symbols in the written expression are added to \code{Sensitive} if there is a direct information flow from a sensitive symbol in the written expression or if there is an indirect information flow from a sensitive branching.
Otherwise, the symbols in the written expression are safe, so they are removed from \code{Sensitive}.
The \code{isSensitExpr} and \code{infoFlow} auxiliary functions exemplify the use of the \code{getExprSymbs} primitive, by capturing used and called symbols in \code{isSensitExpr} and defined symbols in \code{infoFlow}.
Function \code{append\_ScopeSensitiveBranches} takes an expression and the \code{Sensitive} and \code{ScopeSensitiveBranches} static aspects as inputs;
it appends to \code{ScopeSensitiveBranches} a Boolean indicating whether the expression is sensitive or not. 

The third part contains $\aspectPointcut$ constructs that determine how static aspect values are updated, based on the visited statement.
More precisely, the first argument of a  $\aspectPointcut$ is a statement label (\Cref{sec:symbStates}) and its other arguments store the expressions used at the statement.
For instance, \code{pointcut(Assign, left, right)} (\Cref{line:travSensitive:Assign}) involves the \labelAssign statement label, indicating an assignment, and two variables \code{left} and \code{right} which capture the LHS and the RHS of the assignment.
Then the code within the scope of a $\aspectPointcut$ construct constitutes its advice code.
For instance, the block of code associated with \labelEnterProcedure at lines~\ref{line:travSensitive:EnterProcedure:start}--\ref{line:travSensitive:EnterProcedure:end} initializes the static aspect values when the traversal starts.
In particular, the \code{getDescrSymbs} primitive (\Cref{line:travSensitive:EnterProcedure:start}) is used to initialize the \code{Sensitive} variable with the set of source symbols obtained from \code{labeled\_symbols}.

We provide in \Cref{fig:sad:confidentiality} of Appendix~\ref{sec:secondSadExample} (provided as supplementary material) another example of static aspect definition to demonstrate how the \code{fromTraversal} construct is used to import static aspects from past traversals.

 \section{Evaluation}
\label{sec:evaluation}

We have implemented \aspectTool in \python 3.9.6; the \aspectLg parser and lexer code was generated using ANTLR 4.11.0.
In this section, we report our results on the evaluation of the
implementation of our approach.
To ensure \aspectTool can be used by developers to prevent security vulnerabilities in \python projects, we evaluate its ability to detect as many vulnerabilities as possible while reporting as few false alarms as possible.
We also assess the efficiency of \aspectTool in detecting security vulnerabilities.
To summarize, we investigate the following research questions (RQs):
\begin{enumerate}[RQ1:]
\item What is the effectiveness of \aspectTool in detecting security
  vulnerabilities, compared to baseline tools?
We answer this RQ in two steps, to reflect the two settings with which
the baseline tools can be run:
\begin{enumerate}[RQ{1}.1:]
\item What is the effectiveness of \aspectTool in detecting security
  vulnerabilities compared to baseline tools, \emph{when the latter
    do not use any additional information}?
\item What is the effectiveness of \aspectTool in detecting security
  vulnerabilities compared to baseline tools, \emph{when the latter
    use custom rules}?
\end{enumerate}
\item What is the efficiency of \aspectTool in detecting security
  vulnerabilities, compared to baseline tools?
\end{enumerate}

\subsection{Benchmarks}
\label{sec:benchmarks}

The evaluation of the \aspectTool approach requires benchmarks satisfying the following criteria:
\begin{enumerate}
\item The source code of the programs in the benchmark should be available, as the \aspectTool approach relies on static analysis of source code; programs distributed only in binary form are excluded.
\item The source code should be written in \python, as the current implementation of \aspectTool only supports \python.
\item The source code should contain security vulnerabilities to determine whether \aspectTool or alternative approaches can detect and locate them.
\item A vulnerable version of the source code should come with a fixed
  version, to determine whether \aspectTool or alternative approaches
  raise any alarm even when no vulnerability is present.
\end{enumerate}

Despite the popularity and importance of \python, there is a lack of a comprehensive \python benchmark for program analysis~\cite{BKP24};
most studies rely on synthetic benchmarks, which often fail to capture
the behavior of vulnerabilities in the wild~\cite{ZWL+26}. For these
reasons, there is a need to evaluate security tools against
vulnerabilities from real-world \python packages.
We considered the DyPyBench and DepHealth datasets.
The former encompasses 50 Python projects~\cite{BKP24}, which were cloned from the latest commit as of 2023-01-18.
However, from commit descriptions, pull requests, and release notes, it appears that no code changes in the dataset are security-related and thus do not meet our third criterion.
The latter extracted security vulnerabilities affecting PyPi packages~\cite{ACS23} from the Snyk vulnerability database~\cite{Snyk} and satisfies our criteria.

From the DepHealth dataset, we selected entries that point to a commit (allowing us to analyze the source code) and are associated with a CVE identifier, resulting in 134 vulnerabilities.
We manually reviewed all of them, using vulnerability descriptions and references from the CVE website\footnote{\url{https://www.cve.org}}
as well as code changes, commit messages, and developer discussions to locate and identify each vulnerability.
Among the 134 vulnerabilities, we ignored 26, obtaining a dataset of 108 vulnerabilities, each associated with its corresponding \python vulnerability-fixing commit.
The 26 entries we ignored are:
\begin{enumerate}[(a)]\item 17 entries that were written in unsupported languages such as C/C++ and JavaScript;
\item 6 redundant entries, i.e., which are either i) CVE/Snyk duplicates or ii) pointing to the same commit and vulnerability (even if the CVE is different);\item the entries for CVE-2015-5607 and CVE-2016-6298, because the corresponding code change consists only of the introduction of a new procedure, which is not called in the rest of the code, making it impossible to compare the procedure fix before and after the commit;
\item the entry for CVE-2014-0012 because the vulnerability was fixed in another, unmentioned commit.
\end{enumerate}

Some of the remaining 108 vulnerabilities were written in \python 2, while \aspectTool relies on the \python 3 \pyCode{ast.parse} method to build SCFGs.
 We used the \texttt{2to3 [filename].py -w} command line to automatically translate the source code from \python 2 to \python 3 when necessary.
In this case, we retained the generated \texttt{[filename].py.bak} files in the dataset so readers can check the original file.

We remark that each vulnerability is characterized by a unique CVE identifier and is associated with a CWE identifier~\cite{Cwe}, from the NIST database\footnote{\url{https://nvd.nist.gov/vuln}}
when available or from the Snyk database~\cite{Snyk} otherwise.
This CWE identifier is the \emph{vulnerability type} of the vulnerability.
For instance, vulnerability CVE-2016-2513 has vulnerability type
CWE-200, i.e., it corresponds to an exposure of sensitive information.

\subsection{Baseline tools}
\label{sec:baselines}

For fair comparison with \aspectTool, the baseline tools must satisfy the following criteria:
\begin{enumerate}
\item Like \aspectTool, they should be static analyzers. In this way, we exclude dynamic approaches that require program execution, such as concolic testing.
\item They should support the analysis of \python code, enabling them to analyze source code from our dataset.
\item They should be able to detect and locate security
  vulnerabilities (so that their vulnerability detection capabilities can be compared). This means excluding static approaches not explicitly designed for vulnerability analysis, e.g., symbolic execution or linting tools. 
\item They should be maintained (i.e., they should have received at
  least a new commit within the last year). 
\item They should be freely available to ensure the reproducibility of our experiments.
\item They should provide command-line interface (CLI) support for our experiments; this means excluding tools that require a graphical interface.
\item They should not provide results obtained by simply combining results from multiple tools. \end{enumerate}
Based on the above, we identified the following four state-of-the-art tools for comparison with \aspectTool.

1) \bandit (8k stars on \github)~\cite{Bandit} is a tool designed to find common security issues in \python code.
\bandit has been used to detect security
smells~\cite{RRW19} and in the huskyCI~\cite{HuskyCI}
framework. Moreover, a fork has been recently used for malware detection~\cite{VNM23}.

2) \semgrep (15k stars on \github)~\cite{SemGrep} is an open-source static analyzer for finding bugs, detecting vulnerabilities in third-party dependencies, and enforcing code standards.
Both \bandit and \semgrep are used in frameworks like GitLab Static Application Security Testing (SAST)~\cite{GitLabSAST}
and Automated Security Helper~\cite{Ash}.

3) \pysa~\cite{Pysa} is a security-focused static analyzer that reasons about data flows in \python applications.
It is built on top of Pyre (7k stars on \github)\footnote{\url{https://github.com/facebook/pyre-check}}
a type checker for \python, and has been used in several studies~\cite{YLX+21,RBO+24}.

4) \joern (3k stars on \github)~\cite{Joern} is an open-source code analysis platform based on code property graphs (CPGs) and has been used in several studies to detect security vulnerabilities in C/C++ code~\cite{MRC+20,ZLS+19,LZX+16}; nevertheless, it can be applied to \python code.
\joern comes pre-packaged with the code scanner \joernScan, targeting security vulnerabilities.

\subsection{Settings}

\subsubsection{System configuration}
\label{sec:results:config}

We performed all experiments in a Python virtual environment on a MacBook Pro running macOS Sequoia 15.3.2 with a 2.4 GHz 8-Core Intel Core i9 processor.

\subsubsection{Settings for \aspectTool}
\label{sec:settings:saga}

Static aspect definitions are necessary to run \aspectTool and, thus, measure its effectiveness and efficiency.
The first author used the \aspectLg language to express static aspects of interest for security vulnerability detection, obtaining the \aspectLg library of five static aspect definitions 
(Appendix~\ref{sec:sable:sadLibrary}, provided as supplementary material),
associated each vulnerability in the dataset with a static aspect definition, and wrote the corresponding source code annotations (\Cref{sec:staticAspects:expert}).

\subsubsection{Settings for baseline tools}
\label{sec:settings:tools}

We used \bandit version 1.7.5, which comes with built-in rules\footnote{\url{https://bandit.readthedocs.io/en/latest/plugins/index.html\#plugin-id-groupings}}.
We opted for \semgrep version 1.52.0, specifically the \semgrep OSS (Community Edition) version, since it is open source and supports intra-procedural analysis (as \aspectTool) using built-in rules\footnote{\url{https://semgrep.dev/docs/semgrep-pro-vs-oss}}.
We selected the version of \pysa distributed with Pyre 0.9.25, which includes pre-written models (similar to \aspectTool code annotations) and rules (similar to \aspectLg static-aspect definitions) for much of the Python standard library and many open-source libraries~\cite{PysaTutorial}.
We ran \pysa with a default configuration file indicating the source code path and pre-written models and rules, as well as a \pysa file fixing an issue from Pyre when type-checking the \texttt{numpy} package present in our virtual environment\footnote{\url{https://github.com/facebook/pyre-check/issues/915}}.
Finally, we used \joern version 4.0.402, which comes with a default set of queries, the Joern Query Database\footnote{\url{https://github.com/joernio/joern/tree/master/querydb}}.

\subsection{RQ1: \aspectTool effectiveness}
\label{sec:results:effectiveness}

To answer RQ1, we evaluate SAGA's capability to detect security
vulnerabilities without reporting too many false alarms, compared to
the baseline tools.
We performed an empirical study for RQ1 in two steps:
\begin{enumerate}[RQ{1}.1:]
\item We applied \aspectTool and the baseline tools \quotes{out-of-the-box} on the whole dataset of 108 vulnerabilities, i.e., without any particular configuration or files beyond those strictly necessary to run the tool. \item Baseline tools \quotes{out-of-the-box} may not fully realize their potential; at the same time, writing custom rules for all of them and the entire dataset was intractable.
Hence, we used the results obtained from RQ1.1 to determine the scope of RQ1.2, namely, in terms of the sample dataset on which to pursue the analysis (as detailed in \Cref{sec:effectiveness:two:method}) and the custom rules to be written (when possible) for each baseline tool.
We then applied \aspectTool and the baseline tools, using custom rules, to the sample dataset.
\end{enumerate}

In the following, we first detail our choice of performance metrics
for detection effectiveness. Then, we describe the methodology and
report the results for each step.

\subsubsection{Performance Metrics}
\label{sec:metrics}

To answer RQ1, we determine the effectiveness of \aspectTool and the baseline tools in detecting security vulnerabilities.
Their primary output consists of vulnerability locations, i.e., line numbers in the source code where vulnerabilities were detected.
For each commit fixing a vulnerability, we consider \emph{the code in the revision before the commit} as vulnerable and \emph{the code in the revision after the commit} as secure.
We also use the vulnerability description and the code changes to manually identify the ground-truth vulnerability locations in the code before and after the commit.

Since the code before commit is vulnerable, each detected ground-truth location is a \emph{true positive} (TP), while each undetected ground-truth location is a \emph{false negative} (FN). However, any other detected location is a \emph{false positive} (FP).
Since the code after commit is secure, each detected location is also an FP, while each undetected ground-truth location is a \emph{true negative} (TN).
This method (called \emph{strict locations} in the following) considers detections correct that match exactly the ground-truth locations we manually identified, but baseline tools may disagree on the exact location of the vulnerability while still correctly detecting it and thus bringing attention to security issues in the source code.
This stems from differences across tools in security policies, their assumptions, and approximations. 
For instance, \pysa may report any line number in the dataflow between a source and a sink~\cite{PysaTutorial}, whereas we consider the information flow policy to be violated at the sink location.

For the sake of avoiding any favorable bias toward \aspectTool, and thus remaining conservative in our comparisons, we consider an alternative method (called \emph{relaxed locations} in the following) such that the baseline tools are considered correct if they detected vulnerabilities, wherever they are, while \aspectTool is still required to precisely match the ground-truth locations.
More precisely, detections by a baseline tool in the code before commit are TPs if they do not exceed the number of ground-truth locations, while detections in excess are FPs.
However, if there are fewer detections than ground-truth locations, the difference is the number of FNs.
Moreover, each detection by a baseline tool in the code after commit is also an FP, while the number of TNs is the difference between the number of ground-truth locations and the number of detections.

We detail the computation of the number of TPs, FNs, FPs, and TNs, as well as our choice of metrics, in Appendix~\ref{sec:appendix:metrics} (provided as supplementary material).
We use these numbers to report, for \aspectTool and each baseline tool, its \emph{sensitivity} (also called recall) and \emph{specificity}, as well as (for the sake of comparison with other studies) its \emph{precision}.

Nevertheless, developers cannot anticipate the vulnerabilities they
will encounter; moreover, they require the best possible tool to address the issues they will face.
Hence, in addition to the previous \emph{per-location} metrics, we also propose a \emph{per-vulnerability} analysis, enabling us to compare tools at the vulnerability level.
For each metric (sensitivity, specificity, and precision), we compare pairs of tools to determine which ranked best more often than the others, using a a Wilcoxon signed-rank test~\cite{Dem06,AB14}.
This test provides a $\pValue$-value indicating how likely the observation of the samples is, assuming the null hypothesis that both tools have similar performances.
To assess practical significance, we also consider a metric $\effectSize$ for effect size that ranges from $0$ to $1$, where $\effectSize = 0$ indicates that the first tool performs better in every case and $\effectSize = 1$ indicates that the second tool performs better in every case.
Again, we detail the computations in Appendix~\ref{sec:appendix:metrics}.

\subsubsection{RQ1.1 Methodology}
\label{sec:effectiveness:one:method}

To assess the effectiveness of \aspectTool in detecting security vulnerabilities, we ran it for the 108 vulnerabilities in our dataset with our library of static aspect definitions; we also ran the baseline tools \quotes{out-of-the-box} on the full dataset.

\subsubsection{RQ1.1 Results}
\label{sec:effectiveness:one:results}

We present, in \Cref{tab:performance:full}, the metric scores obtained by \aspectTool and the baseline tools globally across all vulnerability locations.
We report the results for both the strict- and relaxed-location methods. \textbf{\aspectTool 
perfectly detected all vulnerabilities in the dataset, achieving 100\% sensitivity and 99.15\% specificity, with only one false positive}.

\begin{table}[t]
\footnotesize
\centering
\setlength\tabcolsep{2.0pt}
\caption{RQ1.1 results: metrics for \aspectTool and baseline tools (full dataset, strict = strict locations, relax = relaxed locations).}
\label{tab:performance:full}
\begin{tabular}{ccccccc}
\toprule
    & \multicolumn{6}{c}{Metrics} \\
\cmidrule{2-7}
    & \multicolumn{2}{c}{sensitivity} & \multicolumn{2}{c}{specificity} & \multicolumn{2}{c}{precision} \\
\cmidrule(lr){2-3}
\cmidrule(lr){4-5}
\cmidrule(lr){6-7}
\multicolumn{1}{c}{Tools}
    & strict & relax
    & strict & relax
    & strict & relax \\
\midrule
\multicolumn{1}{c}{\aspectTool}
    & 100.0\% & 100.0\%
    & 99.15\% & 99.15\%
    & 99.17\% & 99.17\% \\
\multicolumn{1}{c}{\bandit}
    & 11.76\% & 59.66\% 
    & 18.39\% & 10.4\% 
    & 2.79\% & 14.14\% \\
\multicolumn{1}{c}{\semgrep}
    & 3.36\% & 28.57\% 
    & 42.07\% & 39.81\% 
    & 2.48\% & 21.12\% \\
\multicolumn{1}{c}{\pysa}
    & 0.0\% & 0.84\% 
    & 95.12\% & 95.87\% 
    & 0.0\% & 16.67\% \\
\multicolumn{1}{c}{\joern}
    & 0.0\% & 31.93\% 
    & 33.52\% & 29.2\% 
    & 0.0\% & 16.38\% \\
\bottomrule
\end{tabular}
\end{table}
 \begin{table}[t]
\footnotesize
\centering
\setlength\tabcolsep{1.5pt}
\caption{RQ1.1 results: comparison of sensitivity/recall values for each vulnerability (full dataset).}
\label{tab:comparison:sensitivity}
\begin{tabular}{cc|c|c|c|c|c|}
\cline{3-7}
    & & \aspectTool & \bandit & \semgrep & \pysa & \joern \\
\hline
\multicolumn{1}{|c|}{\multirow{2}{*}{\aspectTool}} & $\pValue$
	 & 
	 & \cellcolor{gray!65}\num{1.6e-09}
	 & \cellcolor{gray!92}\num{2.9e-18}
	 & \cellcolor{gray!100}\num{4.5e-25}
	 & \cellcolor{gray!89}\num{3.8e-17}\\
\multicolumn{1}{|c|}{} & $\effectSize$
	 & 
	 & \cellcolor{gray!65}$\Uparrow$ 0.82
	 & \cellcolor{gray!92}$\Uparrow$ 0.96
	 & \cellcolor{gray!100}$\Uparrow$ 1.0
	 & \cellcolor{gray!89}$\Uparrow$ 0.95\\
\hline
\multicolumn{1}{|c|}{\multirow{2}{*}{\bandit}} & $\pValue$
	 & \cellcolor{gray!65}\num{1.6e-09}
	 & 
	 & \cellcolor{gray!51}\num{2.5e-06}
	 & \cellcolor{gray!85}\num{3.0e-15}
	 & \cellcolor{gray!42}\num{8.1e-05}\\
\multicolumn{1}{|c|}{} & $\effectSize$
	 & \cellcolor{gray!65}$\Downarrow$ 0.18
	 & 
	 & \cellcolor{gray!51}$\Uparrow$ 0.75
	 & \cellcolor{gray!85}$\Uparrow$ 0.92
	 & \cellcolor{gray!42}$\Uparrow$ 0.71\\
\hline
\multicolumn{1}{|c|}{\multirow{2}{*}{\semgrep}} & $\pValue$
	 & \cellcolor{gray!92}\num{2.9e-18}
	 & \cellcolor{gray!51}\num{2.5e-06}
	 & 
	 & \cellcolor{gray!49}\num{3.4e-06}
	 & \cellcolor{white!6}\num{5.5e-01}\\
\multicolumn{1}{|c|}{} & $\effectSize$
	 & \cellcolor{gray!92}$\Downarrow$ 0.04
	 & \cellcolor{gray!51}$\Downarrow$ 0.25
	 & 
	 & \cellcolor{gray!49}$\Uparrow$ 0.74
	 & \cellcolor{white!6}0.47\\
\hline
\multicolumn{1}{|c|}{\multirow{2}{*}{\pysa}} & $\pValue$
	 & \cellcolor{gray!100}\num{4.5e-25}
	 & \cellcolor{gray!85}\num{3.0e-15}
	 & \cellcolor{gray!49}\num{3.4e-06}
	 & 
	 & \cellcolor{gray!53}\num{6.0e-07}\\
\multicolumn{1}{|c|}{} & $\effectSize$
	 & \cellcolor{gray!100}$\Downarrow$ 0.0
	 & \cellcolor{gray!85}$\Downarrow$ 0.08
	 & \cellcolor{gray!49}$\Downarrow$ 0.26
	 & 
	 & \cellcolor{gray!53}$\Downarrow$ 0.24\\
\hline
\multicolumn{1}{|c|}{\multirow{2}{*}{\joern}} & $\pValue$
	 & \cellcolor{gray!89}\num{3.8e-17}
	 & \cellcolor{gray!42}\num{8.1e-05}
	 & \cellcolor{white!6}\num{5.5e-01}
	 & \cellcolor{gray!53}\num{6.0e-07}
	 & \\
\multicolumn{1}{|c|}{} & $\effectSize$
	 & \cellcolor{gray!89}$\Downarrow$ 0.05
	 & \cellcolor{gray!42}$\Downarrow$ 0.29
	 & \cellcolor{white!6}0.53
	 & \cellcolor{gray!53}$\Uparrow$ 0.76
	 & \\
\hline
\end{tabular}
\end{table}
 
The only false positive reported by \aspectTool occurred for vulnerability CVE-2014-1829, where the vulnerability was detected but also the fix, because of dynamic properties of the code that could not be assumed by a static analyzer (namely, a conditional branch and a \python binding).
This is detailed in Appendix~\ref{sec:appendix:fp}, provided as supplementary material.

The gap between the results reported for the strict- and relaxed-locations (\Cref{tab:performance:full}) for the baseline tools reflects disagreements over precisely locating security vulnerabilities.
In the rest of this subsection, we focus on the relaxed-location
results to remain conservative to the advantage of baseline tools in our comparison.
\textbf{Baseline tools achieved up to 59.66\% sensitivity and 95.87\%
  specificity; with 100.0\% sensitivity and 99.15\% specificity,
  \aspectTool outperformed them}.

We also used the relaxed-location results to compute the metrics independently for each vulnerability.
In \Cref{tab:comparison:sensitivity,tab:comparison:specificity,tab:comparison:precision}, tools in each row are compared with tools in each column.
To increase readability, we left the diagonal cells empty, since comparing a tool with itself is meaningless.
$\pValue$ denotes the statistical significance and $\effectSize$ the effect size.
When $\pValue > 0.05$, we consider the performance metrics from the two tools not significantly different, and hence the cell is left blank.
Otherwise, we denote the best tool with a symbol, and the cell is colored in a shade of gray according to the effect size, where light shades of gray indicate effect sizes close to $0.5$ and dark shades indicate effect sizes close to either $0$ or $1$.
Since we consider sensitivity/recall, specificity, and precision, the larger the better.
A cell with the $\Uparrow$ (resp. $\Downarrow$) symbol indicates that the tool in the row is better (resp. worse), i.e., with a larger metric, than the tool in the column.

\begin{table}[t]
\footnotesize
\centering
\setlength\tabcolsep{1.5pt}
\caption{RQ1.1 results: comparison, for each vulnerability, of specificity values (full dataset).}
\label{tab:comparison:specificity}
\begin{tabular}{cc|c|c|c|c|c|}
\cline{3-7}
    & & \aspectTool & \bandit & \semgrep & \pysa & \joern \\
\hline
\multicolumn{1}{|c|}{\multirow{2}{*}{\aspectTool}} & $\pValue$
	 & 
	 & \cellcolor{gray!82}\num{2.1e-14}
	 & \cellcolor{gray!48}\num{4.8e-06}
	 & \cellcolor{white!0}\num{1.0e+00}
	 & \cellcolor{gray!53}\num{6.0e-07}\\
\multicolumn{1}{|c|}{} & $\effectSize$
	 & 
	 & \cellcolor{gray!82}$\Uparrow$ 0.91
	 & \cellcolor{gray!48}$\Uparrow$ 0.74
	 & \cellcolor{white!0}0.5
	 & \cellcolor{gray!53}$\Uparrow$ 0.76\\
\hline
\multicolumn{1}{|c|}{\multirow{2}{*}{\bandit}} & $\pValue$
	 & \cellcolor{gray!82}\num{2.1e-14}
	 & 
	 & \cellcolor{gray!48}\num{7.2e-06}
	 & \cellcolor{gray!83}\num{9.7e-15}
	 & \cellcolor{gray!40}\num{1.9e-04}\\
\multicolumn{1}{|c|}{} & $\effectSize$
	 & \cellcolor{gray!82}$\Downarrow$ 0.09
	 & 
	 & \cellcolor{gray!48}$\Downarrow$ 0.26
	 & \cellcolor{gray!83}$\Downarrow$ 0.08
	 & \cellcolor{gray!40}$\Downarrow$ 0.3\\
\hline
\multicolumn{1}{|c|}{\multirow{2}{*}{\semgrep}} & $\pValue$
	 & \cellcolor{gray!48}\num{4.8e-06}
	 & \cellcolor{gray!48}\num{7.2e-06}
	 & 
	 & \cellcolor{gray!49}\num{3.4e-06}
	 & \cellcolor{white!6}\num{5.5e-01}\\
\multicolumn{1}{|c|}{} & $\effectSize$
	 & \cellcolor{gray!48}$\Downarrow$ 0.26
	 & \cellcolor{gray!48}$\Uparrow$ 0.74
	 & 
	 & \cellcolor{gray!49}$\Downarrow$ 0.26
	 & \cellcolor{white!6}0.53\\
\hline
\multicolumn{1}{|c|}{\multirow{2}{*}{\pysa}} & $\pValue$
	 & \cellcolor{white!0}\num{1.0e+00}
	 & \cellcolor{gray!83}\num{9.7e-15}
	 & \cellcolor{gray!49}\num{3.4e-06}
	 & 
	 & \cellcolor{gray!53}\num{6.0e-07}\\
\multicolumn{1}{|c|}{} & $\effectSize$
	 & \cellcolor{white!0}0.5
	 & \cellcolor{gray!83}$\Uparrow$ 0.92
	 & \cellcolor{gray!49}$\Uparrow$ 0.74
	 & 
	 & \cellcolor{gray!53}$\Uparrow$ 0.76\\
\hline
\multicolumn{1}{|c|}{\multirow{2}{*}{\joern}} & $\pValue$
	 & \cellcolor{gray!53}\num{6.0e-07}
	 & \cellcolor{gray!40}\num{1.9e-04}
	 & \cellcolor{white!6}\num{5.5e-01}
	 & \cellcolor{gray!53}\num{6.0e-07}
	 & \\
\multicolumn{1}{|c|}{} & $\effectSize$
	 & \cellcolor{gray!53}$\Downarrow$ 0.24
	 & \cellcolor{gray!40}$\Uparrow$ 0.7
	 & \cellcolor{white!6}0.47
	 & \cellcolor{gray!53}$\Downarrow$ 0.24
	 & \\
\hline
\end{tabular}
\end{table}
 \begin{table}[t]
\footnotesize
\centering
\setlength\tabcolsep{1.5pt}
\caption{RQ1.1 results: comparison, for each vulnerability, of precision values (full dataset).}
\label{tab:comparison:precision}
\begin{tabular}{cc|c|c|c|c|c|}
\cline{3-7}
    & & \aspectTool & \bandit & \semgrep & \pysa & \joern \\
\hline
\multicolumn{1}{|c|}{\multirow{2}{*}{\aspectTool}} & $\pValue$
	 & 
	 & \cellcolor{gray!100}\num{1.2e-19}
	 & \cellcolor{gray!100}\num{4.5e-21}
	 & \cellcolor{gray!100}\num{7.0e-25}
	 & \cellcolor{gray!100}\num{6.2e-21}\\
\multicolumn{1}{|c|}{} & $\effectSize$
	 & 
	 & \cellcolor{gray!100}$\Uparrow$ 1.0
	 & \cellcolor{gray!100}$\Uparrow$ 1.0
	 & \cellcolor{gray!100}$\Uparrow$ 1.0
	 & \cellcolor{gray!100}$\Uparrow$ 1.0\\
\hline
\multicolumn{1}{|c|}{\multirow{2}{*}{\bandit}} & $\pValue$
	 & \cellcolor{gray!100}\num{1.2e-19}
	 & 
	 & \cellcolor{gray!28}\num{8.6e-03}
	 & \cellcolor{gray!83}\num{3.5e-14}
	 & \cellcolor{gray!38}\num{6.2e-04}\\
\multicolumn{1}{|c|}{} & $\effectSize$
	 & \cellcolor{gray!100}$\Downarrow$ 0.0
	 & 
	 & \cellcolor{gray!28}$\Uparrow$ 0.64
	 & \cellcolor{gray!83}$\Uparrow$ 0.92
	 & \cellcolor{gray!38}$\Uparrow$ 0.69\\
\hline
\multicolumn{1}{|c|}{\multirow{2}{*}{\semgrep}} & $\pValue$
	 & \cellcolor{gray!100}\num{4.5e-21}
	 & \cellcolor{gray!28}\num{8.6e-03}
	 & 
	 & \cellcolor{gray!48}\num{7.1e-06}
	 & \cellcolor{white!6}\num{5.8e-01}\\
\multicolumn{1}{|c|}{} & $\effectSize$
	 & \cellcolor{gray!100}$\Downarrow$ 0.0
	 & \cellcolor{gray!28}$\Downarrow$ 0.36
	 & 
	 & \cellcolor{gray!48}$\Uparrow$ 0.74
	 & \cellcolor{white!6}0.53\\
\hline
\multicolumn{1}{|c|}{\multirow{2}{*}{\pysa}} & $\pValue$
	 & \cellcolor{gray!100}\num{7.0e-25}
	 & \cellcolor{gray!83}\num{3.5e-14}
	 & \cellcolor{gray!48}\num{7.1e-06}
	 & 
	 & \cellcolor{gray!51}\num{1.4e-06}\\
\multicolumn{1}{|c|}{} & $\effectSize$
	 & \cellcolor{gray!100}$\Downarrow$ 0.0
	 & \cellcolor{gray!83}$\Downarrow$ 0.08
	 & \cellcolor{gray!48}$\Downarrow$ 0.26
	 & 
	 & \cellcolor{gray!51}$\Downarrow$ 0.24\\
\hline
\multicolumn{1}{|c|}{\multirow{2}{*}{\joern}} & $\pValue$
	 & \cellcolor{gray!100}\num{6.2e-21}
	 & \cellcolor{gray!38}\num{6.2e-04}
	 & \cellcolor{white!6}\num{5.8e-01}
	 & \cellcolor{gray!51}\num{1.4e-06}
	 & \\
\multicolumn{1}{|c|}{} & $\effectSize$
	 & \cellcolor{gray!100}$\Downarrow$ 0.0
	 & \cellcolor{gray!38}$\Downarrow$ 0.31
	 & \cellcolor{white!6}0.47
	 & \cellcolor{gray!51}$\Uparrow$ 0.76
	 & \\
\hline
\end{tabular}
\end{table}
 
\textbf{For sensitivity/recall} (\Cref{tab:comparison:sensitivity})\textbf{, \aspectTool outperformed all the baseline tools}, with small p-values and large effect sizes. \bandit detected more vulnerabilities than the other baseline tools, likely due to its long history as the primary security scanning tool for Python, e.g., through more complete built-in rules.
\semgrep and \joern obtained similar results, while \pysa
obtained the worst sensitivity, a surprising result given
the sources and sinks provided in the pre-written models, but which
likely did not cover the projects in our dataset.

For specificity (\Cref{tab:comparison:specificity}), \aspectTool and
\pysa obtained similar performance, since they both obtained perfect
results for all vulnerabilities but one (CVE-2014-1829 for \aspectTool
and CVE-2016-9964 for \pysa),
hence the effect size of
$0.5$.
\textbf{\aspectTool and \pysa outperformed the other
  tools in terms of specificity}; however, for the latter, this was mostly due to a lack of detection.
Again, \semgrep and \joern obtained similar results.
Finally, \bandit had the worst specificity across vulnerabilities due to its high number of false positives.

Finally, for precision (\Cref{tab:comparison:precision}), \aspectTool outperformed the baseline tools across all vulnerabilities, as indicated by small p-values and an effect size of $1$ in all cases.

In short, as opposed to \bandit (the second-best tool in
terms of sensitivity but the worst in terms of specificity), and \pysa
(the second-best tool in terms of specificity but the worst in terms of sensitivity), \textbf{\aspectTool dominated the baseline tools} in the sense of a Pareto front, i.e., for each metric, \aspectTool was either better or as good as the best alternative.

\subsubsection{RQ1.2 Methodology}
\label{sec:effectiveness:two:method}

Using baseline tools \quotes{out-of-the-box} as we did for RQ1.1 may not fully realize their potential.
We mitigated this issue for RQ1.2 by writing custom rules for the baseline tools.
Nevertheless, the task of writing custom rules for four very different baseline tools and 108 vulnerabilities proved intractable, so we restricted ourselves to a sample of vulnerabilities rather than the full dataset.
We first describe the procedure we followed to select vulnerabilities to obtain a representative sample; then we describe how we obtained the custom rules.

\paragraph{Sample dataset}

We used the results from RQ1.1 to determine the \emph{sample dataset} for RQ1.2.
First, we selected all vulnerabilities for which \aspectTool returned a false positive or a false negative to compare \aspectTool's limitations with those of the other tools.
Specifically, \aspectTool obtained a false positive only for
CVE-2014-1829, which was then selected.

We recall that a static aspect definition characterizes the way, in terms of information policy, in which the vulnerability can occur in the source
code.
We want our sample dataset to be diverse enough in terms
of static aspect definitions. Therefore, for the sake of diversity, we selected all vulnerabilities for which a static aspect definition was used by only a few ($< 5$) vulnerabilities.
The coverage of vulnerabilities by static aspect definitions is as follows: 
51 by \sourceTainting,
27 by \checkEndProc, 
26 by \checkCalls,
3 by \involvedSymbols, and
1 by \contextualValue.
All 108 vulnerabilities were covered by one of the five static aspect definitions provided in the \aspectLg library.
Only three static aspect definitions were necessary to cover 104 (96.30\%) of the 108 vulnerabilities, suggesting that similar data flows occur across vulnerability types.
CVE-2015-3010 is the only vulnerability that uses the \contextualValue static aspect definition, while CVE-2011-4030, CVE-2015-7315, and CVE-2016-9243 are the only vulnerabilities that use the \involvedSymbols static aspect definition; therefore, these four vulnerabilities were selected for the sample.

\begin{table}[t!]
\centering
\footnotesize
\caption{Vulnerability types covered by \aspectLg (ST = \sourceTainting, 
CE = \checkEndProc, 
CC = \checkCalls, 
IS = \involvedSymbols, 
CV = \contextualValue
).} 
\label{tab:cwe}
\begin{tabular}{lcccccc}
\toprule
    & \multicolumn{6}{c}{Number of vulnerabilities}\\
\cmidrule{2-7}
Vuln. type
    & ST & CE & CC & IS & CV & Total\\
\midrule
CWE-20
    & 5 & 4 & 4 & 1 & 0 & 14\\
CWE-79
    & 10 & 3 & 1 & 0 & 0 & 14\\
CWE-200
    & 4 & 3 & 5 & 0 & 1 & 13\\
CWE-264
    & 4 & 2 & 4 & 1 & 0 & 11\\
CWE-94
    & 2 & 0 & 5 & 0 & 0 & 7\\
CWE-22
    & 0 & 4 & 0 & 0 & 0 & 4\\
CWE-287
    & 2 & 1 & 1 & 0 & 0 & 4\\
CWE-59
    & 2 & 0 & 1 & 0 & 0 & 3\\
CWE-78
    & 2 & 0 & 1 & 0 & 0 & 3\\
CWE-310
    & 2 & 0 & 1 & 0 & 0 & 3\\
CWE-399
    & 0 & 3 & 0 & 0 & 0 & 3\\
CWE-284
    & 0 & 1 & 0 & 1 & 0 & 2\\
CWE-295
    & 2 & 0 & 0 & 0 & 0 & 2\\
CWE-611
    & 2 & 0 & 0 & 0 & 0 & 2\\
CWE-19
    & 1 & 0 & 0 & 0 & 0 & 1\\
CWE-21
    & 0 & 0 & 1 & 0 & 0 & 1\\
CWE-77
    & 0 & 1 & 0 & 0 & 0 & 1\\
CWE-89
    & 1 & 0 & 0 & 0 & 0 & 1\\
CWE-93
    & 0 & 1 & 0 & 0 & 0 & 1\\
CWE-119
    & 0 & 1 & 0 & 0 & 0 & 1\\
CWE-203
    & 1 & 0 & 0 & 0 & 0 & 1\\
CWE-254
    & 1 & 0 & 0 & 0 & 0 & 1\\
CWE-269
    & 0 & 1 & 0 & 0 & 0 & 1\\
CWE-276
    & 0 & 1 & 0 & 0 & 0 & 1\\
CWE-285
    & 1 & 0 & 0 & 0 & 0 & 1\\
CWE-330
    & 1 & 0 & 0 & 0 & 0 & 1\\
CWE-346
    & 1 & 0 & 0 & 0 & 0 & 1\\
CWE-361
    & 0 & 0 & 1 & 0 & 0 & 1\\
CWE-362
    & 0 & 1 & 0 & 0 & 0 & 1\\
CWE-379
    & 1 & 0 & 0 & 0 & 0 & 1\\
CWE-384
    & 1 & 0 & 0 & 0 & 0 & 1\\
CWE-400
    & 1 & 0 & 0 & 0 & 0 & 1\\
CWE-532
    & 1 & 0 & 0 & 0 & 0 & 1\\
CWE-601
    & 1 & 0 & 0 & 0 & 0 & 1\\
CWE-755
    & 1 & 0 & 0 & 0 & 0 & 1\\
CWE-804
    & 0 & 0 & 1 & 0 & 0 & 1\\
CWE-918
    & 1 & 0 & 0 & 0 & 0 & 1\\
\bottomrule
\end{tabular}
\end{table}
 
We also recall that each vulnerability in the dataset is associated
with a vulnerability type, which characterizes an effect, e.g., CWE-20
for an improper input validation. We also want our sample dataset to be diverse enough in terms of
vulnerability types. Therefore, to ensure diversity, we completed the sample by selecting one vulnerability from each CWE not yet covered.
More precisely, for each uncovered CWE, we determined the set of vulnerabilities in our dataset associated with that CWE and sorted these sets by size.
Then, from the smallest to the largest set, one vulnerability was selected to ensure the proportion of static aspect definitions in the sample was as close as possible to that in the whole dataset.
\Cref{tab:cwe} summarizes the static aspect definitions used for each vulnerability type.
We obtained a sample of 38 vulnerabilities, since there are 37 CWEs and CWE-200 is covered by both CVE-2014-1829 (\aspectTool's false positive) and CVE-2015-3010 (the only vulnerability for \contextualValue).

\paragraph{Custom rules}

To support vulnerability detection across a wide range of projects, \aspectTool includes a library of five static aspect definitions. While static aspect definitions cover information flows in a general way, rules for the baseline tools are usually designed for specific vulnerability types.
Thus, the fact that the baseline tools missed some vulnerabilities in
the scope of RQ1.1
may not be caused by an intrinsic deficiency  in vulnerability
detection, but simply by the lack of rules covering vulnerability types in our dataset.
To better compare \aspectTool and baseline tools' detection capabilities on the sample dataset, we wrote custom rules for the \bandit, \semgrep, \pysa, and \joern baseline tools.

\bandit plugins are \python files that specify the AST node types \bandit must inspect and the code patterns that indicate a vulnerability.
\semgrep rules are YAML files containing code patterns indicating vulnerabilities.
\pysa models specify sources and sinks in source code, while rules indicate which information flows from source to sink constitute a vulnerability.
\joern rules are written in a DSL embedded in Scala; they contain queries to the CPG and custom code that determine whether a vulnerability occurred.

A dedicated developer, with a background in static analysis and code-level security, completed the task of becoming familiar with the baseline tools and writing custom rules for the sample dataset within three months.
To ensure a fair comparison between the tools, the static aspect definitions and the custom rules were written at the same abstraction level, so that each tool has access to the same information (when possible for that tool) regarding sources, sinks, data flows, etc.
We obtained 4 \bandit plugins, 38 \semgrep rules, 6 \pysa models and rules, and 38 \joern queries.
A \bandit rule is used for 9 vulnerabilities, while each other \bandit rule is used for only one vulnerability.
All \joern queries are collected in a single file, but each query is used for only one vulnerability.
\semgrep and \pysa use one rule per vulnerability.

\subsubsection{RQ1.2 Results}
\label{sec:effectiveness:two:results}

\Cref{tab:sample:performance} details the per-location scores for the relaxed-location method.
With the new custom rules, the sensitivity of all baseline tools increased compared to \Cref{tab:performance:full}, ranging from 72.09\% to 81.4\% for \semgrep, \bandit, and \joern, while \pysa reached only 18.6\% because it performs only source-to-sink analysis.
\aspectTool's specificity slightly decreased, as its only false positive was retained, while the sample contains fewer vulnerabilities than the full dataset.
\bandit and \pysa's specificity decreased slightly as we wrote custom rules to detect more vulnerabilities, sometimes at the cost of more false positives.
In summary, \textbf{baseline tools achieved up to 81.4\% sensitivity and 89.36\% specificity. Hence, with 100.0\% sensitivity and 97.67\% specificity, \aspectTool outperformed them}.

\Cref{tab:sample:comparison:sensitivity,tab:sample:comparison:specificity,tab:sample:comparison:precision} compare \aspectTool and baseline tools for the per-vulnerability analysis.
Again, \textbf{for sensitivity/recall} (\Cref{tab:sample:comparison:sensitivity})\textbf{, \aspectTool outperformed all the baseline tools}, with small p-values and large effect sizes. In contrast with \Cref{tab:comparison:sensitivity}, \bandit, \semgrep, and \joern obtained similar results.
Again, \pysa obtained the worst sensitivity scores.
\textbf{For specificity} (\Cref{tab:sample:comparison:specificity}),
\textbf{\aspectTool and \pysa} obtained similar performances and
\textbf{outperformed the other tools}; however, for the latter, this was mostly due to a lack of detection.
Finally, for precision (\Cref{tab:sample:comparison:precision}), \aspectTool outperformed the baseline tools for all vulnerabilities, as indicated by the small p-values and effect sizes.
In short, again, \textbf{\aspectTool dominated the baseline tools} by being better or as good as the best alternative for each metric.

\begin{table}[t]
\footnotesize
\centering
\caption{RQ1.2 results: performance metrics for \aspectTool and the considered baseline tools (sample dataset).}
\label{tab:sample:performance}
\begin{tabular}{cccc}
\toprule
    & \multicolumn{3}{c}{Performance metrics} \\
\cmidrule{2-4}
\multicolumn{1}{c}{Tools}
    & sensitivity & specificity & precision \\
\midrule
\multicolumn{1}{c}{\aspectTool}
    & 100.0\% & 97.67\% & 97.73\% \\
\multicolumn{1}{c}{\bandit}
    & 76.74\% & 3.73\% & 11.34\% \\
\multicolumn{1}{c}{\semgrep}
    & 72.09\% & 63.49\% & 57.41\% \\
\multicolumn{1}{c}{\pysa}
    & 18.6\% & 89.36\% & 61.54\% \\
\multicolumn{1}{c}{\joern}
    & 81.4\% & 67.31\% & 67.31\% \\
\bottomrule
\end{tabular}
\end{table}
 \begin{table}[t]
\footnotesize
\centering
\setlength\tabcolsep{2.0pt}
\caption{RQ1.2 results: comparison of sensitivity/recall values for each vulnerability (sample dataset).}
\label{tab:sample:comparison:sensitivity}
\begin{tabular}{cc|c|c|c|c|c|}
\cline{3-7}
    & & \aspectTool & \bandit & \semgrep & \pysa & \joern \\
\hline
\multicolumn{1}{|c|}{\multirow{2}{*}{\aspectTool}} & $\pValue$
	 & 
	 & \cellcolor{gray!45}\num{1.0e-02}
	 & \cellcolor{gray!45}\num{1.0e-02}
	 & \cellcolor{gray!97}\num{1.7e-08}
	 & \cellcolor{gray!37}\num{3.3e-02}\\
\multicolumn{1}{|c|}{} & $\effectSize$
	 & 
	 & \cellcolor{gray!45}$\Uparrow$ 0.73
	 & \cellcolor{gray!45}$\Uparrow$ 0.73
	 & \cellcolor{gray!97}$\Uparrow$ 0.99
	 & \cellcolor{gray!37}$\Uparrow$ 0.69\\
\hline
\multicolumn{1}{|c|}{\multirow{2}{*}{\bandit}} & $\pValue$
	 & \cellcolor{gray!45}\num{1.0e-02}
	 & 
	 & \cellcolor{white!0}\num{1.0e+00}
	 & \cellcolor{gray!79}\num{1.1e-05}
	 & \cellcolor{white!10}\num{5.9e-01}\\
\multicolumn{1}{|c|}{} & $\effectSize$
	 & \cellcolor{gray!45}$\Downarrow$ 0.27
	 & 
	 & \cellcolor{white!0}0.5
	 & \cellcolor{gray!79}$\Uparrow$ 0.89
	 & \cellcolor{white!10}0.45\\
\hline
\multicolumn{1}{|c|}{\multirow{2}{*}{\semgrep}} & $\pValue$
	 & \cellcolor{gray!45}\num{1.0e-02}
	 & \cellcolor{white!0}\num{1.0e+00}
	 & 
	 & \cellcolor{gray!79}\num{1.1e-05}
	 & \cellcolor{white!10}\num{5.9e-01}\\
\multicolumn{1}{|c|}{} & $\effectSize$
	 & \cellcolor{gray!45}$\Downarrow$ 0.27
	 & \cellcolor{white!0}0.5
	 & 
	 & \cellcolor{gray!79}$\Uparrow$ 0.89
	 & \cellcolor{white!10}0.45\\
\hline
\multicolumn{1}{|c|}{\multirow{2}{*}{\pysa}} & $\pValue$
	 & \cellcolor{gray!97}\num{1.7e-08}
	 & \cellcolor{gray!79}\num{1.1e-05}
	 & \cellcolor{gray!79}\num{1.1e-05}
	 & 
	 & \cellcolor{gray!88}\num{1.0e-06}\\
\multicolumn{1}{|c|}{} & $\effectSize$
	 & \cellcolor{gray!97}$\Downarrow$ 0.01
	 & \cellcolor{gray!79}$\Downarrow$ 0.11
	 & \cellcolor{gray!79}$\Downarrow$ 0.11
	 & 
	 & \cellcolor{gray!88}$\Downarrow$ 0.06\\
\hline
\multicolumn{1}{|c|}{\multirow{2}{*}{\joern}} & $\pValue$
	 & \cellcolor{gray!37}\num{3.3e-02}
	 & \cellcolor{white!10}\num{5.9e-01}
	 & \cellcolor{white!10}\num{5.9e-01}
	 & \cellcolor{gray!88}\num{1.0e-06}
	 & \\
\multicolumn{1}{|c|}{} & $\effectSize$
	 & \cellcolor{gray!37}$\Downarrow$ 0.31
	 & \cellcolor{white!10}0.55
	 & \cellcolor{white!10}0.55
	 & \cellcolor{gray!88}$\Uparrow$ 0.94
	 & \\
\hline
\end{tabular}
\end{table}
 \begin{table}[t]
\footnotesize
\centering
\setlength\tabcolsep{2.0pt}
\caption{RQ1.2 results: comparison, for each vulnerability, of specificity values (sample dataset).}
\label{tab:sample:comparison:specificity}
\begin{tabular}{cc|c|c|c|c|c|}
\cline{3-7}
    & & \aspectTool & \bandit & \semgrep & \pysa & \joern \\
\hline
\multicolumn{1}{|c|}{\multirow{2}{*}{\aspectTool}} & $\pValue$
	 & 
	 & \cellcolor{gray!87}\num{6.3e-07}
	 & \cellcolor{white!27}\num{1.2e-01}
	 & \cellcolor{white!0}\num{1.0e+00}
	 & \cellcolor{gray!36}\num{4.4e-02}\\
\multicolumn{1}{|c|}{} & $\effectSize$
	 & 
	 & \cellcolor{gray!87}$\Uparrow$ 0.94
	 & \cellcolor{white!27}0.64
	 & \cellcolor{white!0}0.5
	 & \cellcolor{gray!36}$\Uparrow$ 0.68\\
\hline
\multicolumn{1}{|c|}{\multirow{2}{*}{\bandit}} & $\pValue$
	 & \cellcolor{gray!87}\num{6.3e-07}
	 & 
	 & \cellcolor{gray!88}\num{1.3e-06}
	 & \cellcolor{gray!91}\num{2.8e-07}
	 & \cellcolor{gray!76}\num{2.4e-05}\\
\multicolumn{1}{|c|}{} & $\effectSize$
	 & \cellcolor{gray!87}$\Downarrow$ 0.06
	 & 
	 & \cellcolor{gray!88}$\Downarrow$ 0.06
	 & \cellcolor{gray!91}$\Downarrow$ 0.04
	 & \cellcolor{gray!76}$\Downarrow$ 0.12\\
\hline
\multicolumn{1}{|c|}{\multirow{2}{*}{\semgrep}} & $\pValue$
	 & \cellcolor{white!27}\num{1.2e-01}
	 & \cellcolor{gray!88}\num{1.3e-06}
	 & 
	 & \cellcolor{white!27}\num{1.2e-01}
	 & \cellcolor{white!13}\num{4.7e-01}\\
\multicolumn{1}{|c|}{} & $\effectSize$
	 & \cellcolor{white!27}0.36
	 & \cellcolor{gray!88}$\Uparrow$ 0.94
	 & 
	 & \cellcolor{white!27}0.36
	 & \cellcolor{white!13}0.57\\
\hline
\multicolumn{1}{|c|}{\multirow{2}{*}{\pysa}} & $\pValue$
	 & \cellcolor{white!0}\num{1.0e+00}
	 & \cellcolor{gray!91}\num{2.8e-07}
	 & \cellcolor{white!27}\num{1.2e-01}
	 & 
	 & \cellcolor{gray!36}\num{4.4e-02}\\
\multicolumn{1}{|c|}{} & $\effectSize$
	 & \cellcolor{white!0}0.5
	 & \cellcolor{gray!91}$\Uparrow$ 0.96
	 & \cellcolor{white!27}0.64
	 & 
	 & \cellcolor{gray!36}$\Uparrow$ 0.68\\
\hline
\multicolumn{1}{|c|}{\multirow{2}{*}{\joern}} & $\pValue$
	 & \cellcolor{gray!36}\num{4.4e-02}
	 & \cellcolor{gray!76}\num{2.4e-05}
	 & \cellcolor{white!13}\num{4.7e-01}
	 & \cellcolor{gray!36}\num{4.4e-02}
	 & \\
\multicolumn{1}{|c|}{} & $\effectSize$
	 & \cellcolor{gray!36}$\Downarrow$ 0.32
	 & \cellcolor{gray!76}$\Uparrow$ 0.88
	 & \cellcolor{white!13}0.43
	 & \cellcolor{gray!36}$\Downarrow$ 0.32
	 & \\
\hline
\end{tabular}
\end{table}
 \begin{table}[t]
\footnotesize
\centering
\setlength\tabcolsep{2.0pt}
\caption{RQ1.2 results: comparison, for each vulnerability, of precision values (sample dataset).}
\label{tab:sample:comparison:precision}
\begin{tabular}{cc|c|c|c|c|c|}
\cline{3-7}
    & & \aspectTool & \bandit & \semgrep & \pysa & \joern \\
\hline
\multicolumn{1}{|c|}{\multirow{2}{*}{\aspectTool}} & $\pValue$
	 & 
	 & \cellcolor{gray!100}\num{7.5e-08}
	 & \cellcolor{gray!69}\num{1.6e-04}
	 & \cellcolor{gray!98}\num{1.6e-08}
	 & \cellcolor{gray!59}\num{1.2e-03}\\
\multicolumn{1}{|c|}{} & $\effectSize$
	 & 
	 & \cellcolor{gray!100}$\Uparrow$ 1.0
	 & \cellcolor{gray!69}$\Uparrow$ 0.84
	 & \cellcolor{gray!98}$\Uparrow$ 0.99
	 & \cellcolor{gray!59}$\Uparrow$ 0.79\\
\hline
\multicolumn{1}{|c|}{\multirow{2}{*}{\bandit}} & $\pValue$
	 & \cellcolor{gray!100}\num{7.5e-08}
	 & 
	 & \cellcolor{gray!75}\num{5.1e-05}
	 & \cellcolor{gray!48}\num{9.2e-03}
	 & \cellcolor{gray!86}\num{4.2e-06}\\
\multicolumn{1}{|c|}{} & $\effectSize$
	 & \cellcolor{gray!100}$\Downarrow$ 0.0
	 & 
	 & \cellcolor{gray!75}$\Downarrow$ 0.12
	 & \cellcolor{gray!48}$\Uparrow$ 0.74
	 & \cellcolor{gray!86}$\Downarrow$ 0.07\\
\hline
\multicolumn{1}{|c|}{\multirow{2}{*}{\semgrep}} & $\pValue$
	 & \cellcolor{gray!69}\num{1.6e-04}
	 & \cellcolor{gray!75}\num{5.1e-05}
	 & 
	 & \cellcolor{gray!76}\num{3.5e-05}
	 & \cellcolor{white!11}\num{5.7e-01}\\
\multicolumn{1}{|c|}{} & $\effectSize$
	 & \cellcolor{gray!69}$\Downarrow$ 0.16
	 & \cellcolor{gray!75}$\Uparrow$ 0.88
	 & 
	 & \cellcolor{gray!76}$\Uparrow$ 0.88
	 & \cellcolor{white!11}0.45\\
\hline
\multicolumn{1}{|c|}{\multirow{2}{*}{\pysa}} & $\pValue$
	 & \cellcolor{gray!98}\num{1.6e-08}
	 & \cellcolor{gray!48}\num{9.2e-03}
	 & \cellcolor{gray!76}\num{3.5e-05}
	 & 
	 & \cellcolor{gray!89}\num{1.1e-06}\\
\multicolumn{1}{|c|}{} & $\effectSize$
	 & \cellcolor{gray!98}$\Downarrow$ 0.01
	 & \cellcolor{gray!48}$\Downarrow$ 0.26
	 & \cellcolor{gray!76}$\Downarrow$ 0.12
	 & 
	 & \cellcolor{gray!89}$\Downarrow$ 0.05\\
\hline
\multicolumn{1}{|c|}{\multirow{2}{*}{\joern}} & $\pValue$
	 & \cellcolor{gray!59}\num{1.2e-03}
	 & \cellcolor{gray!86}\num{4.2e-06}
	 & \cellcolor{white!11}\num{5.7e-01}
	 & \cellcolor{gray!89}\num{1.1e-06}
	 & \\
\multicolumn{1}{|c|}{} & $\effectSize$
	 & \cellcolor{gray!59}$\Downarrow$ 0.21
	 & \cellcolor{gray!86}$\Uparrow$ 0.93
	 & \cellcolor{white!11}0.55
	 & \cellcolor{gray!89}$\Uparrow$ 0.95
	 & \\
\hline
\end{tabular}
\end{table}

\subsection{RQ2: \aspectTool efficiency}
\label{sec:results:efficiency}

To answer RQ2, we determine how fast \aspectTool detects security vulnerabilities, compared to alternatives.

\subsubsection{Methodology}

We gathered vulnerability detection information using our \code{gather\_results} function, which calls each tool for all considered vulnerabilities. 
Using the \code{time} package, we obtained the start and end times (in seconds) of \code{gather\_results} and recorded the difference as the tool's execution time for the dataset analysis.
To address nondeterminism caused by the machine's workload, we ran each tool 5 times and reported the median.
As our empirical study for vulnerability detection was
performed in two steps, we ran the tools on both the full (step 1) and
sample (step 2) datasets, reporting the results for each configuration.

Finally, unless specified otherwise, \pysa includes in its analysis all the procedures contained in the virtual environment while the other tools focus on the procedure of interest; for the sake of fair comparison, in step 2 we manually updated each Pyre configuration file in the sample dataset to exclude from the analysis the procedures contained in the virtual environment but not in the source code.

\subsubsection{Results}

\begin{table}[t]
\footnotesize
\centering
\caption{RQ2 results: median execution times (in seconds, for five runs) of \aspectTool and the baseline tools (Full = full dataset and baseline tools out of the box, Sample = sample dataset and custom rules for baseline tools).}
\label{tab:exeTimes}
\begin{tabular}{cccccc}
\toprule
    & \multicolumn{5}{c}{Tools} \\
\cmidrule{2-6}
\multicolumn{1}{c}{Configurations}
    & \aspectTool & \bandit & \semgrep & \pysa & \joern \\
\midrule
\multicolumn{1}{c}{Full}
    & 30.7 & 78.9 & 2633.58 & 15722.16 & 2305.59 \\
\multicolumn{1}{c}{Sample}
    & 11.58 & 28.18 & 211.73 & 772.31 & 966.3 \\
\bottomrule
\end{tabular}
\end{table}
 
\Cref{tab:exeTimes} shows the results for both configurations.
In both cases, \textbf{\aspectTool outperformed all the baseline tools};
It took \SI{30.7}{\s} for the full dataset and \SI{11.58}{\s} for the sample dataset, 
while the baseline tools took between \SI{78.9}{\s} and \SI{15772.16}{\s} (i.e., between 2.5x and 512.1x longer than \aspectTool) for the full dataset
and between \SI{28.18}{\s} and \SI{966.3}{\s} (i.e., between 2.4x and 83x longer) for the sample dataset.

\bandit was the second-best tool in terms of execution time, as it performs a simple analysis that does not account for control or data flows.
\pysa was the slowest tool for the full dataset because, when used \quotes{out
  of the box}, it includes in its analysis all the procedures
contained in the virtual environment.
The execution times of \joern are quite high, as building and analyzing CPGs are time-consuming.

 \section{Discussion}
\label{sec:discussion}

\subsection{Practical implications}
Using \aspectTool requires developers to use static aspect definitions.
Although writing such a definition is a one-time effort, it comes with some costs, as static aspect definitions may be long (e.g., those used in our evaluation  range from 73 to 128 lines
of code), and require more effort (a few hours, based on our experience) than
custom rules used in tools such as \bandit, \semgrep, \pysa, or \joern.
Nevertheless, a custom rule is usually specific to a vulnerability type, while a static aspect definition can usually detect multiple vulnerability types.

We tried to overcome this limitation by providing a library of static
aspect definitions.
In our results (\Cref{sec:effectiveness:two:method}), only three static aspect definitions were necessary to cover 104 (96.30\%) of the 108 vulnerabilities, and five static aspect definitions were necessary to cover all of them, indicating (in an indirect way) the expressiveness of the \aspectLg language.
From these results, we expect the provided library to generalize well.
Hence, when facing a new vulnerability, a developer will probably only need to select the appropriate static aspect definition and write the corresponding source annotation, which requires less effort (a few minutes, based on our experience, once the vulnerability is understood) than writing a new custom rule.

In short, \aspectTool facilitates vulnerability detection by separating the code-specific part (in source code annotations) from the general part (in static aspect definitions) of the rules, and by factorizing similar vulnerability-detection methods into reusable definitions.

\subsection{Threats to Validity}
\label{sec:threats}
The results presented in this section depend on the vulnerability dataset used for evaluating \aspectTool and may not generalize to vulnerabilities in the wild.
This stems from a lack of comprehensive \python benchmarks for program analysis~\cite{BKP24}.
We selected all vulnerabilities that satisfied our evaluation criteria (\Cref{sec:benchmarks}), resulting in a dataset of 108 vulnerabilities. These vulnerabilities were all manually validated by reviewing code changes and contextual information, including commit descriptions and issue discussions.
This dataset covers 37 vulnerability types (\Cref{tab:cwe}), giving us
confidence that our results can generalize.

For RQ1, the evaluation of the effectiveness of the baseline tools in detecting and locating vulnerabilities depends on the ground truth for the vulnerability locations.
We mitigated this threat by using the relaxed-locations method, i.e.,
the output of the baseline tools is considered correct if the tools
detected the vulnerabilities, independently of the reported location.
For RQ1.1, baseline tools are used \quotes{out-of-the-box} (\Cref{sec:baselines}), which may artificially decrease their sensitivity, as rules for detecting vulnerabilities in our dataset may not be available.
We mitigated this threat by writing, for RQ1.2, \bandit plugins, \semgrep rules, \pysa models and rules, and \joern queries for a sample dataset.

For RQ1.2, the experiment results depend on the selection of vulnerabilities and the custom rules.
For the former, we obtained a representative sample by selecting vulnerabilities corresponding to \aspectTool's false positives and negatives, covering all vulnerability types present in the entire dataset, and with a proportion of static aspect definitions similar to that of the entire dataset.
For the latter, to avoid bias in the results, custom rules for the baseline tools were written (over a three-month period) by a developer not involved in writing the \aspectLg library of static aspect definitions.
Nevertheless, our results may not fully reflect the capabilities of the baseline tools. 
More generally, one could increase the sensitivity by writing very specific rules for each
baseline tool, leveraging the \emph{perfect knowledge of the
  vulnerability before it is detected}, so that they are tailored to match a very precise part of
the source code.
For a fair comparison, the static aspect definitions and the custom rules were written at the same abstraction level, so that each tool has access to the same information.

\subsection{Tools Limitations}
\label{sec:limitations}
The baseline tools considered in the evaluation suffer from the
following limitations.
\bandit analyzes each AST node in isolation, which prevents it from considering important features like data-flow analysis, control-flow awareness (e.g., a statement is in a branch), statement sequencing, or detecting the absence of a code pattern (e.g., no sanitization function was called).
\pysa is limited to source-to-sink taint analysis, which prevents it from detecting other code patterns.
Also, it cannot detect the absence of a desired data flow (e.g., from a safe function to a sink).
\semgrep is restricted to specific and rigid pattern-matching definitions, which prevents it from detecting vulnerabilities that do not fit the required templates.
\joern's CPGs contain more information than \aspectTool's SCFGs, but they are heavy in memory and require more time to be traversed and analyzed than SCFGs.
CPG queries offer a large diversity of primitives to explore relevant nodes in a given order, extract relevant information, and write custom code to process it.
Thus, in an iterative exploration of the source code (e.g., in interactive mode),
all the vulnerabilities could be detected.
Nevertheless, we wrote custom rules at the same abstraction level for all the tools (\Cref{sec:threats}), rather than tailoring them.
Also, at the time of evaluation, \joern lacked \python support for \labelFor and comment nodes.

\aspectTool has none of these restrictions, enabling a versatile static analysis which is not limited to taint analysis, e.g., it can handle order-dependent calls (\involvedSymbols checks if a function is called after another one) and stateful properties (\contextualValue remembers values set by specific functions, to be compared later to expected ones).
However, it performs only intra-procedural analysis and does not handle dynamic properties like \python binding (\Cref{sec:effectiveness:one:results}).
The latter is a common issue for static analyzers, e.g., a variable
type may not be correctly inferred even by general program analysis frameworks like Scalpel\footnote{\url{https://python-scalpel.readthedocs.io/en/latest/user-guide/middle-level-modules/Type-Inference.html}}.

 \section{Related Work}
\label{sec:related}

Our approach is related to work in the areas of \emph{source code representation}, \emph{deep learning} and \emph{static analysis} for vulnerability detection.

\subsection{Source code representation}

A recent study reports that graph-based representations outperform sequence- or tree-based representations on tasks such as code classification, clone detection, and vulnerability detection. This warrants better tools to facilitate the construction of code property graphs for programming languages such as Java and \python~\cite{SLX+22}.
Graph construction from \python AST can be done in various ways, for instance, using the JARVIS prototype~\cite{HYC+24}, which evaluates expressions in the graph to perform inter-procedural analysis.
In the \aspectTool approach, we focus on intra-procedural analysis and extract AST information, such as defined and used variables, program locations, and statement labels, for use in static aspect analysis.

\subsection{Deep learning for vulnerability detection}

The same study (\cite{SLX+22}) reports that while deep learning (DL)-based vulnerability detectors perform well in code classification and clone detection tasks, their vulnerability detection performance is less satisfactory.
These DL-based techniques usually have limited ability to locate vulnerabilities at the code-line level, are unable to recognize previously unseen vulnerability-fixing patterns, and are susceptible to selecting features unrelated to vulnerabilities~\cite{CZW+24}.
Hence, \citeauthor{CZW+24} suggests using DL models to identify variables and statements within the program that correlate with vulnerabilities, then static analysis techniques to determine data- or control-flow between these variables and statements.
This suggestion is consistent with \aspectTool's approach, while replacing variable identification using DL models with direct code annotation by security experts or developers.
The same study also notes that manual rule-definition processes can be time-consuming and labor-intensive, underscoring the need for a dedicated DSL such as \aspectLg.

This suggestion was applied in a recent study combining large language
models (LLMs) and static analysis for vulnerability detection in \java
applications~\cite{LDN25}.
The IRIS prototype leverages CodeQL static analysis for each CWE to identify candidate APIs; LLMs to label candidate APIs as sources or sinks; CodeQL to reason about information flows and determine candidate paths; and LLMs to label paths as true or false positives.
They evaluated their approach on a dataset of 120 manually validated vulnerabilities, demonstrating an improvement over static analysis alone and reducing the need for code annotation.

Regarding vulnerability detection in \python source code, the Bidirectional Long Short-Term Memory (BiLSTM) model achieved results (precision from 92.5\% to 97.8\% and recall from 90.7\% to 96.9\%)~\cite{FP24} close to \aspectTool (precision 99.2\% and recall 100\%).
While the \aspectTool approach requires domain knowledge to annotate source code, it does not require labeled datasets for security vulnerabilities, which are often unavailable for \python programs.
Moreover, their approach covered only seven vulnerability types. In comparison, the \aspectTool approach covered 37 vulnerability types, highlighting the efficiency of static analysis techniques for reasoning about information flows that lead to security vulnerabilities.

\subsection{Static analysis for vulnerability detection}

Many tools leverage static analysis to detect security vulnerabilities.
The Andromeda~\cite{TPC+13} commercial tool performs on-demand static analysis on the necessary parts of the program representation to efficiently detect insecure information flows.
Another work~\cite{ZZ13} performs a static analysis separately on the string- and non-string-related parts of the source code.
SFlowInfer~\cite{HDM14} uses a type-based taint analysis with three types: safe, tainted, and an intermediate polymorphic type that can be safe or tainted depending on the evocation context.
JOACO~\cite{TSBB18,TSBB20} extracts security slices of the source code to efficiently determine the information policy violations.

Too often, these works~\cite{ZZ13,HDM14,LK22} focus only on integrity policy violations when tampered data reaches a private sink, overlooking confidentiality when sensitive data reaches a public sink.
This contrasts with \aspectTool, which can simultaneously and flexibly analyze multiple static aspects (sensitive, tampered, etc.) of the source code.

Some works focus on both integrity and confidentiality policies.
For instance, DoubleX~\cite{FSBS21} can detect whether sensitive APIs are executed with attacker-controllable data or could exfiltrate sensitive user data. JOACO~\cite{TSBB18,TSBB20} considers a security lattice with both confidentiality and integrity levels, while Andromeda~\cite{TPC+13} considers information flows in general.

Nevertheless, these tools capture only insecure information flows from sources to sinks.
Moreover, amongst them, only a few~\cite{TPC+13,TSBB18} also address downgraders that allow the endorsement of untrusted inputs or the declassification of confidential information, enabling more realistic information flows~\cite{CMA17}.
Again, this contrasts with \aspectTool, which can handle downgraders or capture intermediate aspects (e.g., sensitive branching) that do not require a sink to leak sensitive information (e.g., side-channel attacks~\cite{Koc96,KJJ99,BCO04}).

Remote side-channel attacks can be detected using the Clockwork~\cite{BBRS20} monitor, which tracks both sensitive branching and public sinks.
Nevertheless, it does not track integrity information, unlike \aspectTool, which can capture all the aforementioned aspects, as well as additional ones.

Another criticism can be raised for some tools~\cite{TSBB18,FSBS21} that consider loop traversal only once.
This may lead to under-approximations, which can be avoided by iterating until a fixed point is reached, as done in \aspectTool (Appendix~\ref{sec:staticAspects:traversal}).

Moreover, the aforementioned tools target PHP~\cite{ZZ13}, Java~\cite{TPC+13,HDM14,TSBB18,TSBB20}, or JavaScript~\cite{TPC+13,BBRS20,LK22} applications,
while we focus on \python source code.
As detailed in \Cref{sec:baselines}, Bandit~\cite{Bandit},
SemGrep~\cite{SemGrep}, Pysa~\cite{Pysa}, and \joern~\cite{Joern} are
\python static analyzers that have been used to detect security
vulnerabilities in several works~\cite{RRW19,YLX+21,VNM23,RBO+24}
and which are integrated in several
tools~\cite{HuskyCI,GitLabSAST,Ash}.
As demonstrated in \Cref{sec:results:effectiveness}, \aspectTool outperforms all these tools in terms of vulnerability detection and execution time.

Recently, \citet{ZWL+26} conducted an empirical study of security-oriented static analysis, comparing tools (including \bandit, \semgrep, and \pysa) on both synthetically generated and manually verified \python datasets.
Since results on the latter were worse than on the former, they concluded results obtained on synthetic datasets may not generalize in the wild.
On the real-world datasets, baseline tools achieved up to 40.7\% sensitivity, similar to our relax-location results (up to 59.66\%).
When combined, the tools achieved around 65\% sensitivity, thereby performing worse than \aspectTool (100\%).
In terms of performance, they concluded that taint-analysis tools like \pysa require more time than pattern-based tools like \bandit, matching our own results, except for  \aspectTool, which performed better despite performing an in-depth analysis.

 \section{Conclusion and Future Work}
\label{sec:conclusion}

\python is the most popular programming language; as such, \python projects involve an increasing number and diversity of security vulnerabilities.
In this paper, we have presented the \aspectTool approach, which efficiently detects security vulnerabilities in \python projects, in a versatile way using static aspects.
Such static aspects are defined using the \aspectLg domain-specific language, based on data- and control-flow information extracted by a source code parser and represented as an SCFG.
\aspectLg is embedded in \python, giving security experts access to the expressive power of a general-purpose programming language to define how static aspects evolve during an SCFG traversal, as opposed to state-of-the-art tools which are restricted to specific detection patterns or pre-defined primitives.
We provide a library comprising five static aspect definitions; moreover, security experts may use \aspectLg to define custom static aspects.

We have evaluated \aspectTool on a large dataset of 108 manually verified security vulnerabilities, affecting 82 PyPI packages and spanning 37 CWEs, and compared it with four state-of-the-art baseline tools.
Only three static aspect definitions were sufficient to cover 104 (96.30\%) of the 108 vulnerabilities, indicating that they are versatile and can be used by developers to detect a range of vulnerabilities.
\aspectTool obtained 100\% sensitivity and 99.15\% specificity, with only one false positive, outperforming the baseline tools, which achieved up to 59.66\% sensitivity and 95.87\% specificity.
This analysis was performed in less than \SI{31}{\s}, i.e., between 2.5 and 512.1 times faster than the baseline tools.

As part of future work, we intend
to perform a user study on the cost of using \aspectTool and \aspectLg, extend the current framework to support inter-procedural analysis, and include DL-based techniques to reduce the need for code annotation.

\section*{Acknowledgments}

The authors would like to thank:
Joshua Heneage Dawes, for his contribution to the approach and his
support during the early stages of this work;
Alexander Vatov, for the implementation of the \scfgPackage and \scfgPython packages and his support during the development of \aspectTool;
Marwa Madni for writing the custom rules to detect vulnerabilities
with the baseline tools;  and the \pysa~\cite{PysaIssueA,PysaIssueB} and \joern~\cite{JoernDiscord} communities for their help to set up the baseline tools.
This work was supported by the H2020 \cosmos European project, grant agreement No. 957254. Lionel Briand was supported by Discovery Grant and Canada Research Chair
programs of the Natural Sciences and Engineering Research Council of Canada (NSERC) and the Research Ireland grant
13/RC/2094-2.

\bibliographystyle{IEEEtranN}

\clearpage
\appendices

\section*{Appendix}
We provide in this appendix a formal definition for the edges of an SCFG (\Cref{sec:scfgEdges}), as well as a second example of static aspect definition (\Cref{sec:secondSadExample}), to illustrate an import from a past traversal and another example of an alarm.
We also provide a description of our static aspect library (\Cref{sec:sable:sadLibrary}) and \aspectLg semantics (\Cref{sec:staticAspects:semantics}), including how the order between several SCFG traversals is determined, how SCFG annotations are updated during a traversal, and a detailed traversal example.
We finally provide a detailed description of the metrics used to evaluate SAGA effectiveness, i.e., to answer RQ1 (\Cref{sec:appendix:metrics}), and a description of the only false positive reported by \aspectTool (\Cref{sec:appendix:fp}) for our dataset.

\section{Formal Definition of SCFG Edges}
\label{sec:scfgEdges}

The function $\edgesRecName(\progStmt, \symbStateScope,$ $\symbStateNext)$
introduced in \Cref{sec:scfg}
takes as arguments the current (sequence of) statement(s) $\progStmt$, 
a symbolic state $\symbStateScope$ indicating the current loop scope in the procedure, and
a symbolic state $\symbStateNext$ indicating the next symbolic state to consider after dealing with $\progStmt$;
It returns a set of edges;
It is defined in \Cref{table:scfg:edges} by induction on $\progStmt$, with each case definition $\eqRule{i}$ being numbered for easier reference.

\begin{figure*}[ht!]
\centering
\scriptsize
$$
\begin{array}{l}
\edgesRec{\progStmt}{\symbStateScope}{\symbStateNext}\\
\left\{
\begin{array}{r@{}l@{}lr}

\eqRule{1}
& \edgesRec{\progStmt_1}{\symbStateScope}{\edgesRecState{\progStmt_2}}
& \text{if } \progStmt = \srcSeq{\progStmt_1}{\progStmt_2}\\

&\cup \edgesRec{\progStmt_2}{\symbStateScope}{\symbStateNext}\\
\\

\eqRule{2}
&\set{\tuple{\symbStateStmt, \symbState_1}}{} \cup \edgesRec{\progStmt_1}{\symbStateScope}{\symbStateEndStmt}
& \text{if } \progStmt = \srcIf\ \srcExpr\srcSeqC\, \srcSeqL\progStmt_1\srcSeqR\\

& \cup \set{\tuple{\symbStateStmt, \symbStateEndStmt}, \tuple{\symbStateEndStmt, \symbStateNext}}{}
& \text{or } \progStmt = \srcStmtElse{\progStmt_1}\\
\\

\eqRule{3}
&\bigcup\limits_{1 \le i \le n} \set{\tuple{\symbStateStmt, \symbState_i}}{} \cup \edgesRec{\progStmt_i}{\symbStateScope}{\symbStateEndStmt}
& \text{if } \progStmt = \srcWith\ \srcExpr_1, \dots, \srcExpr_n \srcSeqC \srcSeqL \progStmt_1 \srcSeqR \hspace{0.7cm} (n = 1)\\

& \cup \set{\tuple{\symbStateEndStmt, \symbStateNext}}{}
& \text{or } \progStmt = \srcIf\ \srcExpr\srcSeqC\, \srcSeqL\progStmt_1\srcSeqR\, \srcElse\srcSeqC\, \srcSeqL \progStmt_2 \srcSeqR \hspace{0.46cm} (n = 2)\\
\\

\eqRule{4}
&\bigcup\limits_{1 \le i \le n} \set{\tuple{\symbStateStmt, \symbStateCase_i}, \tuple{\symbStateCase_i, \symbState_i}}{}
& \text{if } \progStmt = \srcMatch\ {\srcExpr_0}\srcSeqC\, \srcSeqL{\srcStmtCase{\srcExpr_1}{\progStmt_1}}\\

&\quad\quad\quad \cup \edgesRec{\progStmt_i}{\symbStateScope}{\symbStateEndStmt}
&\quad \dots\, \srcStmtCase{\srcExpr_n}{\progStmt_n}\srcSeqR\\

&\cup \set{\tuple{\symbStateStmt, \symbStateEndStmt}, \tuple{\symbStateEndStmt, \symbStateNext}}
&\\
\\

\eqRule{5}
&\bigcup\limits_{1 \le i \le n} \set{\tuple{\symbStateStmt, \symbStateExcept_i}, \tuple{\symbStateExcept_i, \symbState_i}}{}
& \text{if } \progStmt = \srcStmtTry{\progStmt_0}\\

&\quad\quad\quad \cup \edgesRec{\progStmt_i}{\symbStateScope}{\symbStateEndExcept}
&\quad \srcExcept\ \srcExpr_1\srcSeqC\,\srcSeqL \progStmt_1 \srcSeqR\, \dots\, \srcStmtExcept{\srcExpr_n}{\progStmt_n}\\

& \cup \set{\tuple{\symbStateStmt, \symbState_0}, \tuple{\symbStateEndStmt, \symbStateNext}}{}
&\quad \bnfStmtQuestion{\srcStmtElse{\progStmt_{n+1}}}\ \bnfStmtQuestion{\srcStmtFinally{\progStmt_{n+2}}}\\

& \cup \left\{
\begin{array}{l}
\edgesRec{\srcSeq{\progStmt_0}{\srcStmtElse{\progStmt_{n+1}}}}{\symbStateScope}{\symbStateEndExcept}\\
\edgesRec{\progStmt_0}{\symbStateScope}{\symbStateEndExcept}
\end{array}
\right.
&
\begin{array}{l}
\text{in case of } \srcElse \text{ statement}\\
\text{otherwise}
\end{array}
\\
& \cup \set{\tuple{\symbStateFinally_{n+2}, \symbState_{n+2}}}{} \cup \edgesRec{\progStmt_{n+2}}{\symbStateScope}{\symbStateEndStmt}
&
\begin{array}{l}
\text{in case of } \srcFinally \text{ statement}
\end{array}
\\
\\

\eqRule{6}
&\set{\tuple{\symbStateStmt, \symbState_1}}{} \cup \edgesRec{\progStmt_1}{\symbStateStmt}{\symbStateStmt}
& \text{if } \progStmt = \srcStmtWhile{\srcExpr}{\progStmt_1}\ \bnfStmtQuestion{\srcStmtElse{\progStmt_2}}\\

&\cup \set{\tuple{\symbStateStmt, \symbStateEndStmt}}{}
& \text{or } \progStmt = \srcStmtFor{\srcVar}{\srcExpr}{\progStmt_1}\ \bnfStmtQuestion{\srcStmtElse{\progStmt_2}}\\

& \cup \left\{
\begin{array}{l}
\set{\tuple{\symbStateEndStmt, \progSymbState{\srcStmtElse{\progStmt_2}}}}{}\\
\quad \cup \edgesRec{\srcStmtElse{\progStmt_2}}{\symbStateScope}{\symbStateNext}\\
\set{\tuple{\symbStateEndStmt, \symbStateNext}}{}
\end{array}
\right.
&
\begin{array}{l}
\\
\text{in case of } \srcElse \text{ statement}\\
\text{otherwise}
\end{array}
\\
\\

\eqRule{7}
&\set{\tuple{\symbStateStmt, \symbStateScope}}{}
& \text{if } \progStmt = \srcContinue\\
\\

\eqRule{8}
&\set{\tuple{\symbStateStmt, \progEndState{\symbStateScope}}}{}
& \text{if } \progStmt = \srcBreak\\
\\

\eqRule{9}
&\set{\tuple{\symbStateStmt, \symbStateEnd}}{}
& \text{if } \progStmt = \srcStmtReturn{\srcExpr}\\

&
& \text{or } \progStmt = \srcStmtRaise{\srcExpr}\\

\eqRule{10}
&\set{\tuple{\symbStateStmt, \symbStateNext}}{}
& \text{otherwise}\\
\end{array}
\right.
\end{array}
$$
where $\symbStateStmt = \progSymbState{\progStmt}$,
$\symbStateEndStmt = \progEndState{\symbStateStmt}$,
$\symbState_i = \progSymbState{\progStmt_i}$,
$\symbStateExcept_i = \progSymbState{\srcStmtExcept{\srcExpr_i}{\progStmt_i}}$,
$\symbStateFinally_i = \progSymbState{\srcStmtFinally{\progStmt_i}}$,
$\symbStateCase_i = \progSymbState{\srcStmtCase{\srcExpr_i}{\progStmt_i}}$,
$\symbStateEndExcept = \symbStateFinally_{n+2}$ in case of $\srcFinally$ statement and $\symbStateEndStmt$ otherwise,
and the function $\edgesRecState{.}$ is defined by induction on sequences of statements:
$
\edgesRecState{\progStmt}
\eqdef
\left\{
\begin{array}{l@{}l}
\edgesRecState{\progStmt_1}
   &{} \text{ if } \progStmt = \srcSeq{\progStmt_1}{\progStmt_2}\\
\progSymbState{\progStmt}
   &{} \text{ otherwise}\\
\end{array}
\right.
$
 \caption{Inductive definition of function $\edgesRecName$}
\label{table:scfg:edges}
\end{figure*}

The first case $\eqRule{1}$ corresponds to sequences of statements, cases $\eqRule{2}$ to $\eqRule{9}$ to individual control-flow statements, and case $\eqRule{10}$ to the other statements (e.g., assignments).
$\srcExpr$ denotes any valid \python expression,
$\bnfStmtQuestion{x}$ means that item $x$ is optional\footnote{These brackets should not be confused with the brackets \character{[}\ \dots\ \character{]} used in \python list indices or dictionary keys.},
and $\bnfStmtStar{x}$ means that item $x$ appears zero or more times\footnote{These parentheses should not be confused with the parentheses \bnfFunc{\dots} used for tuples or the arguments of a \python function or class.}.
For the sake of clarity, since $\bnfStmtQuestion{\srcStmtAs{\srcName}}$ expressions in $\srcWith$ and $\srcExcept$ statements and $\bnfStmtQuestion{\srcIf\ \srcExpr}$ expressions in $\srcCase$ statements do not impact symbolic state succession,
they are omitted from the definition.
$\srcAsync\ \srcWith$ (resp. $\srcAsync\ \srcFor$) statements are treated like $\srcWith$ (resp. $\srcFor$) statements.
Finally, the body of a $\srcFinally$ statement is executed even if a $\srcReturn$ or $\srcRaise$ statement is reached in the body of 	a corresponding $\srcTry$ or $\srcExcept$ statement.
This behavior is reflected when generating SCFGs; however, it is not illustrated in the figure for the sake of clarity.

\paragraph*{Application to the Running Example}

We detail how to compute $\edgesRec{\progStmt}{\symbState_{\ref*{line:runningExample:notSensitive}}}{\symbState_{\ref*{line:runningExample:notSensitive}}}$, where the fragment $\progStmt$ used in the main paper is:
\begin{lstlisting}[style=neutral]
x = x**2 \end{lstlisting}

The fragment is a sequence of statements $\pyCode{x = x**2 \% p}
\srcSeqN \srcIf\ \dots$ so, according to case $\eqRule{1}$ in \Cref{table:scfg:edges}, we have
$\edgesRec{\progStmt}{\symbState_{\ref*{line:runningExample:notSensitive}}}{\symbState_{\ref*{line:runningExample:notSensitive}}} = \edgesRecName(\pyCode{x = x**2 \% p}, \symbState_{\ref*{line:runningExample:notSensitive}}, \edgesRecState{\srcIf\ \dots}) \cup \edgesRec{\srcIf\ \dots}{\symbState_{\ref*{line:runningExample:notSensitive}}}{\symbState_{\ref*{line:runningExample:notSensitive}}}$.

First, the leftmost statement of the sequence $\srcIf\ \dots$ is associated with $\symbState_{\ref*{line:runningExample:sensitBranching}}$, hence $\edgesRecState{\srcIf\ \dots} = \symbState_{\ref*{line:runningExample:sensitBranching}}$.
Moreover, \pyCode{x = x**2 \% p} is an assignment associated with $\symbState_{\ref*{line:runningExample:square}}$.
So, according to case $\eqRule{10}$ in \Cref{table:scfg:edges},
$\edgesRec{\pyCode{x = x**2 \% p}}{\symbState_{\ref*{line:runningExample:notSensitive}}}{\symbState_{\ref*{line:runningExample:sensitBranching}}} = \set{\tuple{\symbState_{\ref*{line:runningExample:square}}, \symbState_{\ref*{line:runningExample:sensitBranching}}}}{}$.
Second, $\srcIf\ \dots$ is an instance of conditional branching with:
\begin{itemize}
    \item  state $\symbState_{\ref*{line:runningExample:sensitBranching}}$,
    \item  a branch \pyCode{x = g*x} associated with $\symbState_{\ref*{line:runningExample:indirect}}$,
    \item  a branch \pyCode{y = g*x} associated with $\symbState_{\ref*{line:runningExample:dummy}}$,
    \item an ending state $\symbState'_{\ref*{line:runningExample:sensitBranching}}$.
\end{itemize}

So, according to case $\eqRule{3}$ in \Cref{table:scfg:edges}, we have
$
\edgesRec{\srcIf\ \dots}{\symbState_{\ref*{line:runningExample:notSensitive}}}{\symbState_{\ref*{line:runningExample:notSensitive}}} =
\set{\tuple{\symbState_{\ref*{line:runningExample:sensitBranching}}, \symbState_{\ref*{line:runningExample:indirect}}},
\tuple{\symbState_{\ref*{line:runningExample:sensitBranching}}, \symbState_{\ref*{line:runningExample:dummy}}}}{} \cup \edgesRec{\pyCode{x = g*x}}{\symbState_{\ref*{line:runningExample:notSensitive}}}{\symbState'_{\ref*{line:runningExample:sensitBranching}}}
\cup \edgesRec{\pyCode{y = g*x}}{\symbState_{\ref*{line:runningExample:notSensitive}}}{\symbState'_{\ref*{line:runningExample:sensitBranching}}} \cup \set{\tuple{\symbState'_{\ref*{line:runningExample:sensitBranching}}, \symbState_{\ref*{line:runningExample:notSensitive}}}}{}
$.
These branches are assignments, so according to case
$\eqRule{10}$ in \Cref{table:scfg:edges},
$\edgesRec{\pyCode{x = g*x}}{\symbState_{\ref*{line:runningExample:notSensitive}}}{\symbState'_{\ref*{line:runningExample:sensitBranching}}} = \set{\tuple{\symbState_{\ref*{line:runningExample:indirect}}, \symbState'_{\ref*{line:runningExample:sensitBranching}}}}{}$ and $\edgesRec{\pyCode{y = g*x}}{\symbState_{\ref*{line:runningExample:notSensitive}}}{\symbState'_{\ref*{line:runningExample:sensitBranching}}} = \set{\tuple{\symbState_{\ref*{line:runningExample:dummy}}, \symbState'_{\ref*{line:runningExample:sensitBranching}}}}{}$.

Therefore, the other edges of the subgraph are
$\tuple{\symbState_{\ref*{line:runningExample:square}}, \symbState_{\ref*{line:runningExample:sensitBranching}}}$, $\tuple{\symbState_{\ref*{line:runningExample:sensitBranching}}, \symbState_{\ref*{line:runningExample:indirect}}}$,
$\tuple{\symbState_{\ref*{line:runningExample:sensitBranching}}, \symbState_{\ref*{line:runningExample:dummy}}}$, $\tuple{\symbState_{\ref*{line:runningExample:indirect}}, \symbState'_{\ref*{line:runningExample:sensitBranching}}}$, $\tuple{\symbState_{\ref*{line:runningExample:dummy}}, \symbState'_{\ref*{line:runningExample:sensitBranching}}}$, and $\tuple{\symbState'_{\ref*{line:runningExample:sensitBranching}}, \symbState_{\ref*{line:runningExample:notSensitive}}}$.

 \section{Second Example of Static Aspect Definition}
\label{sec:secondSadExample}

The traversal defined in \Cref{fig:sad:confidentiality} analyzes confidentiality violations (\Cref{sec:background:security:infoFlow}), i.e., occurrences where a sensitive value is leaked in a sink.
It does so by importing values of the \aspectSensitive static aspect obtained in the traversal defined in \Cref{fig:sad:travSensitive} of the main paper, using sinks annotated by a security expert, and raising an alarm when a sensitive symbol is involved in the same expression as a sink.

This traversal definition demonstrates how static aspects can be imported from past traversals.
More precisely, the \code{fromTraversal} construct is used at \Cref{line:travConfidentiality:importAspect} to import the \code{Sensitive} static aspect previously defined during the \code{travSensitive} traversal.
Then, values of this static aspect at the symbolic state \code{point} can be captured using the \code{getAspect(point, Sensitive)} primitive (\Cref{line:travConfidentiality:getAspect}).
Finally, the value of the \code{point} variable is instantiated by the \code{currentPoint} primitive when used in the scope of a \code{pointcut}, for instance at \Cref{line:travConfidentiality:Exp:currentPoint}.

\begin{figure*}[t!]
\centering
\begin{minipage}{0.95\linewidth}
\begin{lstlisting}[language=aspectDSL,escapechar=@,basicstyle={\scriptsize \ttfamily}]
traversal travConfidentiality:# define the ConfidentialityViolation static aspect

	fromTraversal travSensitive importAspect Sensitive# aspect from previous traversal@\label{line:travConfidentiality:importAspect}@
	sourceAnnotation labeled_symbols@\label{line:travConfidentiality:sourceAnnotation}@
	aspect Sinks aspectType set@\label{line:travConfidentiality:Sinks}@
	aspect ConfidentialityViolation aspectType bool@\label{line:travConfidentiality:ConfidentialityViolation}@
	triggerFrom ConfidentialityViolation atValue True# may trigger an alarm@\label{line:travConfidentiality:triggerFrom}@
	
	utility:
		def isConfidentialityViolation(point, expr, Sensitive, Sinks):@\label{line:travConfidentiality:isConfidentialityViolation}@
			Symbs_involved = getExprSymbs('use', expr) | getExprSymbs('call', expr)
			Symbs_Sensitive = getAspect(point, Sensitive)@\label{line:travConfidentiality:getAspect}@
			Sinks_involved = Sinks & getExprSymbs('call', expr)
			return len(Sinks_involved) > 0 and len(Symbs_involved & Symbs_Sensitive) > 0

	pointcut(EnterProcedure, inputs):
		Sinks = getDescrSymbs('sink', labeled_symbols)# sinks from source annotation@\label{line:travConfidentiality:getDescrSymbs}@
		ConfidentialityViolation = False

	pointcut(Exp, expr):
		ConfidentialityViolation = isConfidentialityViolation(currentPoint, expr, Sensitive, Sinks)@\label{line:travConfidentiality:Exp:currentPoint}@

	pointcut(Assign, left, right):
		ConfidentialityViolation = isConfidentialityViolation(currentPoint, left, Sensitive, Sinks) or isConfidentialityViolation(currentPoint, right, Sensitive, Sinks)

	pointcut(If, cond):
		ConfidentialityViolation = isConfidentialityViolation(currentPoint, cond, Sensitive, Sinks)

	pointcut(While, cond):
		ConfidentialityViolation = isConfidentialityViolation(currentPoint, cond, Sensitive, Sinks)

	pointcut(For, index, bound):
		ConfidentialityViolation = isConfidentialityViolation(currentPoint, bound, Sensitive, Sinks)

	pointcut(EndIf):
		ConfidentialityViolation = False

	pointcut(EndWhile):
		ConfidentialityViolation = False

	pointcut(EndFor):
		ConfidentialityViolation = False

	pointcut(Return, output):
		ConfidentialityViolation = isConfidentialityViolation(currentPoint, output, Sensitive, Sinks)

	pointcut(ExitProcedure):
		ConfidentialityViolation = False
\end{lstlisting}
 \end{minipage}
\caption{Instructions required to determine the values of the \aspectConfidentialityViolation static aspect during an SCFG traversal.}
\label{fig:sad:confidentiality}
\end{figure*}

As in the \travSensitive{} traversal example (\Cref{sec:staticAspects:example}), the \travConfidentiality{} traversal does not specify a custom merging strategy, so the default merging strategy is used (\Cref{fig:sad:merge}).
Moreover, since \travConfidentiality{} imports at \Cref{line:travConfidentiality:importAspect} the $\aspectSensitive$ static aspect from \travSensitive{}, \aspectTool determines that \travSensitive{} should be performed before \travConfidentiality{} (\Cref{sec:staticAspects:dependencies}).

\begin{figure}[t]
\centering
\includegraphics[width=1.0\linewidth]{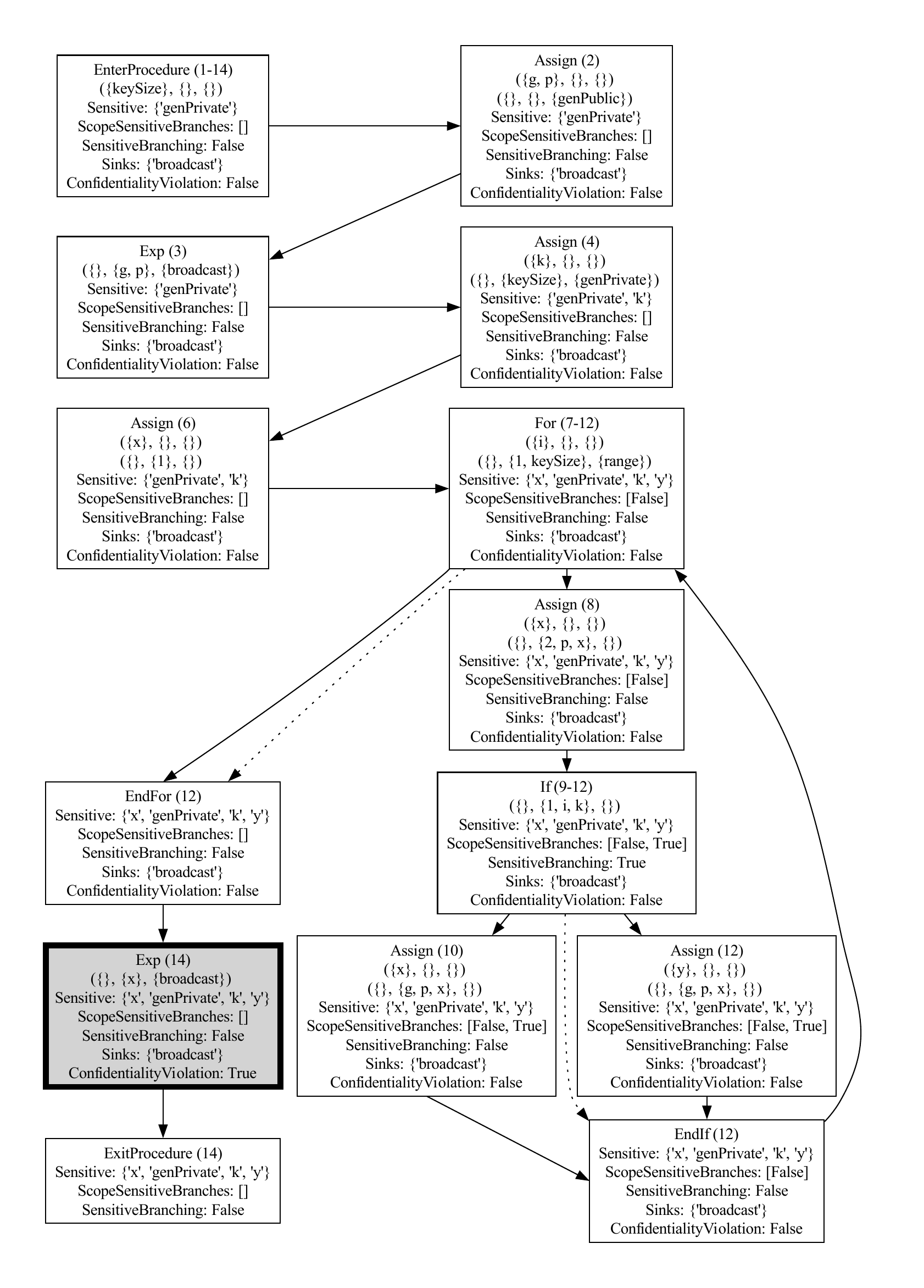}
\caption{Alarm raised during the \travConfidentiality{} traversal (defined in \Cref{fig:sad:confidentiality}).}
\label{fig:scfg:confidentiality:alarm}
\end{figure}

After completing the \travSensitive{} traversal, \Cref{fig:scfg:confidentiality:alarm} illustrates the \travConfidentiality{} traversal.
At the \labelEnterProcedure state, the \aspectSinks set is initialized using the sink symbols (\Cref{sec:staticAspects:expert}), in our example \pyCode{broadcast}.
Then, the SCFG is traversed, until a confidentiality violation is detected, as defined at \Cref{line:travConfidentiality:isConfidentialityViolation} of \Cref{fig:sad:confidentiality}.
At \Cref{line:runningExample:violation} of the source code (\Cref{fig:runningExample} of the main paper), the \pyCode{broadcast} function is called while variable \pyCode{x} is used.
\aspectTool retrieves from the previous \travSensitive{} annotation
that \pyCode{x} is sensitive; 
hence, the visit triggers an alarm.
Finally, the traversal exits the procedure, completing the SCFG annotation.

 \section{Static Aspect Library}
\label{sec:sable:sadLibrary}

To support vulnerability detection in a large variety of \python
projects, we provide a library with five static aspect definitions
(namely, \sourceTainting, \checkEndProc, \checkCalls,
\involvedSymbols, and \contextualValue) which, together, cover the
different ways information flows in security vulnerabilities.
\sourceTainting tracks taint from sensitive or untrusted data to detect information leakage or tampering.
\checkEndProc detects if the source code reduces the attack surface by verifying that critical data is checked in the procedure.
\checkCalls analyzes function calls relevant to security vulnerabilities, e.g., to prevent vulnerable function calls or to ensure sanitization function calls.
\involvedSymbols is a simpler static aspect that detects the presence or the absence of a symbol, when tracking data or control flow is not necessary.
\contextualValue covers cases where some values have to be set in a given context, e.g., a permission before opening or editing a file.
The definitions of these static aspects are long, ranging from 73 to 128 lines of code.
Hence, we provide only code snippets in the supplementary material.
\begin{asparaenum}
\item \emph{\sourceTainting} allows users to annotate sources and
(optionally) sinks; it encodes a taint analysis based on direct information flow (\Cref{sec:background:security:infoFlow}) to trigger an alarm when data tainted by a source is used in a sink or (if none is provided) in an expression returned by a procedure.
If the corresponding option is selected, this analysis may also report instances of sensitive branching (\Cref{sec:background:security:sca}).
The taint analysis can be tailored for different cases: it is possible to annotate sanitization functions or functions that would taint used
symbols (e.g., \pyCode{append}), to remove the taint from a symbol if
it is checked in a condition, to indicate safe symbols (for instance,
safe keys in a tainted dictionary), or, in case tainting is desirable
(e.g., a secret key coming from a generator with sufficient
randomness), to trigger an alarm when a sink or an output does not use
tainted data.
For instance, for CVE-2016-10149, the XML input must be sanitized before parsing to prevent XML external entity attacks.
\Cref{fig:sad:sourceTaintingSource} shows a snippet of a vulnerable source code, the vulnerability being at \Cref{line:sourceTaintingSource:vuln}, where \code{xml\_string} is used in \code{ElementTree.fromstring};
\Cref{fig:sad:sourceTaintingSad} shows a snippet of the
\sourceTainting static aspect definition.
In our CVE-2016-10149 example, 
initially aspect \code{Tainted} contains only the source \code{xml\_string}.
Since the statement is an assignment, the advice code defined at \Cref{line:sourceTaintingSad:assign} is executed.
First, \code{infoFlowUse} (\Cref{line:sourceTaintingSad:infoFlowUse}) is called; since the RHS expression does not involve a function that taints used symbols or a sanitization function, \code{Tainted} is not changed.
Then, \code{isVulnerableSink}
(\Cref{line:sourceTaintingSad:isVulnerableSink}) is called; since in
our example \code{xml\_string} is used in the sink \code{.fromstring}, \code{Vulnerability} becomes $\algoTrue$.
Finally, this update triggers an alarm (\Cref{line:sourceTaintingSad:triggerFrom}), which leads to vulnerability detection.

\begin{figure*}[t!]
\centering
\begin{subfigure}{0.96\textwidth}
\begin{lstlisting}[language=myPython,escapechar=@,basicstyle={\scriptsize \ttfamily}]
def create_class_from_xml_string(target_class, xml_string):
    if not isinstance(xml_string, six.binary_type):
        xml_string = xml_string.encode('utf-8')
    tree = ElementTree.fromstring(xml_string)@\label{line:sourceTaintingSource:vuln}@
    return create_class_from_element_tree(target_class, tree)
\end{lstlisting}
 \caption{Source code snippet.}
\label{fig:sad:sourceTaintingSource}
\end{subfigure}
\begin{subfigure}{0.96\linewidth}
\begin{lstlisting}[language=aspectDSL,escapechar=@,basicstyle={\scriptsize \ttfamily}]
traversal travSourceTainting:
	aspect Sink aspectType set
	aspect Sanitize aspectType set
	aspect PositiveTainting aspectType bool
	aspect SensitiveBranching aspectType bool
	aspect UpdUse aspectType set
	aspect Safe aspectType set
	aspect Tainted aspectType set
	aspect Vulnerability aspectType bool
	triggerFrom Vulnerability atValue True@\label{line:sourceTaintingSad:triggerFrom}@# triggers the alarm
	utility:# functions used in advice code
		# isIntersection takes as input two sets and returns True if the intersection is not empty, getUseCall takes an expression as input and returns its use and call symbols
		def infoFlowUse(expr, UpdUse, Sanitize, Tainted):@\label{line:sourceTaintingSad:infoFlowUse}@
			SymbsUse = getExprSymbs('use', expr)
			SymbsUseCall = getUseCall(expr)
			if isIntersection(SymbsUseCall, UpdUse) and isIntersection(SymbsUseCall, Tainted):
				Tainted = Tainted | SymbsUse
			if isIntersection(SymbsUseCall, Sanitize): 
				Tainted = Tainted - SymbsUse
			return Tainted
		def isVulnerableSink(expr, Safe, Tainted, Sink, PositiveTainting, SensitiveBranching):@\label{line:sourceTaintingSad:isVulnerableSink}@
			Involved = getUseCall(expr) | getExprSymbs('def', expr)
			return isVulnerableExpr(expr, Safe, Tainted, PositiveTainting) and isIntersection(Involved, Sink) and not SensitiveBranching
	pointcut(Assign, left, right):@\label{line:sourceTaintingSad:assign}@# advice code
		Tainted = infoFlowUse(right, UpdUse, Sanitize, Tainted)
		Vulnerability = isVulnerableSink(right, Safe, Tainted, Sink, PositiveTainting, SensitiveBranching)
\end{lstlisting}
 \caption{Code snippet of the static aspect definition.}
\label{fig:sad:sourceTaintingSad}
\end{subfigure}
\caption{Code snippets for the \sourceTainting example.}
\end{figure*}

\item  \emph{\checkEndProc} allows users to annotate symbols that have to be checked in a condition before the end of a procedure; more precisely, it checks all branches reaching the end of the procedure and raises an alarm if there is no branch in which all the annotated symbols have been checked in a condition.
This aspect also enables users to annotate other symbols, for example, to verify whether a given function is called after an annotated symbol is checked in a condition.
For instance, for vulnerability CVE-2017-15612, one wants to verify
that function \code{re.sub} is called after the URL has been checked
in a condition, since this method must be used to remove special
characters from the URL and thus prevent cross-site scripting attacks.
\Cref{fig:sad:checkEndProcSource} shows a snippet of a vulnerable
source code, while \Cref{fig:sad:checkEndProcSad} shows a snippet of the \checkEndProc static aspect definition.
This definition is based on the \code{evalBranch} auxiliary function
(\Cref{line:checkEndProcSadSad:evalBranch}), which determines if the
branch satisfied the expected criterion.
In our CVE-2017-15612 example, this means that if \code{lower\_url} is checked in a condition, then \code{re.sub} must be called in that condition or later in the branch.
Aspect \code{CondSatisfied} is initially $\algoFalse$.
For each statement terminating the procedure (here, \code{return)}, \code{evalBranch} is called to determine if the branch satisfied the criterion (\Cref{line:checkEndProcSadSad:val}).
In that case, aspect \code{CondSatisfied} becomes $\algoTrue$ for the rest of the traversal (\Cref{line:checkEndProcSadSad:condSatisfied}).
In our example, since \code{re.sub} is not called in the source code, \code{CondSatisfied} remains $\algoFalse$ in this traversal.
After all the branches have been traversed, at the end of the procedure, aspect \code{Vulnerability} becomes $\algoTrue$ if \code{CondSatisfied} is $\algoFalse$ (\Cref{line:checkEndProcSadSad:exit}), triggering the alarm (\Cref{line:checkEndProcSadSad:triggerFrom}).

\begin{figure*}[t!]
\centering
\begin{subfigure}{0.96\textwidth}
\begin{lstlisting}[language=myPython,escapechar=@,basicstyle={\scriptsize \ttfamily}]
def escape_link(url):
    lower_url = url.lower().strip('\x00\x1a \n\r\t')
    for scheme in _scheme_blacklist:
        if lower_url.startswith(scheme):#re.sub should be called on lower_url
            return ''
    return escape(url, quote=True, smart_amp=False)
\end{lstlisting}
 \caption{Source code snippet.}
\label{fig:sad:checkEndProcSource}
\end{subfigure}
\begin{subfigure}{0.96\linewidth}
\begin{lstlisting}[language=aspectDSL,escapechar=@,basicstyle={\scriptsize \ttfamily}]
traversal travEndProc:
	aspect Checks aspectType set
	aspect Calls aspectType set
	aspect PreventCheck aspectType set
	aspect Checked aspectType set
	aspect Called aspectType set
	aspect CondSatisfied aspectType bool
	aspect Vulnerability aspectType bool
	triggerFrom Vulnerability atValue True@\label{line:checkEndProcSadSad:triggerFrom}@# triggers the alarm
	utility:# function used in advice code, isEmpty is True if the length of the input is 0
		def evalBranch(Sources, Tainted, Checks, Calls, Checked, Called):@\label{line:checkEndProcSadSad:evalBranch}@
			if isEmpty(Sources):
				if isEmpty(Calls):
					val = Checks <= Checked
				else:
					val = isIntersection(Checks, Checked) and (Calls <= Called)
			else:
				val = isIntersection(Tainted, Checks) and (Tainted & Checks <= Checked)
			return val
	pointcut(Return, output):#aspects Checked, Called, etc. obtained during branch traversal
		Called |= getExprSymbs('call', output)
		val = evalBranch(Sources, Tainted, Checks, Calls, Checked, Called)@\label{line:checkEndProcSadSad:val}@
		CondSatisfied = CondSatisfied or val@\label{line:checkEndProcSadSad:condSatisfied}@
	pointcut(ExitProcedure):# safe if at least one branch checked the condition
		Vulnerability = not CondSatisfied@\label{line:checkEndProcSadSad:exit}@
\end{lstlisting}
 \caption{Code snippet of the static aspect definition.}
\label{fig:sad:checkEndProcSad}
\end{subfigure}
\caption{Code snippets for the \checkEndProc example.}
\end{figure*}

\item \emph{\checkCalls} allows users to annotate functions that have to be called at least once before (if any) other annotated functions are called or (otherwise) before the end of the procedure.
To prevent false positives, this aspect also enables users to annotate
functions so that the advice code checks the calls (and eventually raises an alarm) only if an annotated function has been called before.
For instance, for CVE-2016-2513, the \code{verify} function may leak the number of iterations when verifying the password.
 To prevent a timing attack, the \code{harden\_runtime} function must
 be called, but only if the \code{verify} function was called before.
\Cref{fig:sad:checkCallsSource} shows a snippet of a vulnerable source code, while \Cref{fig:sad:checkCallsSad} shows a snippet of the \checkCalls static aspect definition.
In our CVE-2016-2513 example, the user annotated function \code{harden\_runtime} in aspect \code{Checks} and function \code{verify} in aspect \code{Conditions}.
When the traversal reaches the \code{return} statement, function \code{verify} is present in aspect \code{Called}.
First, function \code{isCheckNotCalledAtExpr} is called (\Cref{line:ccheckCallsSad:isCheckNotCalledAtExpr}), setting \code{CheckNotCalled} to $\algoFalse$ because aspect \code{Events} is empty.
Then, \code{isCheckNotCalled} is called (\Cref{line:ccheckCallsSad:isCheckNotCalled}).
Because aspect \code{Conditions} contains \code{verify} and \code{harden\_runtime} has not been called, \code{CheckNotCalled} becomes $\algoTrue$, which triggers an alarm (\Cref{line:ccheckCallsSad:triggerFrom}).

\begin{figure*}[t!]
\centering
\begin{subfigure}{0.96\textwidth}
\begin{lstlisting}[language=myPython,escapechar=@,basicstyle={\scriptsize \ttfamily}]
def check_password(password, encoded, setter=None, preferred='default'):
    preferred = get_hasher(preferred)
    hasher = identify_hasher(encoded)
    must_update = hasher.algorithm != preferred.algorithm
    is_correct = hasher.verify(password, encoded)@\label{line:checkCallsSource:vuln}@# may leak sensitive timing information
    if setter and is_correct and must_update:
        setter(password)
    return is_correct
\end{lstlisting}
 \caption{Source code snippet.}
\label{fig:sad:checkCallsSource}
\end{subfigure}
\begin{subfigure}{0.96\linewidth}
\begin{lstlisting}[language=aspectDSL,escapechar=@,basicstyle={\scriptsize \ttfamily}]
traversal travCheckCalls:
	aspect Called aspectType set
	aspect Checks aspectType set
	aspect Conditions aspectType set
	aspect Events aspectType set
	aspect CheckNotCalled aspectType bool
	triggerFrom CheckNotCalled atValue True@\label{line:ccheckCallsSad:triggerFrom}@# triggers the alarm
	utility:
		def isCheckNotCalled(Called, Checks, Conditions):
			if isEmpty(Conditions) or isIntersection(Conditions, Called):
				return not (Checks <= Called)
			else:
				return False
		def isCheckNotCalledAtExpr(expr, Called, Checks, Conditions, Events):
			if isIntersection(getExprSymbs('call', expr), Events):
				return isCheckNotCalled(Called, Checks, Conditions)
			else:
				return False
	pointcut(Return, output):
		Called |= getExprSymbs('call', output)
		CheckNotCalled = isCheckNotCalledAtExpr(output, Called, Checks, Conditions, Events)@\label{line:ccheckCallsSad:isCheckNotCalledAtExpr}@
		if isEmpty(Events) and not isIntersection(getUseCall(output), SafeOutputs):
			CheckNotCalled = CheckNotCalled or isCheckNotCalled(Called, Checks, Conditions)@\label{line:ccheckCallsSad:isCheckNotCalled}@
\end{lstlisting}
 \caption{Code snippet of the static aspect definition.}
\label{fig:sad:checkCallsSad}
\end{subfigure}
\caption{Code snippets for the \checkCalls example.}
\end{figure*}

\item \emph{\involvedSymbols} allows users to annotate symbols of
  interest and to indicate if they are expected, in order to trigger
  an alarm (a) if they are not expected but are present in an
  expression or (b) if they are expected and have not been detected during the SCFG traversal.
For instance, for CVE-2011-4030, the \code{\_\_roles\_\_} attribute
should be initialized to prevent the \code{KwAsAttributes} class from
being publishable; in this way, remote attackers are prevented from accessing sub-objects.
\Cref{fig:sad:involvedSymbolsSource} shows a snippet of vulnerable source code, while \Cref{fig:sad:involvedSymbolsSad} shows a snippet of the \involvedSymbols static aspect definition.
Our CVE-2011-4030 example demonstrates class analysis instead of
procedure analysis, where statement labels \labelEnterContainer and
\labelExitContainer replace, respectively,  the usual \labelEnterProcedure and \labelExitProcedure statement labels (\Cref{sec:scfg}).
In this example, a security expert may annotate \code{\_\_roles\_\_} as being expected in the code, i.e., it belongs to aspect \code{Special} and aspect \code{Expected} is $\algoTrue$.
If \code{\_\_roles\_\_} was initialized, then the piece of advice code for the \code{Assign} statement would have added \code{\_\_roles\_\_} to the aspect \code{Involved} (\Cref{line:involvedSymbolsSad:Assign}), but this is not the case in this vulnerable code snippet.
At the end of the class definition, the auxiliary function \code{isVulnerableEnd} is called (\Cref{line:involvedSymbolsSad:vuln}), setting \code{Vulnerability} to $\algoTrue$ because \code{\_\_roles\_\_} was expected but not involved in the code, triggering the alarm (\Cref{line:involvedSymbolsSad:triggerFrom}).

\begin{figure*}[t!]
\centering
\begin{subfigure}{0.96\textwidth}
\begin{lstlisting}[language=myPython,escapechar=@,basicstyle={\scriptsize \ttfamily}]
def escape_link(url):
class KwAsAttributes(Persistent):
    def __init__(self, **kw):
        for key, val in list(kw.items()):
            setattr(self, key, val)
\end{lstlisting}
 \caption{Source code snippet.}
\label{fig:sad:involvedSymbolsSource}
\end{subfigure}
\begin{subfigure}{0.96\linewidth}
\begin{lstlisting}[language=aspectDSL,escapechar=@,basicstyle={\scriptsize \ttfamily}]
traversal travInvolvedSymbs:
	aspect Special aspectType set
	aspect Expected aspectType bool
	aspect Involved aspectType set
	aspect Vulnerability aspectType bool
	triggerFrom Vulnerability atValue True@\label{line:involvedSymbolsSad:triggerFrom}@
	utility:
		def getInvolved(expr):
			return getExprSymbs('def', expr) | getExprSymbs('use', expr) | getExprSymbs('call', expr)
		def isVulnerableExpr(expr, Special, Expected):
			if Expected:
				return False
			else:
				return isIntersection(getInvolved(expr), Special)
		def isVulnerableEnd(Involved, Special, Expected):
			if Expected:
				return not isEmpty(Special - Involved)
			else:
				return False
	pointcut(Assign, left, right):@\label{line:involvedSymbolsSad:Assign}@
		Involved |= getInvolved(right)
		Involved |= getInvolved(left)
		Vulnerability = isVulnerableExpr(right, Special, Expected) or isVulnerableExpr(left, Special, Expected)
	pointcut(ExitContainer):
		Vulnerability = isVulnerableEnd(Involved, Special, Expected)@\label{line:involvedSymbolsSad:vuln}@
\end{lstlisting}
 \caption{Code snippet of the static aspect definition.}
\label{fig:sad:involvedSymbolsSad}
\end{subfigure}
\caption{Code snippets for the \involvedSymbols example.}
\end{figure*}

\item \emph{\contextualValue} allows users to annotate functions that must be called in a context where a given function is used with a given value (expressed as a \python literal).
For instance, for CVE-2015-3010, to prevent local users from reading files, the \code{os.umask} function should have set the permission to \code{0o77} each time the \code{fetch\_file} function is called.
\Cref{fig:sad:contextualValueSource} shows a snippet of vulnerable source code, while \Cref{fig:sad:contextualValueSad} shows a snippet of the \contextualValue static aspect definition.
Each time an assignment is reached during the traversal
(\Cref{line:contextualValueSad:Assign}), the sequence
of functions  \code{updateCurrentValues} and \code{isVulnerability} is called for both the LHS and the RHS.
If a statement \code{oldmask = os.umask(0o77)} was present at the
beginning of the procedure, then function \code{updateCurrentValues}
would have updated aspect \code{CurrentValues} to
$\set{\code{0o77}}{}$
(\Cref{line:contextualValueSad:updateCurrentValues}); however, in this vulnerable code snippet, aspect \code{CurrentValues} stays empty.
Thus, when \code{isVulnerability} is called (\Cref{line:contextualValueSad:isVulnerability}) in an expression where function \code{fetch\_file} is called in the source code (\Cref{line:contextualValueSource:vuln}), aspect \code{Vulnerability} becomes $\algoTrue$, triggering the alarm (\Cref{line:contextualValueSad:triggerFrom}).

\begin{figure*}[t!]
\centering
\begin{subfigure}{0.96\textwidth}
\begin{lstlisting}[language=myPython,escapechar=@,basicstyle={\scriptsize \ttfamily}]
def gatherkeys(args):
    keyring = '/etc/ceph/{cluster}.client.admin.keyring'.format(cluster=args.cluster)
    r = fetch_file(args=args, frompath=keyring, topath='{cluster}.client.admin.keyring'.format(cluster=args.cluster), _hosts=args.mon,)@\label{line:contextualValueSource:vuln}@
    if not r:
        raise exc.KeyNotFoundError(keyring, args.mon)
\end{lstlisting}
 \caption{Source code snippet.}
\label{fig:sad:contextualValueSource}
\end{subfigure}
\begin{subfigure}{0.96\linewidth}
\begin{lstlisting}[language=aspectDSL,escapechar=@,basicstyle={\scriptsize \ttfamily}]
traversal travContextValue:
	aspect ExpectedValues aspectType set
	aspect SetFunctions aspectType set
	aspect CurrentValues aspectType set
	aspect Checks aspectType set
	aspect Vulnerability aspectType bool
	triggerFrom Vulnerability atValue True@\label{line:contextualValueSad:triggerFrom}@
	utility:
		def updateCurrentValues(expr, SetFunctions, CurrentValues):
			SymbsCalled = getExprSymbs('call', expr)
			if isIntersection(SymbsCalled, SetFunctions):
				return getExprSymbs('use', expr)
			else:
				return CurrentValues
		def isVulnerability(expr, Checks, CurrentValues, ExpectedValues):
			SymbsCalled = getExprSymbs('call', expr)
			if isIntersection(SymbsCalled, Checks):
				if isEmpty(CurrentValues):
					return True
				else:
					return isIntersection(CurrentValues, ExpectedValues)
			else:
				return False
	pointcut(Assign, left, right):@\label{line:contextualValueSad:Assign}@
		CurrentValues = updateCurrentValues(left, SetFunctions, CurrentValues)
		Vulnerability = isVulnerability(left, Checks, CurrentValues, ExpectedValues)
		CurrentValues = updateCurrentValues(right, SetFunctions, CurrentValues)@\label{line:contextualValueSad:updateCurrentValues}@
		Vulnerability |= isVulnerability(right, Checks, CurrentValues, ExpectedValues)@\label{line:contextualValueSad:isVulnerability}@
\end{lstlisting}
 \caption{Code snippet of the static aspect definition.}
\label{fig:sad:contextualValueSad}
\end{subfigure}
\caption{Code snippets for the \involvedSymbols example.}
\end{figure*}
  
\end{asparaenum}

 \section{\aspectLg Semantics}
\label{sec:staticAspects:semantics}

In this section, we describe the formal semantics of the \aspectLg language.
Since static aspect values may be determined during multiple SCFG traversals, we first explain how the dependencies between static aspects are leveraged to determine the order of traversals (\Cref{sec:staticAspects:dependencies}).
Then, we describe our representation of static aspect values during a traversal (\Cref{sec:staticAspects:travMap}).
Next, we formally describe how an SCFG traversal is performed
(\Cref{sec:staticAspects:traversal}),  including the application to our running example (\Cref{sec:staticAspects:example}).
Finally, we describe vulnerability alarms and \aspectTool's output (\Cref{sec:staticAspects:alarms}).

\subsection{Order of SCFG Traversals}
\label{sec:staticAspects:dependencies}

We denote by $\semTraversals$ the set of the traversal names encountered while parsing the static aspect definition.
We define $\semDependencies{\cdot}$ as a map taking a traversal name $\semTrav$ as input and returning the set of the traversal names imported in $\semTrav$.
More precisely, if $\semTrav$ contains $n$ constructs $\aspectFromTraversal$ $\semTravImport_1$, \dots, $\aspectFromTraversal$ $\semTravImport_n$, then $\semDependencies{\semTrav} \eqdef \set{\semTravImport_1, \dots, \semTravImport_n}{}$.

The order for the SCFG traversals is determined by \Cref{table:orderTraversal}; it takes as input the set of traversal names $\semTraversals$ and the map $\semDependencies{\cdot}$ and returns $\semOrderedTraversals$, a list of traversal names.
The algorithm builds $\semOrderedTraversals$ as a list where each traversal name is placed after its dependencies.
After initializing this list to empty
(\Cref{line:orderTraversal:init}), the algorithm iterates over the following steps: the \emph{valid} traversal names, i.e., remaining traversal names with dependencies met by the already ordered traversal names, are selected (\Cref{line:orderTraversal:valid}), then added (in any order) to $\semOrderedTraversals$ (\Cref{line:orderTraversal:append}).
In this way, traversal names without dependencies are added first,
followed by traversal names with dependencies already in
$\semOrderedTraversals$, and so on until all traversals names are ordered (\Cref{line:orderTraversal:while}).

\begin{algorithm*}[t]
\caption{Order of Traversals}
\footnotesize
\begin{multicols}{2}
\begin{algorithmic}[1]
    \Require Set of Traversals $\semTraversals$, Map: Traversal $\mapsto$ Set of Traversals $\semDependencies{\cdot}$
    \Ensure List of Traversals $\semOrderedTraversals$
    \State Set of Traversals $\semRemainingTraversals \leftarrow \semTraversals$
    \State $\semOrderedTraversals \leftarrow \dataList{\ }$\label{line:orderTraversal:init}
    \While{$\semRemainingTraversals \neq \varnothing$}\label{line:orderTraversal:while}
        \State Set of Traversals $\semValidTraversals$ $\leftarrow$ $\{\semTrav \in \semRemainingTraversals \knowing$ $\semDependencies{\semTrav} \subseteq set(\semOrderedTraversals)\}$ \label{line:orderTraversal:valid}
        \State $\semRemainingTraversals \leftarrow \semRemainingTraversals \setminus \semValidTraversals$
        \State $\semOrderedTraversals \leftarrow \semOrderedTraversals +  \setToList{\semValidTraversals}$\label{line:orderTraversal:append}
    \EndWhile
    \State \Return $\semOrderedTraversals$
\end{algorithmic}
\end{multicols}
 \label{table:orderTraversal}
\end{algorithm*}

\subsection{Traversal Map}
\label{sec:staticAspects:travMap}

Static aspects at a given symbolic state of the source code are SCFG
annotations (\Cref{sec:staticAspects:annotations}), with each static aspect being characterized by a name and a value.
While the name of a static aspect considered during a traversal is determined by a declaration in the static aspect  definition (\Cref{sec:staticAspects:syntax}), its value may evolve during the traversal, depending on the encountered symbolic state, but also on previous static aspect values.
For instance, both examples of information flow in \Cref{fig:code:infoFlow} involve variable \pyCode{x} being tainted by variable \pyCode{secret}, i.e., \pyCode{x} becomes a sensitive variable only if \pyCode{secret} was already a sensitive variable.

To store static aspect information during an SCFG traversal, we introduce a \emph{traversal map}, which maps a string (the static aspect name) to any value (the static aspect value).
It is thus similar to an SCFG annotation, but it is linked to a
traversal instead of a symbolic state.

\label{sec:staticAspects:merge}

\begin{figure}[t]
\centering
\begin{minipage}{0.95\linewidth}
\begin{lstlisting}[language=aspectDSL,escapechar=@,basicstyle={\scriptsize \ttfamily}]
mergeAspects(travMap1, travMap2):
	travMap0 = {}
	for name in travMap1.keys() & travMap2.keys():
		val1 = deepcopy(travMap1[name])
		val2 = deepcopy(travMap2[name])
		if type(val1) is bool and type(val2) is bool:
			travMap0[name] = val1 or val2
		elif type(val1) is set and type(val2) is set:
			travMap0[name] = val1 | val2
		elif val1 == val2:
			travMap0[name] = val1
		else:
			raise ValueError("Conflict of values: for non-Boolean, non-set values, values for both traversal maps are expected to be the same.")
	for name in travMap1.keys() - travMap2.keys():
		travMap0[name] = deepcopy(travMap1[name])
	for name in travMap2.keys() - travMap1.keys():
		travMap0[name] = deepcopy(travMap2[name])
	return travMap0
\end{lstlisting}
 \end{minipage}
\caption{Default strategy for merging traversal maps from different branches.}
\label{fig:sad:merge}
\end{figure}

A traversal map is updated at each symbolic state visited during the traversal.
When visiting an instance of conditional branching, the current traversal map is duplicated for the traversal of each branch.
We call \emph{join points} the symbolic states where branches meet (e.g., at an \labelEndIf statement).
Two traversal maps \code{travMap1} and \code{travMap2} may come from
two distinct branches so that, at a join point, these traversal maps may have value conflicts.
\aspectTool deals with value conflicts using two merging strategies.
In both cases, \aspectTool merges the traversal maps from the branches into one, which is then used in the rest of the traversal.

The \emph{default merging strategy} is detailed in \Cref{fig:sad:merge}.
In short, if a static aspect takes  a value \code{val1} from
\code{travMap1} different than another value \code{val2} from \code{travMap2}, the algorithm checks their type.
If \code{val1} and \code{val2} are Booleans, then the new value is \code{val1} \pyCode{or} \code{val2}.
If \code{val1} and \code{val2} are sets, then the new value is the union of the sets.
Otherwise, the algorithm returns an error.
We chose \pyCode{or} and union operations to resolve value conflicts because, in a security context, we consider the worst case, e.g., if a variable is sensitive in one branch but not in the other, then it is safer to assume the variable is sensitive in the code following the branching.

The \emph{custom merging strategy} is made available to security experts by the $\aspectMergeAspects$ construct (\Cref{sec:staticAspects:syntax}).
Using a syntax similar to \Cref{fig:sad:merge} but with other choices regarding value conflicts, the expert can specify how to merge \code{travMap1} and \code{travMap2}.
If no custom merging strategy is provided in a static aspect definition, \aspectTool selects the default merging strategy.

\subsection{Formal Description of one SCFG Traversal}
\label{sec:staticAspects:traversal}

Since static aspects are obtained through static analysis, one cannot generally determine the number of loops performed in the procedure of interest.
Hence, the traversal of a loop body is repeated until a
\emph{fixpoint} is reached, i.e., when the static aspect values in the traversal map coincide with the values of the symbolic state annotation.
Thus, during a traversal, a symbolic state corresponding to a loop statement may be visited either from its body (when the traversal is repeated) or from the rest of the program (when the symbolic state is entered).
To distinguish between these cases, we enrich each symbolic state $\symbState$ with an attribute $\ptEnterLoop{\symbState}$, a Boolean indicating if a loop was entered and corresponding to the $\aspectEnterLoop{\symbState}$ primitive (\Cref{sec:staticAspects:syntax}).

\Cref{table:initTraversal} details the initialization and output of a traversal.
It takes as input the procedure of interest $\procedure$ and $\mathit{Aspects}$, the set of the static aspect names introduced in the traversal definition using the $\aspectAspect$ keyword (\Cref{sec:staticAspects:syntax}).
First, $\semStep$ and $\getAlarmsName$ are initialized.
These variables are used to store alarm information; they are detailed at the end of the section.
Then, $\ptEnterLoop{\symbState}$ is initialized for each symbolic state $\symbState$ in $\vertices{\procedure}$ (\Cref{line:initTraversal:enterLoop}), where $\vertices{\procedure}$ denotes the vertices of $\scfg{\procedure}$ (\Cref{sec:scfg}).
The traversal map (\Cref{sec:staticAspects:travMap}) $\semTravMap$ is initialized at Lines~\ref{line:initTraversal:travMapInitStart}--\ref{line:initTraversal:travMapInitEnd}.
The SCFG traversal is performed by executing $\semVisit{\semStartPt}{\pyNone}{\semTravMap}$ (\Cref{line:initTraversal:visit}),
where $\semStartPt$ is the symbolic state indicating the start of $\scfg{\procedure}$ (\Cref{sec:scfg}),
and $\pyNone$ is the corresponding \python object.

\begin{algorithm*}[t]
\caption{Initialization and Output of a Traversal}
\footnotesize
\begin{multicols}{2}
\begin{algorithmic}[1]
    \Require Procedure $\procedure$, Set $\mathit{Aspects}$
    \Ensure Map $\getAlarmsName$
    \State $\semStep \leftarrow 0$\label{line:initTraversal:step}
    \State $\getAlarmsName \leftarrow \{\}$\label{line:initTraversal:alarms}
    \For{$\symbState \in \vertices{\procedure}$}\label{line:initTraversal:enterLoop}
        \If{$\ptStmt{\symbState} \in \set{\labelFor, \labelWhile}{}$}
            \State $\ptEnterLoop{\symbState} \leftarrow \algoTrue$
        \Else
            \State $\ptEnterLoop{\symbState} \leftarrow \algoFalse$
        \EndIf
    \EndFor
    \State $\semTravMap \leftarrow \{\}$\label{line:initTraversal:travMapInitStart}
    \For{$\ptAspectName \in \mathit{Aspects}$}
        \State $\semTravMap(\ptAspectName) = \pyNone$
    \EndFor\label{line:initTraversal:travMapInitEnd}
    \State $\tuple{\symbStateEnd, \semTravMap} \leftarrow \semVisit{\semStartPt(\procedure)}{\pyNone}{\semTravMap}$\label{line:initTraversal:visit}
    \State $\semTravMap \leftarrow \semWeave{\symbStateEnd}{\ptAnnotation{\symbStateEnd}}$\label{line:initTraversal:weave}
    \State $\ptAnnotate{\symbStateEnd}{\semTravMap}$\label{line:initTraversal:annotate}
    \State \Return $\getAlarmsName$\label{line:initTraversal:return}
\end{algorithmic}
\end{multicols}
 \label{table:initTraversal}
\end{algorithm*}

The $\semVisitName$ function is defined in \Cref{table:sable:traversal} using auxiliary functions from \Cref{table:sable:traversalAux},
where statement label $\ptStmt{\symbState}$ and expressions $\ptExprs{\symbState}$ are from \Cref{sec:symbStates},
ending symbolic states $\ptExit{\symbState}$ and symbolic state successors $\ptNexts{\symbState}$ are from \Cref{sec:scfg},
SCFG annotation $\ptAnnotation{\symbState}$ is from \Cref{sec:staticAspects:annotations},
advice code $\getAdvice{\ptStmtLabel}$ is from \Cref{sec:staticAspects:syntax}, $\semMergeMapsName$ is either the function provided by the expert or the default merging function (\Cref{sec:staticAspects:merge}),
and $\Dom{\cdot}$ denotes the domain of a map.

\begin{figure*}[t]
\centering
\footnotesize
$$
\begin{array}{|l|}
\hline

\semVisit{\symbState_1}{\semExitPt}{\semTravMap_1}
\eqdef
\left\{
\begin{array}{ll}
    \tuple{\symbState_1, \semTravMap_1}
        & \text{if } \symbState_1 = \semExitPt\\
    \semTransition{\symbState_2}{\semExitPt}{\ptAnnotation{\symbState_1}}{\semTravMap_2}
        & \text{otherwise}\\
\end{array}
\right.\\
\qquad \text{where }
\semTravMap_2 =
\left\{
\begin{array}{ll}
    \semMerge{\ptAnnotation{\symbState_1}}{\semTravMap_1}
    & \text{if } \symbState_1 = \semEndPt\\
    \semWeave{\symbState_1}{\semTravMap_1} 
    & \text{otherwise}\\
\end{array}
\right.\\
\qquad \text{and } \symbState_2 = \ptAnnotate{\symbState_1}{\semTravMap_2}\\

\hline
\semTransition{\symbState_1}{\semExitPt}{\semOldAnnot}{\semTravMap}
\eqdef
\left\{
\begin{array}{ll}
    \tuple{\symbState_1, \semTravMap}
        & \text{if } \card{\ptNexts{\symbState_1}} = 0\\
    \semVisit{\element{\ptNexts{\symbState_1}}}{\semExitPt}{\semTravMap}
        & \text{if } \card{\ptNexts{\symbState_1}} = 1\\
    \semBranching{\symbState_1}{\semExitPt}{\semOldAnnot}{\semTravMap}
        & \text{if } \card{\ptNexts{\symbState_1}} \ge 2\\
\end{array}
\right.\\

\hline
\semBranching{\symbState_1}{\semExitPt}{\semOldAnnot}{\semTravMap}
\eqdef
\left\{
\begin{array}{ll}
    \semLoopBranching{\symbState_1}{\semExitPt}{\semOldAnnot}{\semTravMap}
        & \text{if } \ptStmt{\symbState_1} \in \set{\labelFor, \labelWhile}{}\\
    \semCondBranching{\symbState_1}{\semExitPt}{\semTravMap}
        & \text{otherwise}\\
\end{array}
\right.\\

\hline
\semLoopBranching{\symbState_1}{\semExitPt}{\semOldAnnot}{\semTravMap}
\eqdef \semVisit{\symbState_2}{\semExitPt}{\semTravMap}\\
\qquad\text{where } \symbState_2=
\left\{
\begin{array}{ll}
    \ptExit{\ptUpdEnterLoop{\symbState_1}{\algoTrue}}
        & \text{if } \semIsFixpoint{\semOldAnnot}{\semTravMap}\\
    \element{\ptNexts{\symbState_3} \setminus \set{\ptExit{\symbState_3}}{}}
        & \text{otherwise}\\
\end{array}
\right.\\
\qquad\text{and } \symbState_3 = \ptUpdEnterLoop{\symbState_1}{\algoFalse}\\

\hline
\semCondBranching{\symbState_1}{\semExitPt}{\semTravMap}
\eqdef
\left\{
\begin{array}{ll}
    \tuple{\symbState_1, \semTravMap}
        & \text{if } \card{\semTravMaps} = 0\\
    \semVisit{\ptExit{\symbState_1}}{\semExitPt}{\semMergeMaps{\semTravMaps}}
        & \text{otherwise}\\
\end{array}
\right.\\
\qquad \text{where } \semBranches = \set{\semVisit{\symbState_2}{\ptExit{\symbState_1}}{\semTravMap}}{\symbState_2 \in \ptNexts{\symbState_1}}\\
\qquad \text{and } \semTravMaps = \set{\semBranchMap}{\tuple{\semBranchPt, \semBranchMap} \in \semBranches \land\ \ptStmt{\semBranchPt} \neq \labelExitProcedure}\\

\hline
\end{array}
$$
 \caption{Inductive definition of function $\semVisitName$, using the auxiliary functions from \Cref{table:sable:traversalAux}}
\label{table:sable:traversal}
\end{figure*}

\begin{figure*}[t]
\centering
\footnotesize
$$
\begin{array}{|l|}
\hline

\element{\set{x}{}} \eqdef x\\

\hline
\ptAnnotate{\symbState_1}{\semTravMap} \eqdef \symbState_2\\
\qquad \text{where } \symbState_2 \text{ has the same attributes as } \symbState_1 \text{ except, for each }\ptAspectName:\\
\qquad \ptAnnotation{\symbState_2}(\ptAspectName)
=
\left\{
\begin{array}{ll}
    \semTravMap(\ptAspectName)
    & \text{if } \ptAspectName \in \Dom{\semTravMap}\\
    \ptAnnotation{\symbState_1}(\ptAspectName)
    & \text{otherwise}\\
\end{array}
\right.\\

\hline
\ptUpdEnterLoop{\symbState_1}{b} \eqdef \symbState_2\\
\qquad \text{where } \symbState_2 \text{ has the same attributes as } \symbState_1 \text{ except } \ptEnterLoop{\symbState_2} = b\\

\hline
\semWeave{\symbState_1}{\semTravMap_1}
\eqdef
\left\{
\begin{array}{ll}
    \semTravMap_2
        & \text{if } \ptStmt{\symbState_1} \in \Dom{\getAdviceName}\\
    \semTravMap_1
        & \text{otherwise}\\
\end{array}
\right.\\
\qquad \text{where } \semTravMap_2 = \exeAdvice{\getAdvice{\ptStmt{\symbState_1}}}{\ptExprs{\symbState_1}}{\srcSymbs}{\semTravMap_1}\\
\qquad \text{and } \exeAdvice{\argAdvice}{\argExprs}{\srcSymbs}{\semTravMap_1} \text{ executes advice code } \argAdvice \text{ for the expressions } \argExprs \text{, the}\\
\qquad \text{source annotation } \srcSymbs \text{(\Cref{sec:staticAspects:expert}), and the traversal map } \semTravMap_1 \text{, then returns the resulting traversal map}\\

\hline
\semIsFixpoint{\semOldAnnot}{\semTravMap} \eqdef
\left(
\begin{array}{r}
    \Dom{\semTravMap} \subseteq \Dom{\semOldAnnot} \land \forall \ptAspectName \in \Dom{\semTravMap}: \\
    \semTravMap(\ptAspectName) = \semOldAnnot(\ptAspectName)
\end{array}
\right)\\

\hline
\semMergeMaps{\semTravMaps_1}
\eqdef 
\left\{
\begin{array}{ll}
    \element{\semTravMaps_1}
        & \text{if } \card{\semTravMaps_1} = 1\\
    \semMerge{\semMergeMaps{\semTravMaps_2}}{\semTravMap}
        & \text{if } \card{\semTravMaps_1} \ge 2\\
\end{array}
\right.\\
\qquad \text{where } \semTravMap \in \semTravMaps_1 \text{ and } \semTravMaps_2 = \semTravMaps_1 \setminus \set{\semTravMap}{}\\

\hline
\end{array}
$$

 \caption{Auxiliary functions for the definition of function $\semVisitName$ in \Cref{table:sable:traversal}}
\label{table:sable:traversalAux}
\end{figure*}

In \Cref{table:sable:traversal}, $\semTransitionName$ determines the next state to visit, based on the number of children.
In case of a symbolic state corresponding to a loop statement, $\semLoopBranchingName$ repeats the traversal of the loop body until a fixpoint is reached.
When the traversal reaches an instance of conditional branching (e.g., an \labelIf statement), the current traversal map is used by $\semCondBranchingName$ to independently traverse the corresponding branches (the order does not matter). Then, traversal maps from branches that did not reach the end of the procedure are merged at the join point of the conditional branching to continue the traversal.
The output of the $\semVisitName$ function is a pair $\tuple{\symbState, \semTravMap}$, where $\symbState$ is the visited symbolic state (used to know the previous state in $\semCondBranchingName$) and $\semTravMap$ is the traversal map updated after visiting $\symbState$.
If a static aspect value $\semTravMap(\ptAspectName)$ does not satisfy the type of $\ptAspectName$ obtained from the $\aspectAspectType$ keyword (\Cref{sec:staticAspects:syntax}), then an error is raised.

In \Cref{table:sable:traversalAux},
$\elementName$ takes a singleton as input and returns its element,
$\ptAnnotateName$ updates the symbolic state annotation by static aspect values from the traversal map,
$\ptUpdEnterLoopName$ updates the $\ptEnterLoopName$ attribute of the symbolic state by the provided Boolean,
$\semWeaveName$ first executes the advice code corresponding to the
current statement label and then returns the resulting traversal map,
$\semIsFixpointName$ determines if a fixpoint was reached by the traversal at a loop statement by comparing the current traversal map with the one obtained the last time the statement was visited,
and $\semMergeMapsName$ is recursively called to merge any number of traversal maps.
The $\ptAnnotate{\symbState_1}{\semTravMap}$ and
$\ptUpdEnterLoop{\symbState_1}{b}$ functions in \Cref{table:sable:traversalAux} return a symbolic state $\symbState_2$ and replace $\symbState_1$ by $\symbState_2$ in the SCFG.
The $\ptAnnotateName$ function only affects static aspects of the symbolic state annotation that are present in the traversal map; in particular, static aspects from previous traversals are not changed.

Since several paths (e.g., from \labelReturn or \labelRaise statements) could lead to $\symbStateEnd$, the last symbolic state of the SCFG (\Cref{sec:scfg}), but they are not introduced by a branching statement (hence, they lack a join point like an \labelEndIf symbolic state), we introduced an exception in the definition of $\semVisitName$ for $\semTravMap_2$ (\Cref{table:sable:traversal}) to ensure that, at the end of the traversal, information from these different branches are merged at $\semEndPt$ as at any other join point.
Then, in \Cref{table:initTraversal}, after the execution of $\semVisitName$, function $\semWeaveName$ is called (\Cref{line:initTraversal:weave}) to obtain the final traversal map $\semTravMap$, which is used by function $\ptAnnotateName$ (\Cref{line:initTraversal:annotate}) to annotate $\symbStateEnd$.

Finally, \aspectTool uses two variables to gather alarm information during an SCFG traversal:
(1) $\semStep$ is an integer representing the current number of steps
that were performed so far by the traversal; 
(2) $\getAlarmsName$ is a map storing information about the raised
alarms\footnote{More precisely, $\getAlarmsName$  maps each symbolic state of the SCFG to a map between step numbers when an alarm was triggered for this state and a map between aspect names and aspect values that triggered the alarm.}.
At the beginning of the traversal, $\semStep$ is initialized to $0$ (\Cref{line:initTraversal:step}) and $\getAlarmsName$ to the empty map (\Cref{line:initTraversal:alarms}).
During the traversal, $\semStep$ is incremented each time the $\semWeaveName$ function is entered and, at the end of $\semWeaveName$, $\getAlarmsName$ is updated based on the alarm triggers $\getTriggersName$ (\Cref{sec:staticAspects:syntax}) as follows: for each $\ptAspectName \in \Dom{\semTravMap}$, if $\semTravMap(\ptAspectName) \in \getTriggers{\ptAspectName}$, then we have $\getAlarmsName(\symbState)(\semStep)(\ptAspectName) = \semTravMap(\ptAspectName)$.
At the end of the traversal, the alarms are returned (\Cref{line:initTraversal:return}) to generate the security report.

\subsection{Traversal Example}
\label{sec:staticAspects:example}

\begin{figure}[t]
\centering
\includegraphics[width=1.0\linewidth]{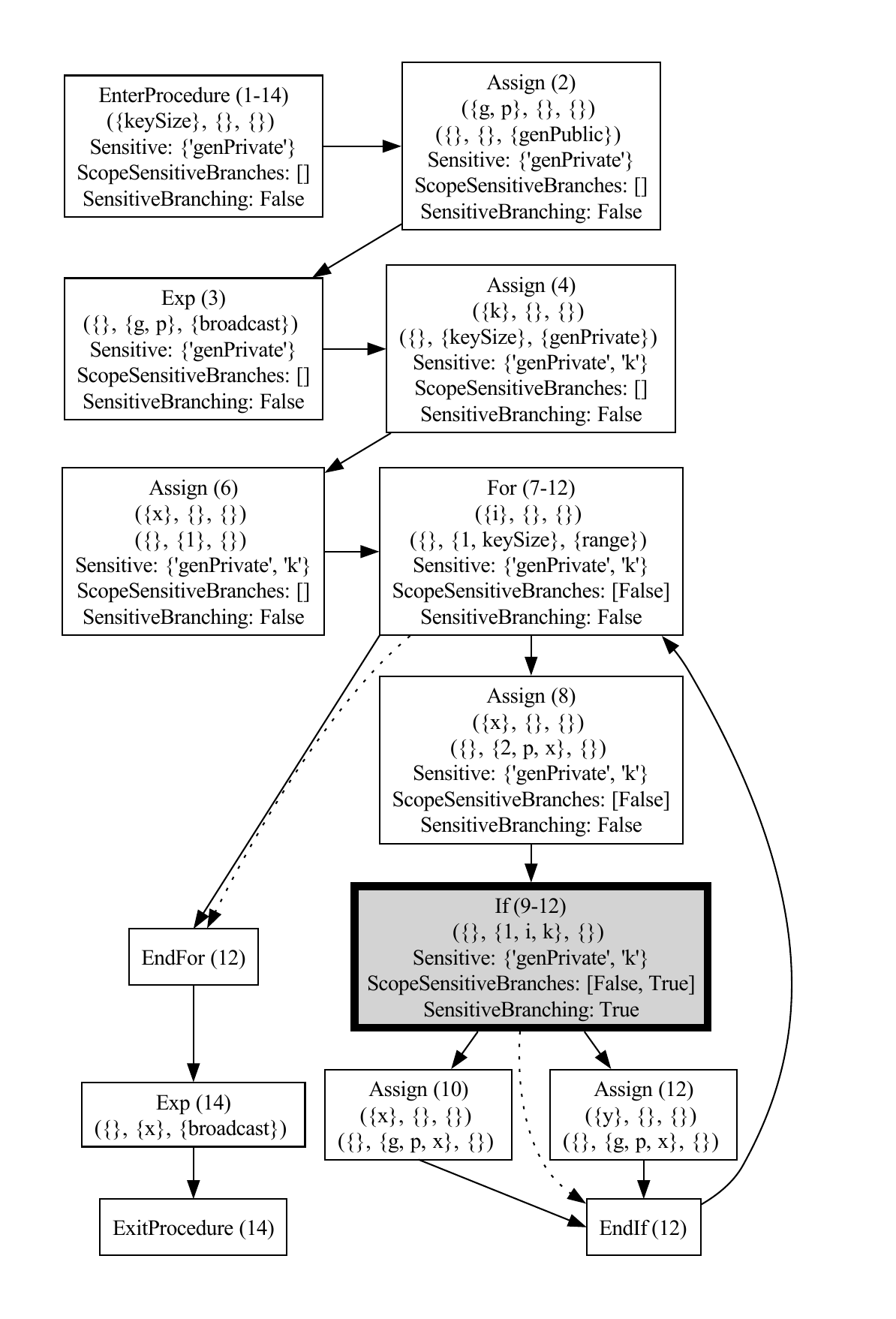}
\caption{First steps of the \travSensitive{} traversal, until the first alarm is raised.}
\label{fig:scfg:sensitive:firstAlarm}
\end{figure}

\begin{figure}[t]
\centering
\includegraphics[width=1.0\linewidth]{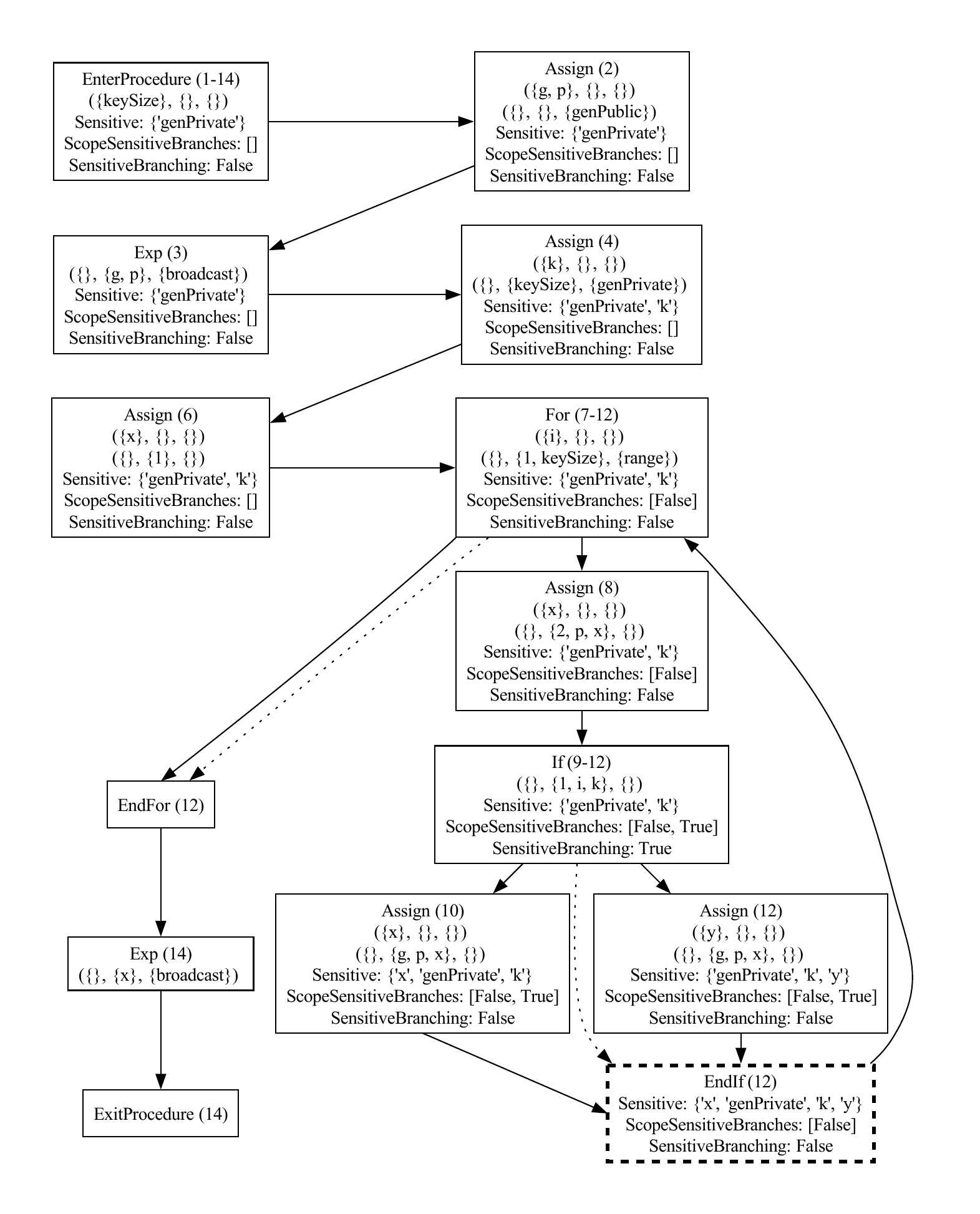}
\caption{\travSensitive{} traversal of an instance of sensitive conditional branching.}
\label{fig:scfg:sensitive:firstLoop}
\end{figure}

\begin{figure}[t]
\centering
\includegraphics[width=1.0\linewidth]{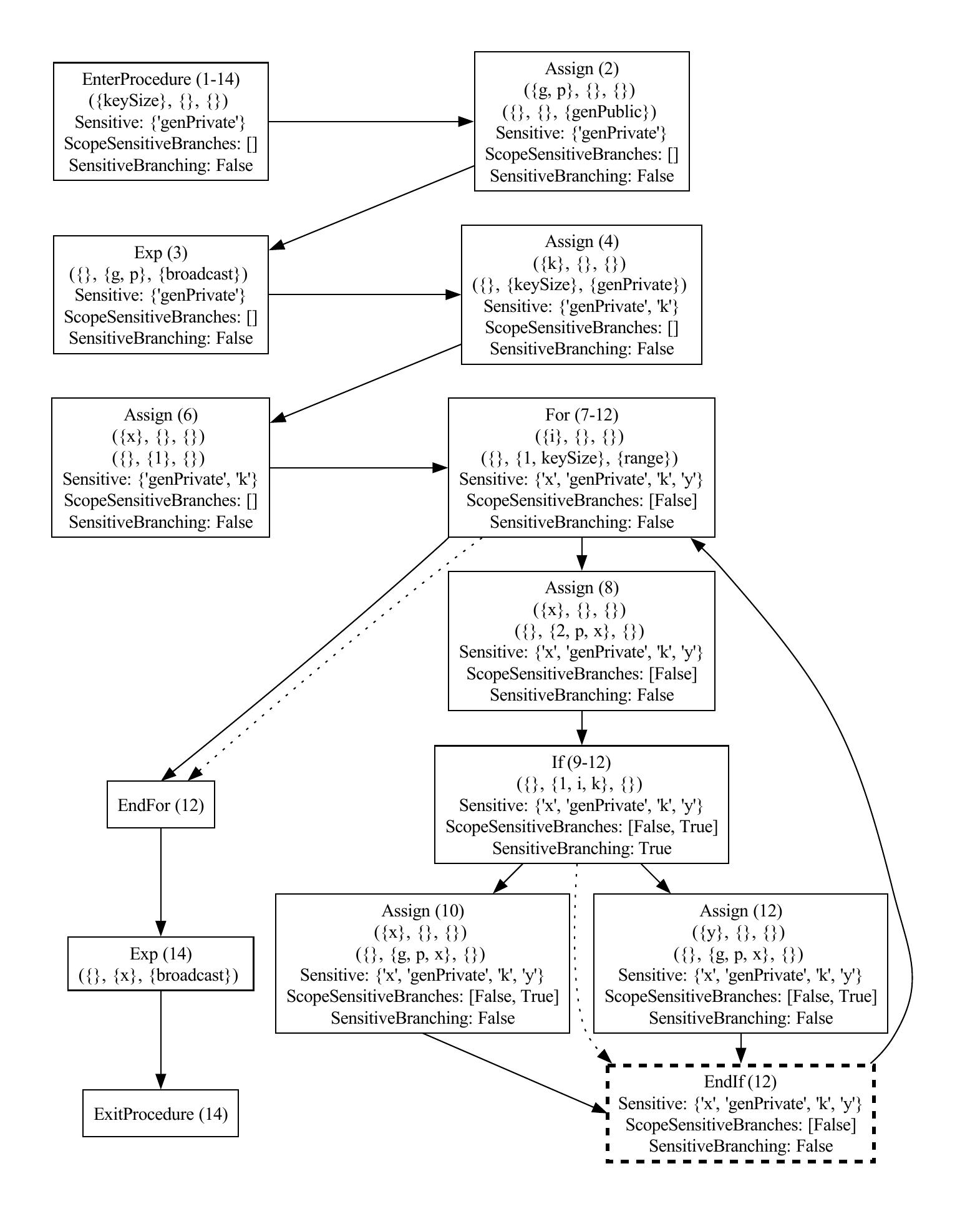}
\caption{Second \travSensitive{} traversal of the loop body.}
\label{fig:scfg:sensitive:secondLoop}
\end{figure}

\begin{figure}[t]
\includegraphics[width=1.0\linewidth]{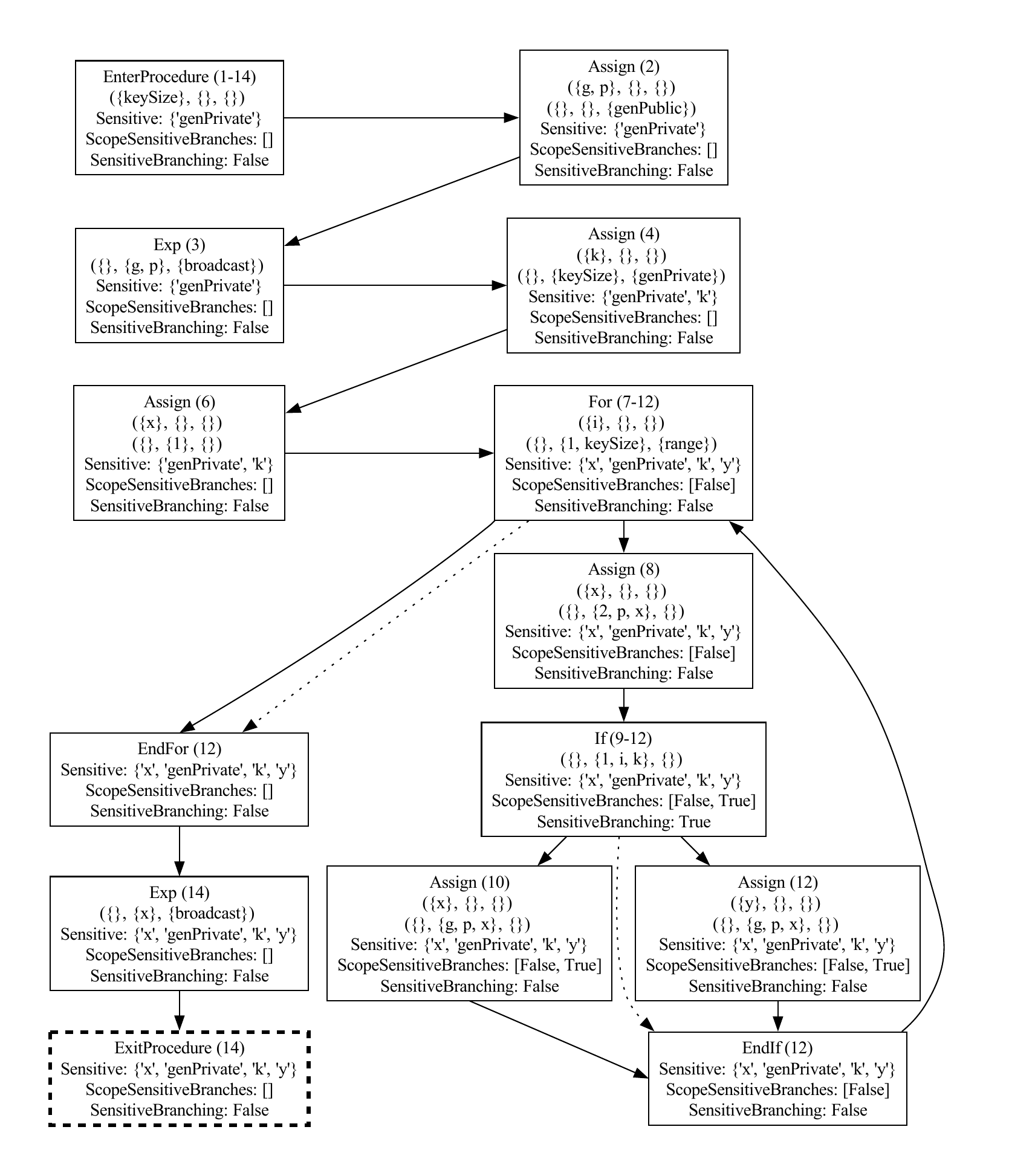}
\caption{End of the \travSensitive{} traversal.}
\label{fig:scfg:sensitive:ExitProcedure}
\end{figure}

Recall that the initial SCFG of the running example is shown in
\Cref{fig:code:runningExampleSCFG}, while the \travSensitive{}
traversal is defined in \Cref{fig:sad:travSensitive}.
Since this traversal does not specify a custom merging strategy, the default merging strategy is used (\Cref{fig:sad:merge}).
\Cref{fig:scfg:sensitive:firstAlarm,fig:scfg:sensitive:firstLoop,fig:scfg:sensitive:secondLoop,fig:scfg:sensitive:ExitProcedure} show the SCFG of the running example annotated with static aspects (\Cref{sec:staticAspects:annotations}) during different steps of the \travSensitive{} traversal.
In these figures, a box with a thick outline indicates the currently
visited symbolic state; grey-filled boxes indicate that an alarm is
raised while boxes with a dashed outline mean that an alarm is not raised.
We remind the reader that the number between parentheses after the statement label is the location of the corresponding statement in the source code (\Cref{sec:symbStates}).


We first detail in \Cref{fig:scfg:sensitive:firstAlarm} the first steps of the \travSensitive{} traversal.
The traversal starts at the \labelEnterProcedure statement (\Cref{line:initTraversal:visit} of \Cref{table:initTraversal}).
The advice code at \Cref{line:travSensitive:EnterProcedure} of \Cref{fig:sad:travSensitive} is executed to initialize the values of the static aspects; in particular, the \aspectSensitive set is initialized using the source symbols (\Cref{sec:staticAspects:expert}), in our example \pyCode{genPrivate}.
The next four steps correspond to expression or assignment statements.
Notably, at \Cref{line:runningExample:source} of the source code, \pyCode{k} becomes sensitive.
The next step corresponds to a \labelFor statement: the advice code at \Cref{line:travSensitive:For} of \Cref{fig:sad:travSensitive} is executed.
Since $\ptEnterLoop{\symbState}$ at the current state $\symbState$ was initialized to $\algoTrue$, the \aspectScopeSensitiveBranches stack is updated.
Since no sensitive variable or function is read, $\algoFalse$ is pushed to the stack.
Moreover, since a fixpoint has not been reached yet  $\ptEnterLoop{\symbState}$ becomes $\algoFalse$ and the traversal proceeds to the loop body.
Finally, after computing the squaring at \Cref{line:runningExample:square} of the source code, the last visited state in \Cref{fig:scfg:sensitive:firstAlarm} is the conditional branching at \Cref{line:runningExample:sensitBranching}.
Since the read variable \pyCode{k} is sensitive, this occurrence of
branching is sensitive; this leads to $\algoTrue$ being  pushed to the stack \aspectScopeSensitiveBranches and an alarm being raised.

We now detail the traversal of the rest of the loop body in \Cref{fig:scfg:sensitive:firstLoop}.
Since this is an occurrence of conditional branching, each branch is explored with the current traversal map; for instance, so far, only \pyCode{genPrivate} is sensitive.
The \srcThen{} branch corresponds to the assignment statement at \Cref{line:runningExample:indirect}.
Since \aspectScopeSensitiveBranches indicates the assignment is in the scope of a sensitive branching, there is an indirect information flow from a sensitive symbol (here, \pyCode{k}) to the written variable (here, \pyCode{x}), which thus becomes sensitive.
Similarly, the \srcElse{} branch corresponds to the assignment statement at \Cref{line:runningExample:dummy}, where \pyCode{y} becomes sensitive.
The last state of \Cref{fig:scfg:sensitive:firstLoop} is the \labelEndIf join point, where traversal maps from both branches are merged, with both \pyCode{x} and \pyCode{y} becoming sensitive.


After the first traversal of the loop body, the next state in \Cref{fig:scfg:sensitive:secondLoop} corresponds to the \labelFor statement.
Since \pyCode{x} and \pyCode{y} were not sensitive the last time this state was visited, the traversal proceeds again to the loop body.
Again, the occurrence of conditional branching at
\Cref{line:runningExample:sensitBranching} is sensitive, meaning that its visit raises an alarm.
Then, the following steps are similar to \Cref{fig:scfg:sensitive:firstLoop}, except that \pyCode{x} and \pyCode{y} are now sensitive until the join point.

Then, the next state in \Cref{fig:scfg:sensitive:ExitProcedure} corresponds to the \labelFor statement.
This time, the static aspect values in the traversal map match the
previous annotation at the symbolic state $\symbState$: this indicates
that a fixpoint has been reached.
Thus, $\ptEnterLoop{\symbState}$ becomes $\algoTrue$ and the \labelEndFor state is visited.
Then, the traversal proceeds until the end of the procedure without raising new alarms.

\subsection{Vulnerability Alarms}
\label{sec:staticAspects:alarms}

The alarms triggered during an SCFG traversal are stored in the map $\getAlarmsName$ (\Cref{sec:staticAspects:traversal}).
$\Dom{\getAlarmsName}$ contains all the visited symbolic states
$\symbState$ that triggered an alarm; the program locations denoted by
$\ptLoc{\symbState}$ (\Cref{sec:symbStates}) enable \aspectTool to locate the detected vulnerabilities in the analyzed source code.
Moreover, for each symbolic state $\symbState \in \Dom{\getAlarmsName}$, $\Dom{\getAlarmsName(\symbState)}$ contains the steps of the SCFG traversal where an alarm was triggered at $\symbState$.
In our running example, this occurred twice (at steps 8 and 14 in \Cref{sec:staticAspects:example}) for the \labelIf statement at \Cref{line:runningExample:sensitBranching} of the source code and once (at step 35 in Appendix~A) when the \pyCode{broadcast} sink is called with the sensitive variable \pyCode{x} at \Cref{line:runningExample:violation}.
\aspectTool reports the detected vulnerability locations as well as the static aspect names and values involved in each alarm.
\Cref{fig:secReport} shows \aspectTool's textual output for our running example.
Knowing the involved steps enables the security expert to investigate why an alarm was triggered.
For instance,
\Cref{line:secReport:firstAlarm} indicates that aspect \aspectSensitBranching{} became $\algoTrue$ at step 8.
The expert can investigate the SCFG annotation at this step (\Cref{fig:scfg:sensitive:firstAlarm}) and the steps before to determine the evolution of the traversal map and  what lead to the vulnerability detection.

\begin{figure}[t]
\centering
\begin{minipage}{0.8\linewidth}
\begin{lstlisting}[language=myPython,escapechar=@,basicstyle={\scriptsize \ttfamily}]
<StaticAspectAnalysis (id 4508616112):
   - procedure = source_code.py:runningExample
   - source_annotation = source_annotation.json
   - static_aspect_definition = static_aspect_definition.sable
   - alarms:
      - line 18
         SensitiveBranching = True at step 8@\label{line:secReport:firstAlarm}@
         SensitiveBranching = True at step 14@\label{line:secReport:secondAlarm}@
      - line 29
         ConfidentialityViolation = True at step 35@\label{line:secReport:thirdAlarm}@>
\end{lstlisting}
 \end{minipage}
\caption{Vulnerability alarms triggered during the SCFG traversal of the running example (\Cref{fig:runningExample}).}
\label{fig:secReport}
\end{figure}

 \section{Performance Metrics}
\label{sec:appendix:metrics}

We detail the computation of the metrics presented in \Cref{sec:metrics}.
For each commit in our benchmark fixing a vulnerability $\vuln$, we denote
$\truthBefore_{\vuln}$ the set of ground-truth vulnerability locations in the code before the commit and $\truthAfter_{\vuln}$ the set of the corresponding locations after the commit.
\aspectTool and the baseline tools are binary classifiers for security, aiming to detect the presence or absence of vulnerabilities in the source code.
Their primary output is a set of vulnerability locations, i.e., line numbers in the source code where vulnerabilities were detected.
For each tool, we denote
$\detectBefore_{\vuln}$ the set of vulnerability locations detected before commit and
$\detectAfter_{\vuln}$ the set of vulnerability locations detected after commit
For evaluation purposes, we consider only code locations that are either detected or in the ground-truth sets, excluding the vast majority of irrelevant locations from the analysis.

\begin{figure*}[t]
\centering
\begin{tikzpicture}[scale=0.9]
\begin{scope}[local bounding box=before]
        \draw (-8.0, 0) ellipse (1.5cm and 0.75cm);
        \draw (-6.0, 0) ellipse (1.5cm and 0.75cm);
        \node (FN) at (-8.5, 0) {FN};
        \node (TP) at (-7.0, 0) {TP};
        \node (FPbefore) at (-5.5, 0) {FP};
    \end{scope}
    \node (truthBefore) at (before.north west) {\footnotesize ground-truth\textcolor{white}{--}};
    \node (detectBefore) at (before.north east) {\footnotesize\ detection};
    \node (beforeCommit) at (before.north) {\footnotesize \begin{tabular}{c}before commit\\[0.5cm]\end{tabular}};
\begin{scope}[local bounding box=after]
        \draw (0.0, 0.0) ellipse (1.5cm and 0.75cm);
        \fill[white] (2.0, 0.0) ellipse (1.5cm and 0.75cm);
        \draw (2.0, 0.0) ellipse (1.5cm and 0.75cm);
        \node (TN) at (-0.5, 0.0) {TN};
        \node (FPafter) at (2.0, 0.0) {FP};
    \end{scope}
    \node (truthAfter) at (after.north west) {\footnotesize ground-truth\textcolor{white}{--}};
    \node (detectAfter) at (after.north east) {\footnotesize\ detection};
    \node (afterCommit) at (after.north) {\footnotesize \begin{tabular}{c}after commit\\[0.5cm]\end{tabular}};
\end{tikzpicture} \caption{Set representation of true positives (TPs), false negatives (FNs), false positives (FPs), and true negatives (TNs) before commit = vulnerable (left) and after commit = secure (right).}
\label{fig:confusion_matrix}
\end{figure*}

\begin{table*}[t]
\footnotesize
\centering
\caption{Confusion matrix for the strict-location (left) and the relaxed-location (right) methods.}
\label{tab:confusionMatrix}
$$
\begin{array}{|r@{}l|@{}r@{}l|}
\hline
    \TP^{\locStrict}_{\vuln} \eqdef
        &{} \card{\truthBefore_{\vuln} \cap \detectBefore_{\vuln}}
    &\TP^{\locRelax}_{\vuln} \eqdef
        &{} \min(\card{\truthBefore_{\vuln}}, \card{\detectBefore_{\vuln}})\\
    \FN^{\locStrict}_{\vuln} \eqdef
        &{} \card{\truthBefore_{\vuln} \setminus \detectBefore_{\vuln}}
    &\FN^{\locRelax}_{\vuln} \eqdef
        &{} \max(\card{\truthBefore_{\vuln}} - \card{\detectBefore_{\vuln}}, 0)\\
    \TN^{\locStrict}_{\vuln} \eqdef
        &{} \card{\truthAfter_{\vuln} \setminus \detectAfter_{\vuln}}
    &\TN^{\locRelax}_{\vuln} \eqdef
        &{} \max(\card{\truthAfter_{\vuln}} - \card{\detectAfter_{\vuln}}, 0)\\
    \FP^{\locStrict}_{\vuln} \eqdef
        &{} \card{\detectBefore_{\vuln} \setminus \truthBefore_{\vuln}} + \card{\detectAfter_{\vuln}}
    &{}\FP^{\locRelax}_{\vuln} \eqdef
        &{} \max(\card{\detectBefore_{\vuln}} - \card{\truthBefore_{\vuln}}, 0) + \card{\detectAfter_{\vuln}}\\
\hline
\end{array}
$$
\end{table*}
 
We illustrate the TPs, FNs, FPs, and TNs in \Cref{fig:confusion_matrix} and we determine the \emph{confusion matrix} (i.e.,  the number of TPs, FNs, FPs, and TNs) for each vulnerability $\vuln$ in \Cref{tab:confusionMatrix}, respectively in the left part for the \emph{strict-locations} method and in the right part for the relaxed-locations method, where $\card{S}$ denotes the cardinality of set $S$.

Independent of the method (strict or relaxed locations) used to determine the confusion matrix, TPs and TNs are correctly classified, while FPs and FNs are errors of the binary classifier.
FPs are referred to as type 1 errors, while FNs are called type 2 errors.
A classifier is \emph{complete} if every vulnerability is detected, while it is \emph{sound} if every detection comes from a vulnerability~\cite{Mey19}.
In other words, a classifier with no type 1 error is sound, while a classifier with no type 2 error is complete.
While it is easy to remove either type 1 errors (by reporting no detection) or type 2 errors (by reporting all code locations as vulnerable), it is usually not possible to achieve both simultaneously.

In a security context, Type 1 errors are false alarms, i.e., no vulnerability was present, but a detection was reported anyway.
Too many false alarms are a problem, since developers do not have time to sift through all the results, making the classifier useless in practice.
Type 2 errors are missed vulnerabilities in the source code that go undetected.
In the wild, those missed vulnerabilities would not be fixed and could
then be exploited by attackers.
Thus, in a security context, type 2 errors are more serious than type 1 errors.
So, there is a trade-off between completeness and soundness; in a security context, completeness is more important, but one cannot be entirely sacrificed to the other.

To estimate this trade-off, we consider a quantitative version of these (qualitative) properties~\cite{Mey19}.
Since soundness expresses that each detection should come from a vulnerability, one can quantify soundness as the number of detected vulnerabilities divided by the number of detections, i.e., \emph{precision}, computed as $\precision{\TP}{\FP} \eqdef \frac{\TP}{\TP + \FP}$ if $\TP + \FP > 0$.
Since we focus on vulnerability-fixing commits, we know there are vulnerability locations to detect.
Hence, in the absence of detection ($\TP + \FP = 0$), precision is assumed to be zero.

Nevertheless, soundness can also be expressed in a logically equivalent way as: each non-vulnerable location should not be detected.
In that case, soundness would be quantified as the number of non-vulnerable locations that were correctly not detected, divided by the number of non-vulnerable locations, i.e., the true negative rate, also known as \emph{specificity}, computed as
$\specificity{\TN}{\FP} \eqdef \frac{\TN}{\FP + \TN}$.
Note that, since the code after the commit is secure, there are always
non-vulnerable code locations ($\FP + \TN > 0$).

Similarly, completeness expresses that each vulnerability should be detected. One can quantify completeness as the number of detected vulnerabilities divided by the number of vulnerable locations, i.e., the true positive rate, also known as \emph{sensitivity} when used with specificity or \emph{recall} when used with precision, and computed as
$\sensitivity{\TP}{\FN}
\eqdef \frac{\TP}{\TP + \FN}$.
Note that, since the code before the commit is vulnerable, there are
always vulnerable code locations ($\TP + \FN > 0$).

In short, specificity and precision quantify performance in terms of type 1 errors, while sensitivity (also called recall) quantifies performance in terms of type 2 errors.

Software engineering researchers have adopted the precision, recall, and $\Fone$ (i.e., the harmonic mean of precision and recall) metrics from the information retrieval community.
These metrics do not consider TNs since negative cases are only taken into account to report FPs, which makes sense when facing a single-class problem, e.g., the retrieval of relevant pages from the Web when the number of irrelevant pages correctly not retrieved is vast, cannot be determined, and is not of interest.
However, these metrics may have unintended consequences when applied to two-class problems, for example, where knowing that a code location was correctly classified as non-vulnerable is crucial, particularly for project resources and software quality.
For instance, a review of software defect prediction studies concluded that 21.95\% of comparisons done in these studies would have had a different result if $\Fone$ was replaced by an unbiased performance metric~\cite{YS21}, e.g., $A$ is better than $B$ with $\Fone$, while $B$ is better than $A$ with the unbiased metric.

Moreover, unlike sensitivity/recall and specificity, precision depends on the \emph{prevalence}, i.e., the proportion of actual positive cases (in our case, vulnerable locations) in the dataset.
For instance, in a balanced dataset (prevalence = 50\%), one may obtain the following results: $\TP = 9$, $\FN = 1$, $\FP = 2$, and $\TN = 8$, leading to sensitivity/recall = 90\%, specificity = 80\%, and precision = 82\%.
Then, by multiplying the number of non-vulnerable locations by 9 to obtain $\TP = 9$, $\FN = 1$, $\FP = 18$, and $\TN = 72$, i.e., a highly-unbalanced dataset (prevalence = 10\%), sensitivity/recall and specificity would not change, but precision would drop to 33\% only because of the large number of FPs.
This is an issue, since datasets are usually not balanced, e.g., a study on software defect prediction reported prevalence rates between 6.94\% and 32.29\%~\cite{WY13}.

In other words, precision captures a mixture of contextual information and tool effectiveness.
In our study, we focus on tool effectiveness using sensitivity and specificity, as they do not depend on prevalence (and are therefore not affected by imbalanced datasets). They quantify completeness and soundness by utilizing all cases (TPs, FNs, FPs, and TNs) encountered in our binary classification problem.
Nevertheless, for the sake of comparison with other studies, we report precision in addition to sensitivity and specificity results.

Thus, to express the global performance of each tool on the considered benchmark, we compute $\sensitivity{\TP}{\FN}$ and $\specificity{\TN}{\FP}$ where,
for $M$ being $\TP$, $\FN$, $\TN$, or $\FP$,
$M \eqdef \sum_{\vuln \in \Vulns}  M_{\vuln}$
and $M_{\vuln}$ denotes the score obtained for the performance metric
$M$
and vulnerability $\vuln$.
Since each case is the sum of program locations per vulnerability, vulnerabilities with more program locations carry more weight than those with fewer.
This is not necessarily an issue, since this number is usually small
for each vulnerability, and program locations reflect the amount of
work required to sift through the tool results in the case
of FPs or to fix the vulnerability in the case of TPs.

Regarding the \emph{per-vulnerability} analysis (\Cref{sec:metrics}), for each considered metric $\metricRandomVar$, we
denote by $\metricRandomVar_i$ the discrete random variable taking values from the metric scores obtained for tool $i$.
After applying tool $i$ to all vulnerabilities and computing the metrics, we have a sample of values for $\metricRandomVar_i$ that we can compare with samples from other tools.
To compare two $\metricRandomVar_1$ and $\metricRandomVar_2$ samples,
we perform a Wilcoxon signed-rank test---a commonly used statistical test in software engineering data analysis~\cite{Dem06,AB14}--- which is:
\begin{itemize}
\item \emph{non-parametric}, i.e., without hypothesis on the data probability distribution,
\item and \emph{paired}, i.e., each value $m_1(\vuln)$ from the $\metricRandomVar_1$ sample is compared with the value $m_2(\vuln)$ from the $\metricRandomVar_2$ sample obtained for the same vulnerability $\vuln$.
\end{itemize}

This test provides a p-value indicating how likely the observation of the samples is, assuming the null hypothesis $\metricRandomVar_1 = \metricRandomVar_2$. If $\pValue \le 0.05$, we reject the null hypothesis in favor of the alternative hypothesis $\metricRandomVar_1 \neq \metricRandomVar_2$.
To assess practical significance, we consider a metric for effect size.
Metrics for this test are often defined in terms of the positive-rank sum $\tPlus$ and the negative-rank sum $\tMinus$~\cite{Dem06,Ker14}.
More precisely, let $d_\vuln$ denote the difference $m_1(\vuln) - m_2(\vuln)$ between $\vuln$ values taken from the $\metricRandomVar_1$ and $\metricRandomVar_2$ samples.
Then, the absolute differences $\abs{d_\vuln}$ are sorted in increasing order and associated with a rank, denoted $\rank{d_\vuln}$.
The positive-rank sum $\tPlus$ and the negative-rank sum $\tMinus$ are defined as:
$$
\begin{array}{r@{}l}
    \tPlus \eqdef
        &{} \sum_{\vuln \in \Vulns, d_\vuln > 0} \rank{d_\vuln} + \frac{1}{2}\sum_{\vuln \in \Vulns, d_\vuln = 0} \rank{d_\vuln}\\
    \tMinus \eqdef
        &{} \sum_{\vuln \in \Vulns, d_\vuln < 0} \rank{d_\vuln} + \frac{1}{2}\sum_{\vuln \in \Vulns, d_\vuln = 0} \rank{d_\vuln}\\
\end{array}
$$
The effect size is $\effectSize \eqdef \frac{\tPlus}{\tPlus + \tMinus}$, so that it ranges from $0$ to $1$, where $\effectSize = 0$ indicates that $\metricRandomVar_1 < \metricRandomVar_2$ in every case and $\effectSize = 1$ indicates that $\metricRandomVar_1 > \metricRandomVar_2$ in every case.

 \section{\aspectTool's False Positive}
\label{sec:appendix:fp}

\begin{figure*}[t]
\centering
\begin{minipage}{0.95\linewidth}
\begin{lstlisting}[language=Python,escapechar=@,basicstyle={\scriptsize \ttfamily}]
class SessionRedirectMixin(object):
    def resolve_redirects(self, resp, req, stream=False, timeout=None, verify=True, cert=None, proxies=None):
        (...)
        headers = prepared_request.headers@\label{line:falsePositive:binding}@
        (...)
        # If we get redirected to a new host, we should strip out any authentication headers.
        original_parsed = urlparse(resp.request.url)
        redirect_parsed = urlparse(url)
        if (original_parsed.hostname != redirect_parsed.hostname and 'Authorization' in headers):@\label{line:falsePositive:conditional}@
            del headers['Authorization']@\label{line:falsePositive:sanitize}@
        (...)
        resp = self.send(@\label{line:falsePositive:send}@prepared_request, stream=stream, timeout=timeout, verify=verify, cert=cert, proxies=proxies, allow_redirects=False,)
        (...)
\end{lstlisting}
\end{minipage}\caption{Relevant lines from the source code for CVE-2014-1829 (the
  only false positive reported by \aspectTool).}
\label{fig:code:falsePositive}
\end{figure*}

\aspectTool reported only one false positive in the fix of vulnerability CVE-2014-1829 (\Cref{fig:code:falsePositive}).
Indeed, the sanitization of the authorization field occurred only in a conditional branch (\Cref{line:falsePositive:sanitize}), hence could not be statically assumed in the rest of the source code.
Moreover, to detect that the request was sanitized before being sent
(\Cref{line:falsePositive:send}), such sanitization had to be
propagated to the \pyCode{prepared\_request} variable due to \python
dynamic binding (\Cref{line:falsePositive:binding}).

\end{document}